\documentclass[lettersize,journal]{IEEEtran}
\IEEEoverridecommandlockouts

\usepackage{cite}
\usepackage{amsmath,amssymb,amsfonts}
\usepackage{algorithmic}
\usepackage{diagbox}
\usepackage{algorithm}
\usepackage{multirow}
\usepackage{caption}
\usepackage{graphicx}
\usepackage{subfigure}
\usepackage{textcomp}
\usepackage{xcolor}
\usepackage{color}
\usepackage{booktabs}
\usepackage{wrapfig}
\usepackage{comment}
\usepackage{multicol}
\def\BibTeX{{\rm B\kern-.05em{\sc i\kern-.025em b}\kern-.08em
		T\kern-.1667em\lower.7ex\hbox{E}\kern-.125emX}}
\begin{document}
\newtheorem{definition}{\it Definition}
\newtheorem{theorem}{\bf Theorem}
\newtheorem{lemma}{\it Lemma}
\newtheorem{corollary}{\it Corollary}
\newtheorem{remark}{\it Remark}
\newtheorem{example}{\it Example}
\newtheorem{case}{\bf Case Study}
\newtheorem{assumption}{\it Assumption}
\newtheorem{property}{\it Property}

\newtheorem{proposition}{\it Proposition}

\newcommand{\hP}[1]{{\boldsymbol h}_{{#1}{\bullet}}}
\newcommand{\hS}[1]{{\boldsymbol h}_{{\bullet}{#1}}}

\newcommand{\ba}{\boldsymbol{a}}
\newcommand{\baq}{\overline{q}}
\newcommand{\bA}{\boldsymbol{A}}
\newcommand{\bb}{\boldsymbol{b}}
\newcommand{\bB}{\boldsymbol{B}}
\newcommand{\bc}{\boldsymbol{c}}
\newcommand{\bp}{\boldsymbol{p}}
\newcommand{\bcO}{\boldsymbol{\cal O}}
\newcommand{\be}{\boldsymbol{e}}
\newcommand{\bh}{\boldsymbol{h}}
\newcommand{\bH}{\boldsymbol{H}}
\newcommand{\bl}{\boldsymbol{l}}
\newcommand{\bm}{\boldsymbol{m}}
\newcommand{\bn}{\boldsymbol{n}}
\newcommand{\bo}{\boldsymbol{o}}
\newcommand{\bO}{\boldsymbol{O}}
\newcommand{\bq}{\boldsymbol{q}}
\newcommand{\br}{\boldsymbol{r}}
\newcommand{\bR}{\boldsymbol{R}}
\newcommand{\bs}{\boldsymbol{s}}
\newcommand{\bS}{\boldsymbol{S}}
\newcommand{\bT}{\boldsymbol{T}}
\newcommand{\bu}{\boldsymbol{u}}
\newcommand{\bv}{\boldsymbol{v}}
\newcommand{\bw}{\boldsymbol{w}}
\newcommand{\bx}{\boldsymbol{x}}
\newcommand{\by}{\boldsymbol{y}}

\newcommand{\balpha}{\boldsymbol{\alpha}}
\newcommand{\bbeta}{\boldsymbol{\beta}}
\newcommand{\bomega}{\boldsymbol{\omega}}
\newcommand{\bOmega}{\boldsymbol{\Omega}}
\newcommand{\bphi}{\boldsymbol{\phi}}
\newcommand{\bvarpi}{\boldsymbol{\varpi}}
\newcommand{\bpi}{\boldsymbol{\pi}}
\newcommand{\bxi}{\boldsymbol{\xi}}

\newcommand{\cA}{{\cal A}}
\newcommand{\bcA}{\boldsymbol{\cal A}}
\newcommand{\cB}{{\cal B}}
\newcommand{\cE}{{\cal E}}
\newcommand{\cG}{{\cal G}}
\newcommand{\cH}{{\cal H}}
\newcommand{\bcH}{\boldsymbol {\cal H}}
\newcommand{\cK}{{\cal K}}
\newcommand{\cO}{{\cal O}}
\newcommand{\cR}{{\cal R}}
\newcommand{\cS}{{\cal S}}
\newcommand{\dcS}{\ddot{{\cal S}}}
\newcommand{\ds}{\ddot{{s}}}
\newcommand{\cT}{{\cal T}}
\newcommand{\cU}{{\cal U}}
\newcommand{\wt}[1]{\widetilde{#1}}

\newcommand{\mA}{\mathbb{A}}
\newcommand{\mE}{\mathbb{E}}
\newcommand{\mG}{\mathbb{G}}
\newcommand{\mR}{\mathbb{R}}
\newcommand{\mS}{\mathbb{S}}
\newcommand{\mU}{\mathbb{U}}
\newcommand{\mV}{\mathbb{V}}
\newcommand{\mW}{\mathbb{W}}

\newcommand{\uq}{\underline{q}}
\newcommand{\ubq}{\underline{\boldsymbol q}}

\newcommand{\red}[1]{\textcolor[rgb]{1,0,0}{#1}}
\newcommand{\gre}[1]{\textcolor[rgb]{0,1,0}{#1}}
\newcommand{\blu}[1]{\textcolor[rgb]{0,0,0}{#1}}

\title{Reasoning over the Air: A Reasoning-based Implicit Semantic-Aware Communication Framework}
\author{Yong~Xiao, \IEEEmembership{Senior~Member,~IEEE}, Yiwei Liao, Yingyu Li, Guangming Shi, \IEEEmembership{Fellow, IEEE}, H. Vincent Poor, \IEEEmembership{Life Fellow, IEEE}, Walid Saad, \IEEEmembership{Fellow, IEEE}, M\'erouane Debbah, \IEEEmembership{Fellow, IEEE}, and Mehdi Bennis, \IEEEmembership{Fellow, IEEE}

\thanks{*This work has been accepted at IEEE Transactions on Wireless Communications. Copyright may be transferred without notice, after which this version may no longer be accessible.

An abridged version of this paper was presented at the Proc. of the IEEE ICC Workshop on Data Driven Intelligence for Networks and Systems, Seoul, South Korea, May 2022\cite{XY2022ReasononAir}. The work of Yong Xiao was supported in part by the National Natural Science Foundation of China under Grant 62071193, in part by the  Major Key Project of Peng Cheng Laboratory under Grant PCL2023AS1-2, and in part by the Key Research and Development Program of Hubei Province of China under Grant 2021EHB015. The work of Yingyu Li was supported in part by the Major Key Project of Peng Cheng Laboratory under grant PCL2023AS1-2, in part by the National Natural Science Foundation of China under Grant 62301516, in part by the Key Research and Development Program of Hubei Province under Grant 2021EHB015, and in part by the National Key Research and Development Program of China under Grant 2022YFB2903201. The work of Guangming Shi was supported in part by the National Natural Science Foundation of China under Grant 62293483 and Grant 61976169, in part by the National Key Research and Development Program of China under grant 2019YFA0706604, and in part by the  Major Key Project of Peng Cheng Laboratory under Grant PCL2023AS1-2.  The work of H. Vincent Poor was supported in part by U.S National Science Foundation under Grants CNS-2128448, ECCS-2335876 and CNS-2225511. {\it (Corresponding author: Guangming Shi)}

%
%

Y. Xiao is with the School of Electronic Information and Communications at the Huazhong University of Science and Technology, Wuhan 430074, China, and also with the Peng Cheng Laboratory, Shenzhen, Guangdong 518055, China, and also with the Pazhou Laboratory (Huangpu), Guangzhou, Guangdong 510555, China (e-mail: yongxiao@hust.edu.cn).

Y. Liao is with the School of Electronic Information and Communications at the Huazhong University of Science and Technology, Wuhan 430074, China (e-mail: liao\_yiwei@hust.edu.cn).

Y. Li is with the School of Mech. Eng. and Elect. Inform. at the China University of Geosciences, Wuhan, China, 430074 (e-mail: liyingyu29@cug.edu.cn). 

G. Shi is with the Peng Cheng Laboratory, Shenzhen, Guangdong 518055, China, also with the School of Artificial Intelligence, the Xidian University, Xi’an, Shaanxi 710071, China, and also with  the Pazhou Laboratory (Huangpu), Guangzhou, Guangdong 510555, China (e-mail: gmshi@xidian.edu.cn). 

V. Poor is with the Department of Electrical and Computer Engineering, Princeton University, Princeton, NJ 08544 (e-mail: poor@princeton.edu).

W. Saad is with the Department of Electrical and Computer Engineering at Virginia Tech, Blacksburg, VA 24061 (e-mail: walids@vt.edu) and also with Cyber Security Systems and Applied AI Research Center, Lebanese American University, Beirut, Lebanon.

M. Debbah is with Khalifa University of Science and Technology, P O Box 127788, Abu Dhabi, UAE (email: merouane.debbah@ku.ac.ae) and also with CentraleSupelec, University Paris-Saclay, 91192 Gif-sur-Yvette, France

M. Bennis is with the University of Oulu, Oulu, Finland, 90014 (e-mail: mehdi.bennis@oulu.fi).\\

Code is available at \texttt{https://github.com/Yiwei-Liao/iSAC}
}
}

\maketitle
\begin{abstract}
Semantic-aware communication is a novel paradigm that draws inspiration from human communication focusing on the delivery of the meaning of messages. It has attracted significant interest recently due to its potential to improve the efficiency and reliability of communication and enhance users' quality-of-experience (QoE). Most existing works focus on transmitting and delivering the explicit semantic meaning that can be directly identified from the source signal. This paper investigates the implicit semantic-aware communication in which the hidden information, e.g., hidden relations, concepts and implicit reasoning mechanisms of users, that cannot be directly observed from the source signal must be recognized  and interpreted by the intended users.  To this end,  a novel implicit semantic-aware communication (iSAC) architecture is proposed for representing, communicating, and interpreting the implicit semantic meaning between source and destination users. A graph-inspired structure is first developed to represent the complete semantics, including both explicit and implicit, of a message. A projection-based semantic encoder is then proposed to convert the high-dimensional graphical representation of explicit semantics into a low-dimensional semantic constellation space for efficient physical channel transmission. To enable the destination user to learn and imitate the implicit semantic reasoning process of source user, a generative adversarial imitation learning-based solution, called G-RML, is proposed. Different from existing communication solutions, the source user in G-RML does not focus only on  sending as much of the useful messages as possible; but, instead, it tries to guide the  destination user to learn a reasoning mechanism to map any observed explicit semantics to the corresponding implicit semantics that are most relevant to the semantic meaning. By applying G-RML, we prove that the destination user can accurately imitate the reasoning process of the source user and automatically generate a set of implicit reasoning paths following the same probability distribution as the expert paths. Compared to the existing solutions, our proposed G-RML requires much less communication and computational resources and scales well to the scenarios involving the communication of rich semantic meanings consisting of a large number of concepts and relations. Numerical results show that the proposed solution achieves up to  92\% accuracy of implicit meaning interpretation. 
\end{abstract}


\section{Introduction}
Recent developments in communication systems witnessed a growing interest in human-oriented services and applications such as extended reality (XR), Tactile Internet, and connected vehicles, most of which are data-hungry and resource-consuming with high reliability and low latency requirements. The traditional data-oriented communication system is now viewed as one of the major obstacles for delivering quality-of-experience (QoE)-demanding services and applications to end-users. This motivates a novel paradigm, called {\em semantic communication},  which allows the meaning of messages to be identified and utilized during communication. Compared to the traditional systems, semantic communication allows communication participants including both information source and destination to exploit commonly-shared human knowledge and experience as well as syntax, semantics, and inference rules to assist the transportation and accurate delivery of the intended meaning. Recent observation suggests that semantic communication could significantly improve efficiency and reliability of communication, enhance users' QoE, and achieve smoother cross-protocol/domain communication\cite{XY2021SemanticCommMagazine,XY2022ITWSSC,XY2022LifeLong,strinati20216g,ZHG2022SemanticCommWCMagazine, XY2023CollaberativeJSAC}.


Most existing works on semantic communication focused on transporting the explicit semantic information, e.g., human-labeled objects or signal features that can be directly identified from the source signals (e.g, image, voice, and text signals) by adopting machine learning algorithms, especially the deep learning (DL)-based algorithms\cite{ZHG2022SemanticCommCL, xie2020lite}.  

In reality, the information communicated between users can be much more than just explicit information. For example, an image showing ``a child is riding a bicycle" consists of explicit objects, ``a child" and ``a bicycle". The relationship (``ride") between these two objects however cannot be directly recognized from the image. In another example introduced in \cite{XY2021SemanticCommMagazine} and \cite{bao2011towards}, a child sends a message to her father asking ``what is a Tweety". The key semantic element of this message ``Tweety" can have multiple interpretations including a smartphone App of a social media website, a canary bird, and a character in a cartoon TV show. To understand the exact meaning of the message, the receiver (the father), must be able to infer the implicit information, i.e., hidden information that cannot be observed from the signal itself, from the context and background of the child, e.g., if the child does not own a smartphone, it is less likely that the word ``Tweety" should be interpreted as a smartphone app.
	
From the above examples, we can observe that, in addition to the explicit information, the content of communication often consists of rich implicit information that is 
of critical importance in understanding and efficiently delivering the semantic meaning of messages at the intended destinations. Compared to the traditional explicit semantic information-based communication solutions, implicit semantic-aware communication has 
multiple unique benefits. First, the implicit 
relationships among concepts and ideas in the communication messages can be utilized to further reduce the redundancy of the messages transmitted . It is also helpful for recovering some missing or corrupted information at the destination user. Second, the implicit information may reflect the truly important meaning information, i.e., real intention, of (human) user's information and, thus, can be used for further enhancing communication QoE and avoid misinterpretation/misunderstanding. 
Finally, implicit semantics including the context and the reasoning mechanism can be used for inferring the meaning of unknown knowledge as well as discovering hidden relationships with previously learned concepts. This can be very useful for developing novel self-learning and adaptive solutions for supporting cognition and alerting of network abnormalities as well as autonomous decision-making under unknown scenarios.     

Despite these promising  potentials, implicit semantic-aware communication has been relatively unexplored in the literature due to the following challenges. 
%
%
First, 
the implicit meaning is generally difficult to represent, recognize, transport, and recover. In particular, the meaning of messages often involves  complex  relations and hidden inference mechanisms that 
cannot be directly identified from the source signal and therefore is very difficult to represent in a simple and comprehensive way.  Second, the implicit meaning can be closely related to user-related information such as users' background and personal preferences, and is hence difficult to estimate and infer. 
Finally, accurately recovering and evaluating the implicit meaning at the destination user is also challenging. 
Most existing works assume that the destination user can have a well-formulated decoding scheme that can directly output the recovered messages based on the input signals. In implicit semantic-aware communication, however, the destination user must jointly consider both the signal received from the channel as well as the background and personal preference of the source user to interpret the semantics of the message.  
A simple and effective solution that can seamlessly combine the received signal with personally relevant information for accurate interpretation of the semantic meaning involving implicit semantic information is currently missing. 

The main contribution of this paper is a novel implicit semantic-aware communication (iSAC) architecture for representing, modeling, communicating, and optimizing the interpretation of the message semantics involving both explicit and implicit meaning. We first propose a novel graph-inspired  representation of semantics, which includes three key components: {explicit semantics (objects and relations), implicit semantics (inffered information behind entities), and reasoning mechanism (user's reasoning rule and/or inference policy).} Different from existing knowledge graph-based solutions that rely on a pre-stored dictionary-based terms, our proposed solution allows biased personal understanding and preference of the source user to be sequantially learned, encoded, and interpreted by the destination user during the semantic communication process. 
We introduce a novel projection-based semantic encoding solution 
to convert explicit semantics into a low-dimensional semantic constellation space in which the Euclidean distance between semantic terms (entities or relations) is proportional to their meaning difference. 
We also introduce a novel generative imitation-based reasoning mechanism learning (G-RML) solution for supporting automatic encoding, transportation, and decoding/interpretation of implicit semantic meaning at the destination user. One of the key differences between traditional communication solutions and G-RML is that, in the latter solution, the source user does not focus on sending as much of the useful message as possible. Instead, in  G-RML, the source user will try guide the destination user to learn a reasoning mechanism to automatically  map explicit semantics to a set of implicit semantics including hidden concepts, objects, and their relationships that are most relevant to the meaning of the source user. 
We summarize our main contributions as follows:

    
    \noindent{\bf New representation of semantics of messages:}
 { a novel solution for representing both explicit and implicit semantics is proposed. 
 A projection-based semantic encoder is developed to 
 reduce the redundancy of the transmit messages and also improve the robustness against semantic misinterpretation caused by the   
 channel corruption. }
 
 \noindent{\bf Novel implicit semantic-aware communication architecture:} We propose a novel architecture, called iSAC, for supporting automatic  interpretation/decoding of the implicit semantics at the destination user, respectively. Our proposed architecture involves a novel semantic distance measure of implicit semantics that can be directly learned by the source user.

 \noindent{\bf New implicit semantic recovery solution:} We introduce a novel generative adversarial imitation-based solution, G-RML, for the 
 destination user to learn a reasoning mechanism 
 that can automatically generate implicit semantic paths. G-RML does not require destination users to know any expert semantic paths or to consult the source user during the communication.  
 
 \noindent{\bf Extensive simulations:} We conduct extensive simulations to evaluate the performance of our proposed solution based on three popular real-world human knowledge datasets.  Numerical results suggest that, the proposed solution achieves up to  92\% accuracy of implicit meaning interpretation. Also, under the considered scenarios, the semantic symbol error rate of the existing communication solution without using any reasoning mechanism is 28.65\%, while the semantic symbol error rate of our proposed solution is 4.94\%, resulting in 83\% reduction. 





\vspace{-0.1in}
\section{Related Works}
\label{Section_RelatedWork}
	

\noindent{\bf Semantic Communication:}
Most existing works on semantic communication focused on the representation, identification, and computation of explicit semantics, such as labels and signal features, from a specific type of signals such as image, voice, video, and text\cite{bourtsoulatze2018deep,lee2019deep,farsad2018deep}. 
More specifically, for image signals, the authors in \cite{bourtsoulatze2018deep} studied image compression tasks and proposed a DL-based joint source and channel  coding solution for image transmission. 
In \cite{Jan2020ImageRetri}, the authors investigated the image retrieval problem in a wireless edge network with power and bandwidth constraints. 
The authors in \cite{lee2019deep} introduced a DL-based joint transmission-recognition scheme for Internet-of-things (IoT) devices that allows 
image data to be uploaded to the closest  edge server for 
image classification. 
For transmitting text, in \cite{farsad2018deep}, the authors developed a neural network-based architecture for joint source-channel coding. 
proposed a transformer-based solution to perform joint channel coding for text transmission. The authors in \cite{seo2021semantics} proposed a novel model, called semantics-native communication, in which the signal lengths in bits related to the expected semantic representation have been derived to quantify the extracted effective semantics of the signal. The authors in \cite{karimzadeh2021common} proposed a curriculum learning framework for goal-oriented semantic communication with a common language between the source and destination users.   
For audio and video transmission, the authors in \cite{Weng2021SpeechSemantic} applied the attention mechanism to extract information of the transmitted speech signals for minimizing the error rate at the semantic level. 
In \cite{Tong2021AudioSemantic}, the authors proposed a wave to vector (wav2vec)-based autoencoder that is capable of extracting the semantic information from audio signals. 
Federated learning was also adopted to enhance the performance of the proposed autoencoder, which enables accurate 
feature 
extraction from the 
audio signals. The authors in \cite{jiang2022SemanticVideoConference} designed a semantic-based video conference network and proposed a semantic error detector to capture the expression changes of the speakers. Recently, semantic communication systems involving multimodal data sources have also been investigated. 
For a detailed survey of recent progress, 
we refer interested readers to \cite{xu2022edge, Gunduz2023JSAC}. 

\noindent{\bf Knowledge Reasoning:}
Recently, knowledge reasoning has been considered as a promising solution to estimate the implicit meaning of messages.  Existing reasoning methods can be classified into three categories: rule-based reasoning, distributed representation-based reasoning and neural network-based reasoning\cite{chen2020review}. Specifically, a rule-based systems represent knowledge as a set of rules specifying what to do or what to conclude in different situations. Such a system mimics the reasoning process of a human expert in solving a knowledge-intensive problem\cite{grosan2011rule}. In \cite{bordes2013translating}, the authors presented a method that models relationships of multi-relational data by interpreting them as translations operating on the low-dimensional embeddings of the entities. The authors in \cite{xiong2017deeppath} proposed a novel reinforcement learning framework for learning multi-hop relational paths, which includes a reward function taking accuracy, diversity, and efficiency into consideration. Most existing methods rely on the concept of knowledge triplets that are usually difficult to be identified and the specific reward values for successful reasoning are difficult to specify in most practical scenarios. Also, there is still a lack of a commonly adopted metric for measuring the semantic distance between different signals involving implicit semantics. 
\vspace{-0.1in}
\section{A Reasoning-based Semantic-Aware Communication Framework}
\label{Section_Primer}

\subsection{Representation of Semantics}
\label{Subsection_SemanticRepresentation}

One of the key challenges in iSAC is to develop a general and comprehensive way to represent the semantic meaning of any given message, including both explicit and implicit semantics. In this paper, we consider a general setting and assume that the intended meaning of the source user can be represented by a sequence of concepts and terms connected with various types of relations 
that reflect its personalized understanding of messages. This aligns with the recent observations in cognitive neuroscience \cite{zhang2020connecting,liuzzi2020general} which report that, instead of communicating individual concepts, e.g., symbols and labels, a human brain has the capability to generate complex semantic meaning by connecting different terms and concepts via various types of relations. Also, motivated by another recent discovery in cognitive neuroscience\cite{zhang2020connecting}, suggesting that human users tend to reason hidden concepts and ideas by following the most directly linked knowledge terms based on a user-related reasoning mechanism, we model the inference process from explicit semantics to the implicit semantics  as a 
sequential decision making process dominated by a reasoning mechanism. 

We formally define the representation of semantics of a message as a triple $\langle \boldsymbol{v}, \boldsymbol{u}_{\boldsymbol{v}}, \Pi \rangle$ where $\boldsymbol{v}$ is the explicit semantics, $\boldsymbol{u}_{\boldsymbol{v}}$ is the implicit semantics that are linked to the explicit semantics $\boldsymbol{v}$, and $\Pi$ is the reasoning mechanism, which are described in details as follows:

\subsubsection{Explicit Semantics} correspond to a set of real-world entities and relations such as ``child", ``bicycle", ``Tweety", ``social media website", ``canary bird" and ``belong to", ``is/are" etc., that can be directly detected in the source signal. 
We can further define the explicit semantics as $\boldsymbol{v} = \langle e^0, r^0 \rangle$ where $e^0$ and $r^0$ are sets of entities and relations that can be directly recognized in the source signal. 

\subsubsection{Implicit Semantics $\boldsymbol{u}_{\boldsymbol{v}}$} correspond to the hidden information, including the hidden relations and entities that cannot be directly identified from the source signal, but are important for the interpretation of semantic meaning reflected by the explicit semantics, 
e.g, a child ``rides" a bicycle, Tweety ``is" a ``canary bird", and Tweety ``is" a ``cartoon character" ``in" a ``TV show". 
In this paper, we consider a sequential inference process in which the relevant hidden entities and relations are sequential decided one iteration at a time, based on the observed explicit semantics. We can further define the implicit semantics as $\boldsymbol{u}_{\boldsymbol{v}} = \langle e^i, r^i \rangle_{i = \{1, 2, \ldots\}}$ where $e^i$ and $r^i$  are the set of hidden entities and relations inferred in the $i$th iteration for 
$i \ge 1$. 

\subsubsection{Reasoning Mechanism $\Pi$}
corresponds to a user-related inference function or rule that characterizes the correlations between explicit and implicit semantics.  
For example, in the previous example, the entity ``Tweety" may link to several other hidden entities such as ``smart phone App", ``canary bird" and ``cartoon character" with corresponding hidden relations. The receiver (the father) therefore needs to identify possible entities and the corresponding relations that are linked to entity ``Tweety" based on the reasoning mechanism learned or updated from the communication history. For example, if, in his/her recent communications with the father, the child mentioned about reading some books about animals, the father would be able to infer the most appropriate meaning of the child's question by finding a reasoning path connecting a sequence of hidden entities and relations, i.e., ``Tweety is a canary bird that is highly likely to have appeared in a book recently read by the child". It can be observed that the reasoning mechanism plays an essential role in inferring the appropriate meaning of the message. Generally speaking, the semantics of a source signal can involve at least one key explicit entity and one or multiple implicit reasoning paths associated with the explicit entity. The implicit reasoning paths generated from the explicit semantics may not be unique and can be closely related to the user's background, environment, and context of communication. We consider a sequential inference process that is dominated by a sequential decision making policy, called the reasoning policy. Let $\Pi$ be the reasoning policy for inferring the most relevant hidden relations and entities based on $e^j$ and  $r^{j}$, e.g., we have $\langle e^{j+1}, r^{j+1}\rangle = \Pi \left \langle e^{j}, r^{j} \right \rangle$. We also use $\hat{\Pi} _{\pi _{\theta}}$ to denote the reasoning mechanism for generating possible semantic paths via sequential inference process based on $\Pi$.

\subsection{Knowledge Base}
One of the key ideas of iSAC is to incorporate a priori knowledge and personal background information including facts, concepts, and experiences (reflecting relations) to assist in the interpretation of the intended meaning. In other words, the knowledge shared among the communication participants (users) is essential in iSAC system. We refer to the collection of all the  knowledge entities and relations that are accessible to each user as the \emph{knowledge base}. The knowledge bases of different users do not have to be the same. In other words, the knowledge base of each user may consist of two parts of elements:
	
\begin{itemize}
	\item[(1)] \emph{Common knowledge:} corresponds to knowledge entities and relations that are shared among all the communication participating users (e.g., source and destination users).
	\item[(2)] \emph{Private knowledge:} involves the entities and relations associated with personal understanding and viewpoints about some real-world concepts and relationships. Generally speaking, the private knowledge entities and relations of different users do not have to be the same.  
\end{itemize}

\subsection{Reasoning Mechanism Modeling and Learning}
	
Interpreting the implicit meaning 
is known to be a notoriously challenging task. To address this challenge, we propose a novel solution, called G-RML, for learning a reasoning mechanism to automatically infer the implicit semantics based on the observed explicit semantic information. 
In G-RML, the implicit semantic inference process is modeled by a sequential decision making process characterized by a reasoning policy. In this policy, the most possible hidden relations, as well as the connected entities, are sequentially extended from the explicit semantics by the destination user. Since the destination user is generally impossible to know a specific reward function characterizing the performance of its inference results, 
G-RML adapts the imitation learning-based solution in which during the training phase,  the source user will guide the destination user to learning a policy that can generate reasoning paths to match the distribution of the real expert paths.    

\section{System Model and Problem Formulation}
\label{Section_SystemModel}

\begin{figure}
\centering
\vspace{-0.2in}
\includegraphics[width=8cm]{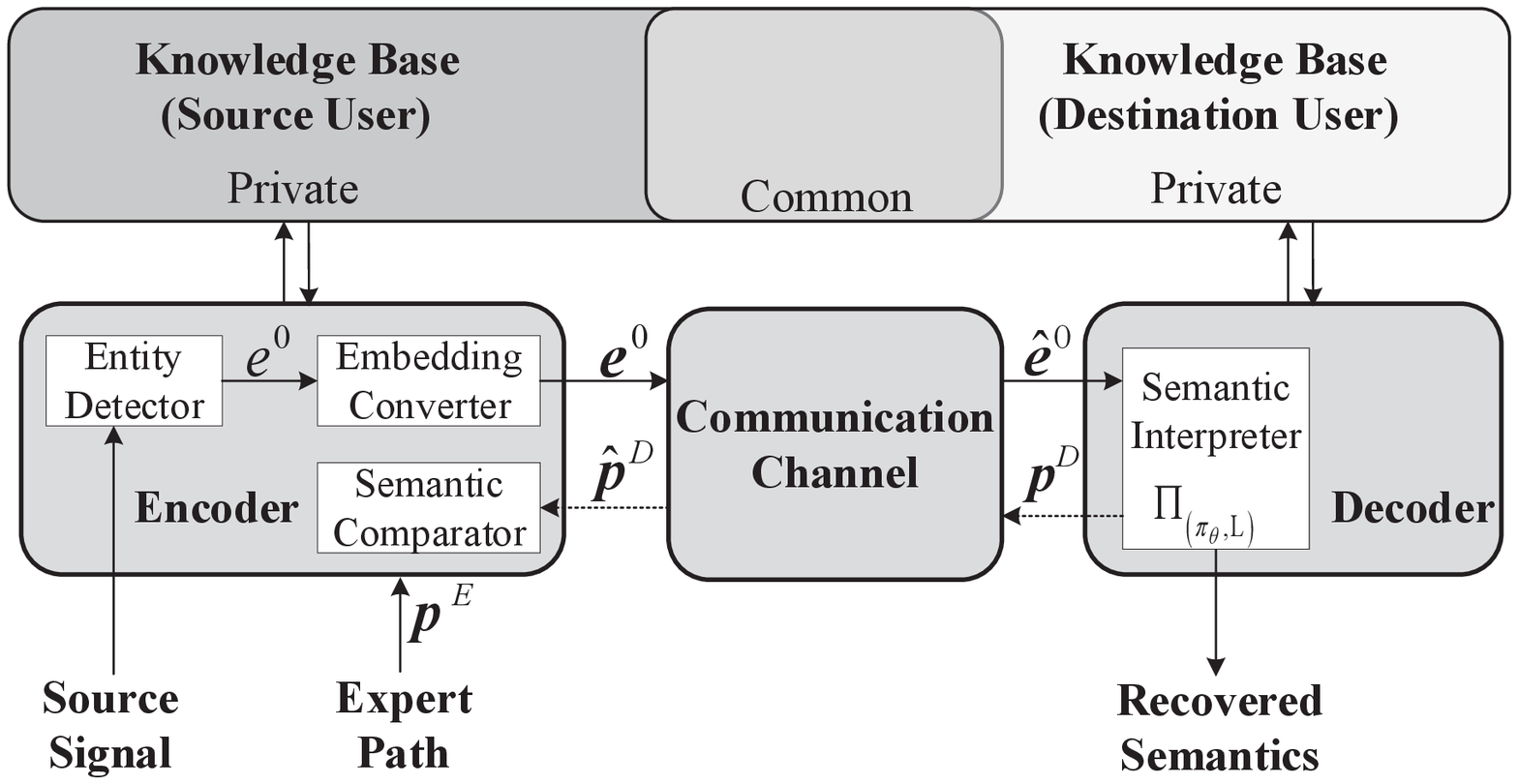}
\caption{\footnotesize An iSAC architecture.}
\label{Fig_systemmodel}
\vspace{-0.25in}
\end{figure}

\subsection{System Model}

We introduce the iSAC architecture, 
as illustrated in Fig. \ref{Fig_systemmodel}(a). It consists of two phrases of implementation: model training phrase and communication phrase. In the training phrase, the source and destination users will jointly train two models, an explicit semantic encoding model for efficient transmission of explicit semantics identified by the source user, and an implicit semantic inference model which allows the destination user to infer implicit semantic meaning of the source user based on the received explicit semantics. 
To train the first model, we design a projection-based semantic encoder to convert the high-dimensional information of the explicit semantics into a compact low-dimensional representation for physical channel transmission. For the training process of the second model, the destination user will construct a semantic interpreter that tries to generate the possible implicit semantics to be sent to the source user for evaluating. The source user will train a semantic comparator to output the semantic distance, a metric for measuring the meaning dissimilarity between the generated implicit semantics sent from the destination user and its local set of expert semantic paths obtained from the communication history. The source user will send the semantic distance to the semantic interpreter at the destination user for the model adjustment. In the communication phrase, the source user will directly send the key explicit semantics detected from the source signal. The destination user can directly apply the trained semantic interpreter to infer the implicit semantics without consulting the source user. \blu{Our proposed architecture can be directly implemented into the existing communication systems. The only requirement is that the semantic encoding and interpretation models need to be pre-trained offline and installed into the existing communication  system in which the traditional encoding solution can be replaced with the pre-trained semantic encoder and also the digital symbol-based decoding solution can be replaced with our proposed pre-trained semantic interpreter.} The detailed procedures of both phrases and their implementations are illustrated in Fig. \ref{Fig_systemmodel}(b).

\noindent{\it (1) Source User Side: }

\noindent{\bf Explicit Semantics Detector:} The encoder should be able to first identify a set of explicit semantic terms, e.g., at least one visible entity and/or relation from the observed source signal. This can be achieved by using some well-trained models such as YOLO \cite{redmon2016look} and wav2letter \cite{audio2019} to identify known labels of objects from various types (e.g., image, voice, and text) of source signals. As mentioned earlier, these labels themselves cannot fully characterize the complete meaning represented by the detected objects. In other words, the meaning of semantic terms identified from the source signal can be linked to a set of hidden features and attributes including the properties as well as the relationships to other terms.

\noindent{\bf (Explicit) Semantic Encoder:} The semantic entities and relations, as well as their corresponding attributes may consist of rich information that are generally inefficient for physical channel transmission. The source user therefore needs to convert the high-dimensional explicit semantics into a compact representation to be sent to the destination user via physical channel. For this purpose, we design an projection-based encoding function to map the explicit semantics into a low-dimensional representation space, called semantic constellation space, that is efficient for transmission in the physical channel. In traditional communication systems, the encoder at the source user consists of two separate coding processes, source encoding and channel encoding with objectives to maximize the total amount of transmitted information and improve the robustness of data delivery against channel corruption, respectively. In contrast, here, our goal is to develop a single projection function $g \left( \cdot \right)$ for semantic encoder to 
achieve both objectives. Suppose $e^0$ and $r^0$ consist of a single entity and a connected relation. We have $e^0 \in \mR^m$ and $r^0 \in \mR^{m'}$ where $m$ and $m'$ are the dimensional sizes of entity and relation, respectively. Let $\bx$ be the coded low-dimensional representation of signals sent to the physical channel. The proposed projection function can be applied to jointly encode both entities and relations, i.e., $\bx = g \left( \langle e^0, r^0 \rangle \right) \in \mR^n$ where $n$ is the dimensional size of the encoded signal $\bx$. It can also be applied to encode the entities and relations in seperate fashions. In this case, we  can express the semantic encoder as a mapping function $\bx = \langle g (e^0), g(r^0) \rangle$ for  $g (e^0) \in\mR^{n}$ and $g(r^0) \in \mR^{n'}$ and $n$ and $n'$ are the dimensional sizes of the projected semantic entities and relations, respectively. We have  $n, n' \ll m, m'$ . 

\noindent{\bf Semantic Comparator:} One of the key differences between semantic communication and traditional communication is that the delivered result of a message is generally not binary (successful or failure of delivery) but can be characterized by the {\it semantic distance}, a continuous valued metric characterizing how far the implicit semantic meaning $\eta^D$ interpreted by the destination user is different from the truly intended meaning of the source user reflected in the expert semantic paths. Let  $\Gamma\left(\eta, {\hat \eta}\right)$ be the semantic distance between the original meaning $\eta$ of source user and the meaning $\hat \eta$ recovered by the destination user. 
Different from the explicit semantics $\langle e^{0}, r^0 \rangle$ that can be directly identified from the source signal, the implicit semantic meaning is generally nondeterministic and may consist of a set of possible reasoning path, each includes a sequence of entities and relations inferred from the explicit semantics, i.e., $\eta = \langle{e^{0}, r^0, e^1, r^1, e^2, r^2, \ldots}\rangle$. Therefore, in this paper, we develop a statistic-based distance metric to evaluate the semantic distance between two sets of semantic reasoning paths, include those are generated by the destination user and the sampled expert semantic paths. 

\noindent{\it (2) Physical Communication Channel:}

We consider a  discrete memoryless channel with a limited capacity and 
the received signal at the destination user can therefore be written as 
\begin{eqnarray}
{\by} = {\boldsymbol H} \bx + {\boldsymbol \delta},
\label{eq_ReceivedEmbeddingGaussianNoise}
\end{eqnarray} 
where $\bx$ is the  signal encoded by the semantic encoder of the source user, ${\boldsymbol H}$ is the fading coefficient of the channel, and $\boldsymbol \delta$ is the additive Gaussian noise. We assume ${\boldsymbol H}$ and $\boldsymbol \delta$ follow stationary probability distribution functions during both training and communication phrases. In this paper, we assume the source user cannot know the channel fading coefficient. The destination user therefore needs to learn a semantic interpretation network based on the noisy version of received explicit semantics. 


\noindent{\it (3) Destination User Side:}

\noindent{\bf Semantic Interpreter:} 
The destination user 
will be able to recover the most likely sequences of hidden relations and entities $\hat \eta$  
by extending various reasoning paths from its received explicit semantics. 
The semantic interpreter should be able to learn a reasoning mechanism $\hat{\Pi} _{\pi _{\theta}}$ with parameter $\theta$ that can sequentially decide the possible relations to extend the reasoning paths $\hat{\eta}=\langle{e^0, r^0, {\hat r}^1, {\hat e}^1, {\hat r}^2, {\hat e}^2, \ldots}\rangle$ to best interpret the true meaning of the source user.

\subsection{Problem Formulation}	
The main objective is to develop a novel solution that allows the destination user to automatically output a set of estimated reasoning paths $\hat \eta$ with the minimum semantic distance from the original semantic paths $\eta$ of the source signal. Formally, we aim at solving the following problem
\begin{eqnarray}
\mbox{(P1)} \;\;\;\; \min_{\theta} \Gamma_{\theta}\left(\eta, {\hat \eta}\right) \nonumber
\end{eqnarray}
where $\theta$ represents the latent parameters of the semantic interpreter at the destination user.

As mentioned earlier, due to the complexity of the knowledge base and rich meanings that can be represented in the communication messages, it is generally difficult to find a simple and comprehensive solution to solve (P1). To overcome this challenge, we propose a generative  imitation learning-based framework, which will be discussed in detail in the next section. 

\section{iSAC Architecture}
\label{Section_Architecture}

\subsection{Source User Side}
\label{Subsection_ArchitectureSourceUser}
\noindent{\textbf{Semantic Encoding:}} 
The source user first converts the identified explicit semantics into a suitable form for channel transmission. 
The representation of semantics proposed in Section \ref{Subsection_SemanticRepresentation} is efficient for characterizing the complex relations among different semantic entities. It is however inefficient for physical channel transmission due to its redundancy in representing rich attributes of entities and relations. Also, any error in decoding semantic entities and relations will result in failure of recovering the entire semantic message. To address the above issues, we would like to design a semantic encoder with the following ideal properties: (1) {\it Efficient Transmission:} the encoded signal should have a much smaller dimensional size, compared to the original graphical representations of semantics, and (2) {\it Robustness against channel corruption:} we would like to design a projection-based encoding function to separate different semantic terms in the constellation space based on the meaning difference and usage preference of the source user. In this way, even if some semantic terms, e.g., entities or relations, have been corrupted during the channel transmission, the destination user can recover these corrupted terms using the designed projection function. More specifically, we assume the source user can observe a set of preferred entity-and-relation combinations, called the {\em semantic preferred set}, e.g., consisting of meaningful combinations of semantic terms that are most frequently observed from the source user's communication history. We can then design a projection function to separate different combinations of semantic terms in the constellation space according to the observed usage preference, i.e., terms that are more frequently observed to be used together should be more closely located in the constellation space, compared to the rarely observed semantic term combinations.  
 
%

To achieve property (1), we propose a projection-based semantic encoding function to convert the high-dimensional representations of explicit semantic terms into the low-dimensional semantic constellation representations. 
More specifically, we adopt an energy-based projection function, TransE \cite{bordes2013translating}, to convert high-dimensional graphical representations of semantic entities and relations into low-dimensional constellation subspaces, which are referred to as entity and relation constellation spaces, respectively. Let $g(e)$ and $g(r)$ be the representations (e.g., phrase and amplitude) of entity $e$ and relation $r$ in their corresponding low-dimensional constellation spaces, respectively, for $g(e) \in \mathbb{R}^{n}$, $g(r) \in \mathbb{R}^{n'}$, and $n$ and $n'$ are dimensional sizes of entity and relation constellation spaces, for $n, n' \ll m, m'$.

To achieve property (2), we first define the energy, a value measuring the possibility, of a combination of an initial entity $e^i$, a connected relation $r^{j}$, and a possible tail entity $e^k$ in the semantic constellation space as ${p}\left(g(e^i),g(r^{j}),g(e^k) \right) = \|g(e^i)+g(r^{j})-g(e^k)\|_{l_1/l_2}$ where $\| \cdot \|_{l_1/l_2}$ means either the $l_1$ or the $l_2$-norm is applicable depending on the distance measure in the constellation space. For example, an $n$-dimensional encoded signal can be sent either via $n$ real-valued signals with $l_1$ norm distance metric, e.g., using the amplitude modulation scheme, or $n/2$ complex-valued signals with $l_2$ norm distance metric using phase and amplitude modulation schemes. Generally speaking, the higher the energy of ${p}\left(g(e^i),g(r^{j}),g(e^k) \right)$, the higher the possibility that entity $e^i$ and relation $r^{j}$ lead to the resulting entity $e^k$. We can then train the semantic encoding (projection) function to separate the semantic preferred (most frequently observed) and unpreferred (rarely observed) combinations of terms by minimizing the following loss function: 

\begin{eqnarray}
\begin{aligned}
{\cal L} = \sum\limits_{\scriptstyle \langle e^i,r^{j},e^k \rangle\in {\cK}^+, \hfill\atop \scriptstyle \langle e^i,r^{j'},e^{k'} \rangle\in {\cK}^-}  &\max\left\{0, d + {p}\left(g(e^i),g(r^{j}),g(e^k) \right) \right. \\ 
&\left.- {p}\left(g(e^i),g(r^{j'}),g(e^{k'}) \right)\right\},
\label{eq_TransELoss}
\end{aligned}
\end{eqnarray}

where ${\cK}^+$ and $\cK^-$ are the sets of semantic preferred and unpreferred entity-relation-entity triplets, respectively. $d$ is the average distance set to separate the preferred and unpreferred entity and relation combinations in semantic constellation space. We can directly extend the above energy function into more general scenarios consisting of semantic preferred paths with various lengths. Let $L$ be the maximum length of a semantic path. We can similarly convert a semantic path $\eta  = \langle e^0, r^0, e^1, r^{1}, e^2, r^{2}, \ldots, e^L, r^L\rangle$ into the form of the path representation in the semantic constellation space, i.e., if we denote $g (\br^{0:L}) = \sum^L_{j=0} g({r}^{j})$, the distance between $g(e^0)+g (\br^{0:L})$ and $g(e^L)$ is given by ${p}\left(g(e^0), g(\br^{0:L}), g(e^L) \right) = \|g(e^0)+g(\br^{0:L})-g(e^L)\|_{l_1/l_2}$. Note that the projection function can be pre-trained by the source user and sent to destination user before the communication phrase. In this way, the destination user does not need to know any semantic preferred or unpreferred set. 
During the recovery of explicit semantics, destination user will always choose the semantic term combinations that minimize the energy.

\begin{figure}
\centering
  \begin{minipage}[t]{0.48\textwidth}
   \centering
   \includegraphics[width=\textwidth]{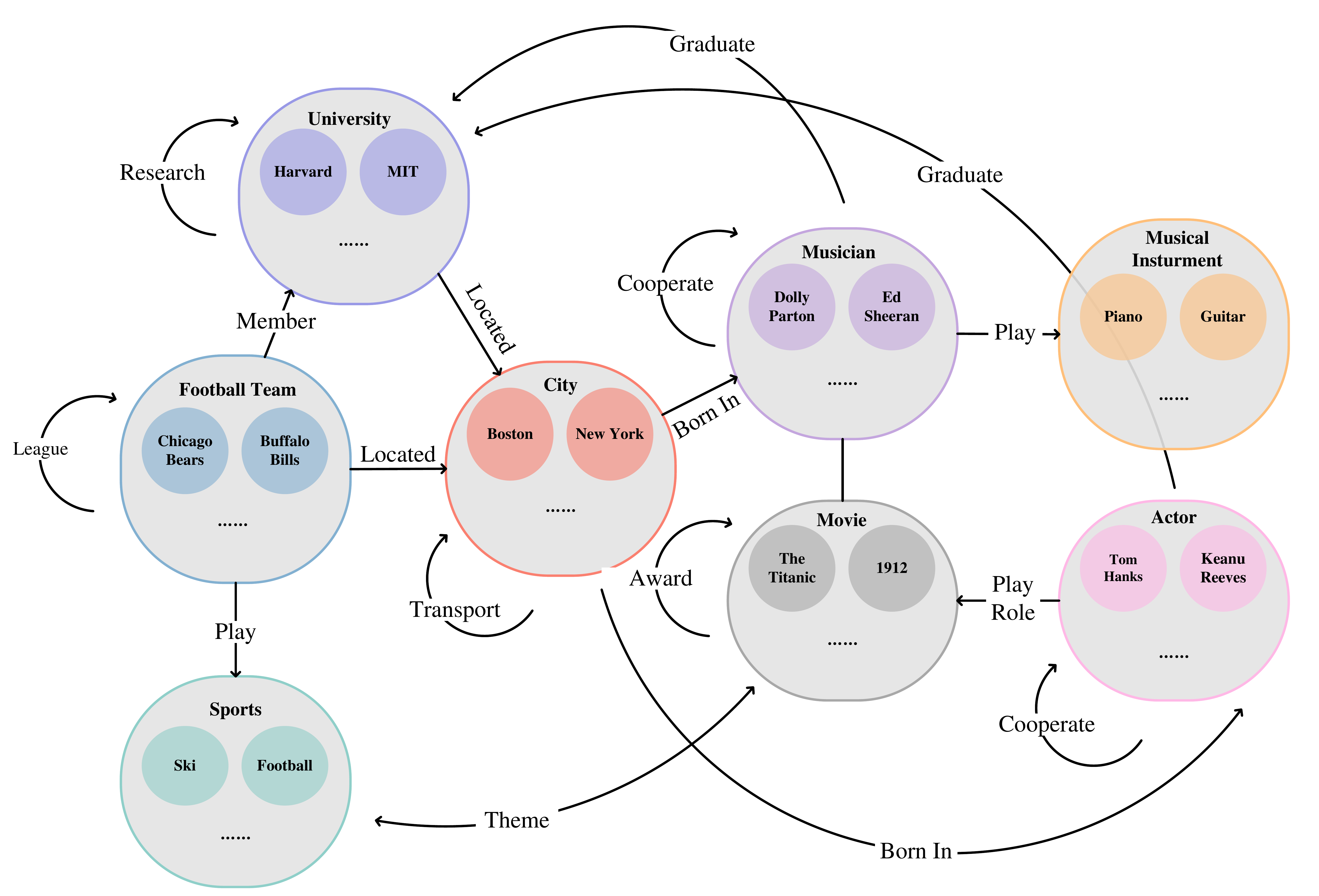}
   \vspace{-0.2in}
	 \caption*{\small{(a)}}
  \end{minipage}
    \begin{minipage}[t]{0.25\textwidth}
	\centering
	\includegraphics[width=\textwidth]{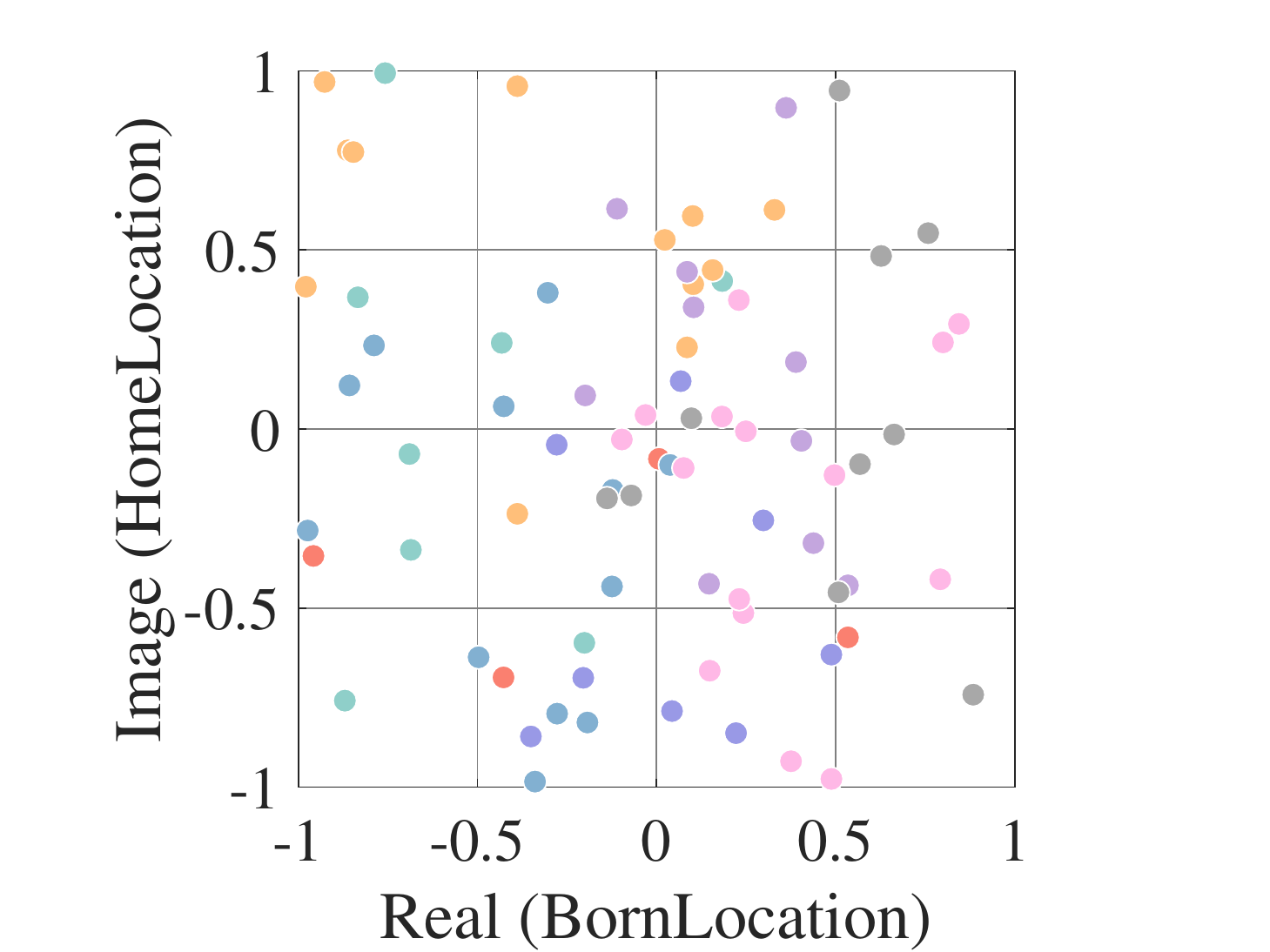}
	\vspace{-0.2in}
	\caption*{\small{(b)}}
  \end{minipage}%
  \begin{minipage}[t]{0.25\textwidth}
	\centering
	\includegraphics[width=\textwidth]{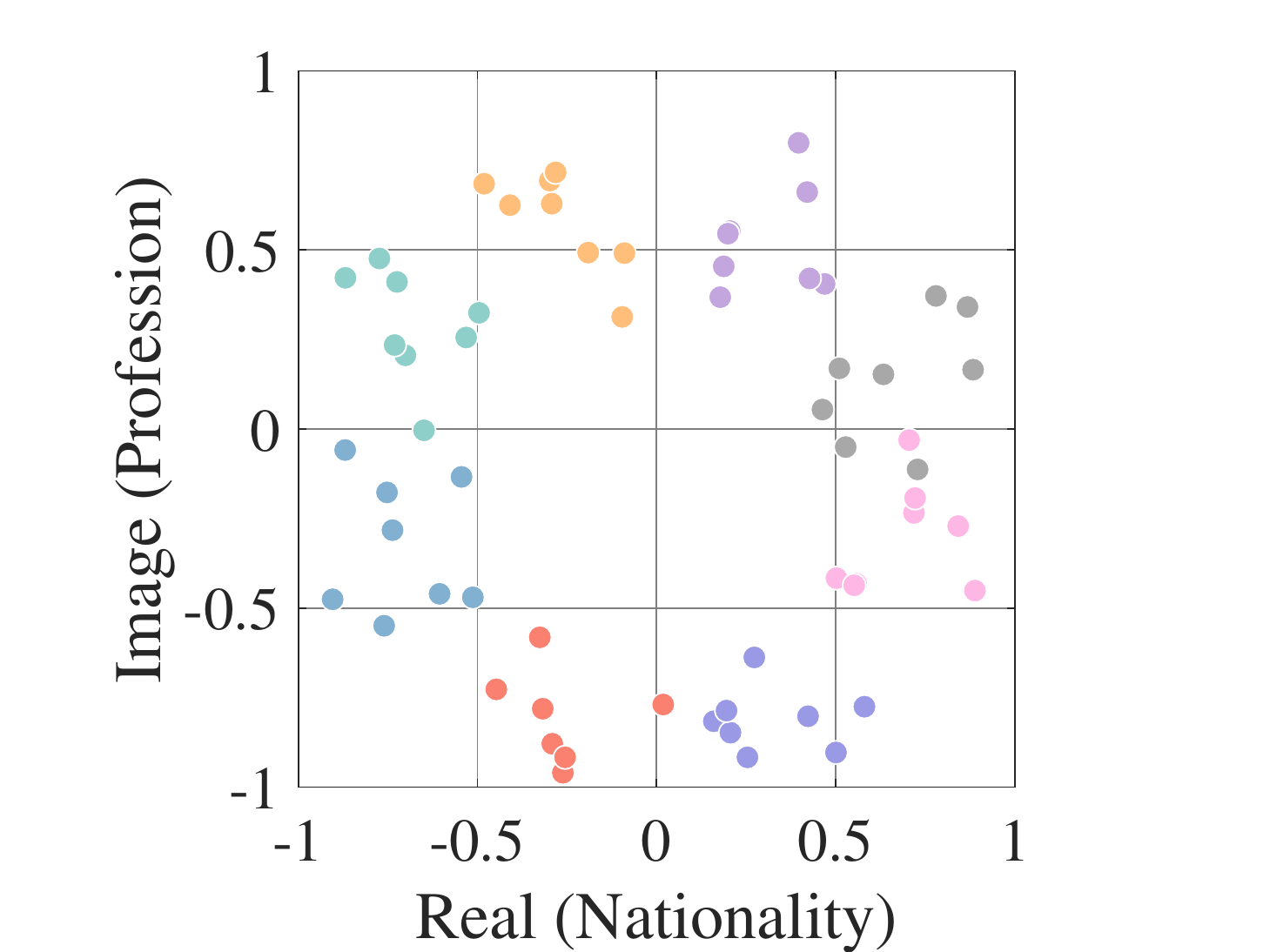}
	\vspace{-0.2in}
	\caption*{\small{(c)}}
\end{minipage}%
  \vspace{-0.05in}
  \caption{\footnotesize{(a) A graphical representation of the semantic meaning of 64 concepts (entities) associated with eight classes (city, sports, university, musician, movie, music instrument, actor and football team) connected by multiple relations, (b) semantic constellation representations of 64 symbols (entities) in two selected dimensions (HomeLocation and BornLocation) where we use dot and cross to represent 64 symbols associated with two classes, (c) semantic constellation of 64 symbols in two selected dimensions (Nationality and Profession).}}
  \label{Fig_SemanticEncoding}
  \vspace{-0.3in}
\end{figure}


To illustrate the performance of our proposed semantic encoding solution, in Fig. \ref{Fig_SemanticEncoding}, we consider the encoding process of a semantic message involving entities and relations associated with eight classes (city, university, drugs, musician, movie, sports, musical instrument and football team) as well as their corresponding relations randomly sampled from the NELL-995 knowledge base dataset where each entity has 200 attributes. The graphical representation of the semantic meaning involved in the given message is illustrated in Fig. \ref{Fig_SemanticEncoding}(a).  We present the projected representation of entities in the semantic constellation space when the imaginary and real parts of the constellation correspond to different dimensions in Fig. \ref{Fig_SemanticEncoding} (b), (c). We can observe that by adopting our proposed encoding (projection) function, different symbols in the semantic constellation space can be separated and the Euclidean distance between symbols depends on their meaning difference (dissimilarity) in the associated dimensions. For example, when the imaginary and real parts of a semantic constellation correspond to Nationality and Profession, respectively, the entities associated with different classes are separated further from each other because entities from different classes have greater degree of differentiation in these two attributes.
However, if the imaginary and real parts correspond to HomeLocation and BornLocations, the constellation points of entities from different classes become almost indistinguishable because the Home locations of these entities may be the same as the born location.

\blu{Our proposed semantic encoding solution has low computational complexity, especially compared to the Graph Convolutional Network (GCN)-based solutions. This is because, in our proposed solution, the loss function only needs to be calculated once for each expert path. More specifically, suppose the dimensional size of each explicit semantic entity is given by $n$ and the total number of expert semantic paths for training the semantic encoding function is given by $K$. The computational complex of training our proposed semantic encoding function is given by $O\left(n K\right)$. The computational complexity of GCN-based solutions, however, depends on the number of CGN layers as well as the non-zero number of edges in the adjacency matrix of all the expert paths\cite{bordes2013translating}, \cite{kipf2016semi}.  
In particular, applying GCN-based solution to train the same semantic encoding function will result in a computational complexity of $O\left(n L A F^2\right)$ where $L$ is the number of layers, $A$ is the total number of semantic entities in the knowledge base, $F$ is the number of node features at every layer. As will be shown in our simulation results, the number of expert semantic paths required to calculate our proposed semantic encoding function is relatively small, e.g., less than 50, which results in a much smaller computational complexity, compared to GCN-based solution which normally involves calculations of large adjacency matrices.} 

\noindent{\textbf{Semantic Distance:}}
The main objective of the destination user is to infer the implicit semantics based on the recovered explicit semantics. 
Different from the explicit semantics that are pre-identified and determined by the source user before encoding, the implicit semantics are generally difficult to be characterized by a deterministic path due to its randomness and polysemy. 
In other words, we cannot use the projection-based energy function developed for the semantic encoder to evaluate the semantic distance between implicit semantics.

Let us now develop a statistic-based distance measure to evaluate the meaning difference between the true implicit semantics and the inferred semantics interpreted by the destination user. 
As mentioned earlier, the implicit semantics expressed by each individual user can be considered as a set of possible semantic paths generated by a hidden expert reasoning mechanism $\Pi$ based on the given explicit semantics. The reasoning mechanism $\Pi$ is defined as a mapping function that maps a set of explicit semantic entities and relations $\boldsymbol{v}$ to a set of semantic paths $\boldsymbol{u_v}$ consisting of a sequence of possible relations, i.e., $\Pi: \boldsymbol{v} \rightarrow \boldsymbol{u_v}$. 
Note that neither source nor destination user can know or directly observe $\Pi$. The source user however can observe a set $\mathcal{K}$ of expert semantic paths, e.g., sampled from the previous communication history, that can be assumed to be generated by $\Pi$. Different expert semantic paths can be initiated  with different combinations of semantic terms. These initial semantic terms can be considered as the explicit semantics and the rest of the terms can be regarded as the implicit semantics generated by $\Pi$. The main objective of the destination user is then to learn an approximated reasoning mechanism $\hat \Pi$ that can match the implicit semantic generation process observed from the expert semantic paths. 
Let $\hat \Pi_{\pi_\theta}$ be the reasoning mechanism learned by the destination user determined by a reasoning policy with parameter $\theta$. We refer to the probability distributions of semantic paths generated by reasoning mechanism $\Pi$ as the occupancy measure of the mechanism $\Pi$, denoted as $\rho_\Pi$. We can then define the semantic distance between the implicit semantics of the source user and that interpreted by the destination user as the (statistic) distance $\Gamma$ between the occupancy measures of semantic paths generated by the two reasoning mechanisms $\Pi$ and $\hat \Pi_{\pi_\theta}$, i.e., $\rho_{\Pi}$ and $\rho_{\hat \Pi_{\pi_\theta}}$ can be written as follows:
\begin{eqnarray}
\begin{aligned}
\rho_{\Pi} &= \Pi 
\sum_{\boldsymbol{u_v} \in {\Delta}|\Pi} \Pr\left( \boldsymbol{u_v} | \Pi \right)   \mbox{ and}\\ \;\;\;
\rho_{{\hat \Pi}_{\theta}} &= {\hat \Pi}_{\pi_\theta} \sum_{\boldsymbol{u_v} \in {\hat \Delta}|\hat \Pi_{\pi_\theta}} \Pr\left( \boldsymbol{u_v} | {\hat \Pi}_{\pi_\theta} \right) 
\end{aligned}
\end{eqnarray}
where ${\Delta}|\Pi$ is the set of 
possible expert semantic paths generated by reasoning mechanism $\Pi$ and ${\hat \Delta}|\hat \Pi_{\pi_\theta}$  is the set of 
estimated implicit semantic paths generated by mechanism $\hat \Pi_{\pi_\theta}$, which will be discussed in detail in Section \ref{Subsection_ArchitectureDestinationUser}. 

In this paper, we adopt the Jensen–Shannon (JS) divergence, one of the most popular statistic distance metrics, to quantify the semantic distance between the occupancy measures of semantic paths generated by different reasoning mechanisms. Our solutions can be directly extended into more general scenarios that use other statistic distance metrics to quantify  semantic distance. We can  write the JS divergence-based  semantic distance between $\Pi$ and $\hat \Pi_{\pi_\theta}$ as
\begin{eqnarray}
\begin{aligned}
\Gamma \left( \Pi, \hat \Pi_{\pi_\theta} \right) &= \mathbb{E}_{\Pi} \left[ \log\left( {\rho_{\Pi} \over \rho_{\Pi} + \rho_{\hat \Pi_{\pi_\theta}}} \right) \right] \\ &+ \mathbb{E}_{\hat \Pi_{\pi_\theta}} \left[  \log\left( {\rho_{\hat \Pi_{\pi_\theta}} \over \rho_{\Pi} + \rho_{\hat \Pi_{\pi_\theta}}} \right) \right] 
\label{eq_JSD}
\end{aligned}
\end{eqnarray}

Based on the above definition, we can rewrite problem (P1) into the following form: 
\begin{eqnarray}
\mbox{(P2)}\;\;\; \min_{\theta} \Gamma \left( \Pi, \hat \Pi_{\pi_\theta} \right)
\end{eqnarray}

\noindent{\textbf{Semantic Comparator:}} 
Since the destination user cannot observe any expert semantic paths, the source user will need to compare the semantic paths generated by the reasoning mechanism learned by the destination user with its locally observed expert paths and send the comparison result to the destination user for model correction and training. 

One straightforward approach is to let the source user know both $\rho_{\Pi}$ and $\rho_{\hat{\Pi} _{\pi _{\theta}}}$. Then, in this case, the source user can directly calculate the semantic distance between $\rho_{\Pi}$ and $\rho_{\hat{\Pi} _{\pi _{\theta}}}$ using (\ref{eq_JSD}) and send the resulting JS divergence value to the destination user. Unfortunately, as mentioned earlier, neither $\Pi$ nor $\hat{\Pi} _{\pi _{\theta}}$ can be known by the source user. To address this issue, 
we adopt a discriminator network at the semantic comparator, denoted as $\varpi_{\phi}$, to distinguish the expert semantic paths from the implicit semantic paths inferred by the destination user. 
More specifically, the semantic comparator trains a neural network to maximize its ability to distinguish the observed expert semantic paths and the  paths generated by the learned reasoning mechanism $\hat{\Pi} _{\pi _{\theta}}$, i.e., the semantic comparator tries to maximize the log likelihood loss function of a classification problem for discriminating the paths sampled from the expert semantic paths and that sampled from the paths generated by reasoning mechanism $\hat{\Pi} _{\pi _{\theta}}$. Let ${\Psi_{\Pi}}$ be the probability distribution of expert semantic paths generated by reasoning mechanism $\Pi$. Also, let ${\Psi_{{\hat \Pi}_{\pi_\theta}}}$ be the probability distribution of interpreted semantic paths generated from reasoning mechanism ${\hat \Pi}_{\pi_\theta}$ learned by the destination user. 
Let $\hat{\mathcal{K}}$ be the set of paths generated by the destination user based on reasoning mechanism $\hat{\Pi} _{\pi _{\theta}}$. 
We write the optimal semantic comparator network as:   
\begin{eqnarray}
\begin{aligned}
\varpi^*_{\phi} &= \arg \underset{\varpi_{\phi}}{\max}\left( \mE_{\eta \sim {\Psi_{{\hat \Pi}_{\pi_\theta}}}} [{\log \varpi_{\phi}\left(\eta   \right)}] \right.  \\ &  \left. +\mE_{\eta \sim {\Psi_{{\Pi}}}} [1- \log \varpi_{\phi}\left( \eta  \right)]\right)
\label{eq_comparator}
\end{aligned}
\end{eqnarray}

We can then prove the following result. 
\begin{proposition}
\label{Proposition_discriminator}
The optimal semantic comparator that solves (\ref{eq_comparator}) is the semantic distance between $\rho_{\Pi}$ and $\rho_{\hat{\Pi} _{\pi _{\theta}}}$ given in (\ref{eq_JSD}).
\end{proposition}
\begin{IEEEproof}
    See Appendix \ref{Proof_Proposition_discriminator}. 
\end{IEEEproof}

The above proposition suggests that the output of the optimal semantic comparator is in fact the value of the semantic distance between the reasoning mechanisms $\Pi$ and $\hat{\Pi} _{\pi _{\theta}}$ given in (\ref{eq_JSD}). In other words, the source user is able to use semantic comparator to directly output the semantic distance between the expert semantic paths and the paths sampled from the reasoning mechanism learned by the destination user without knowing $\Pi$ and $\hat{\Pi} _{\pi _{\theta}}$. This property is critical for the destination user to construct the semantic interpretation network that matches the expert semantic path generation process of the source user without knowing $\Pi$ as will be discussed next.



\subsection{Destination User Side}
\label{Subsection_ArchitectureDestinationUser}
\noindent{\bf Semantic Interpreter:} 
As mentioned earlier, the destination user needs to learn a reasoning mechanism that can infer the implicit semantics based on the explicit semantics received from the source user.  
Let us first define 
the implicit semantic reasoning process from the received explicit semantics of the destination user as a Markov decision process (MDP) problem in which a set of semantic reasoning paths are sequentially generated based on a reasoning mechanism, a policy, that decides a set of relations to extend the currently reasoned paths originated from explicit semantics.  
More formally, we define the implicit semantic path reasoning process as an MDP $\langle {\cal S}, {\cal A}, R, {\Gamma} \rangle$ consisting of the following components:

\begin{itemize}
    \item \textbf{State --} Each reasoning path is sequentially generated by choosing one relation and entity pair at a time. 
    Suppose the maximum length of the inferred reasoning paths is given by $L$. 
    The state of the reasoning process includes the current set of reasoning paths $\boldsymbol{\eta}^t$ as well as its current distance (length) to the initial (visible) entities $t$ for $1\le t\le L$, i.e., the state at the $t$th iteration of a path reasoning is given by $\bs_t = \left(\boldsymbol{\eta}^t, t\right)$. 
    Let ${\cal S}$ be the state space, i.e., we have $\bs_t \in {\cal S}$.   

    \item \textbf{Action --} 
    Given the current state $\bs_t$, the action of the user is to choose the next possible relations to extend the current reasoning paths, i.e., $a_t = \br^t$. Let $\cal A$ be the set of possible relations in the local knowledge base. We have $a_t\in {\cal A}$. 

    \item \textbf{Reward --} The reward function captures the main objective which is to minimize the expected semantic distance between the expert paths and the reasoning paths generated by the semantic interpreter at the destination user, i.e., $\Gamma \left(\Pi, \hat{\Pi} _{\pi _{\theta}} \right)$. As mentioned earlier, most reinforcement learning-based solutions require a specific reward function being carefully defined under given action and state pair. In our setting, however, the destination user only needs to send a set of generated paths to the source user and obtains the semantic distance from the semantic comparator of the source user without knowing $\Pi$. The destination user can apply the projection-based encoding solution developed for the semantic encoder to improve the transmission efficiency for sending the generated paths at the source user. 

 \item \textbf{Policy --} We define the policy $\hat{\Pi} _{\pi _{\theta}}$ as a neural network, called policy network, parameterized by $\theta$, specifying the reasoning mechanism. Our policy network maps the current state $\bs_t$ into 
 the possible relations to extend the paths. For example, if we adopt an $M$-layer neural network, we can write the output of the policy network as
 \begin{eqnarray}
\bl_m = \hat{\Pi} _{\pi _{\theta_m}} \left( \bl_{m-1}; \theta_m \right), \;\;\; m = 1, \ldots, M,
 \end{eqnarray}
	where $\hat{\Pi} _{\pi _{\theta_m}} \left( \bl_{m-1}; \theta_m \right)$ specifies the connections in the $m$th layer of the network with parameters $\theta_m$ with the form $\pi_{\theta_m} \left( \bl_{m-1}; \theta_m \right)  = \sigma\left(\bw_m \bl_{m-1} + \bb_m \right)$ where $\sigma\left(\cdot\right)$ is the activation function (e.g., ReLU, softmax, sigmod, etc.) and $\bw_m$ and $\bb_m$ are the parameters of $\theta_m$, i.e., $\theta_m = \{\bw_m, \bb_m \}$ and $\theta = \langle {\theta_m} \rangle_{m=1, \ldots, M}$. 
 
\end{itemize}

Note that the policy network $\pi_{\theta}$ only specifies the probability distribution of relations extended from a set of current paths. The semantic interpreter however needs to utilize the learned policy to recover a set of possible full reasoning paths, each consists of a sequence of entities and relations. 
We, therefore, define a reasoning mechanism $\hat \Pi_{\left(\pi_{\theta}, L\right)}$ to represent a probability distribution of all the possible paths generated according to the given policy $\pi_\theta$ with the maximum path length constraint $L$. The value of $L$  can be closely related to the depth of meaning that can be expressed by the source user. For example, a shorter reasoning path represents a relatively more straightforward meaning. As the length of the path increases, the chance of disclosing a deeper meaning increases. However, this will also increase the search space and the probability of misrepresenting the true meaning of the source user due to overfitting. 

More formally, the main objective of the semantic interpreter is to minimize the semantic distance between the expert semantic paths and the interpreted paths generated by the policy network of the destination user, i.e., we can write the optimization problem as follows: 
\begin{eqnarray}
\underset{\pi_\theta}{\min} \left( {\mE_{\eta \sim {\Psi_{{\hat \Pi}_{\pi_\theta}}}}} [{\log \varpi^*_{\phi}\left(\eta\right)}] +{\mE_{\eta \sim {\Psi_{{\Pi}}}}}[1- \log \varpi^*_{\phi}\left(\eta\right)]\right)
\label{eq_interpreter}
\end{eqnarray}
where $\varpi^*_{\phi}\left( \cdot \right)$ is the optimal semantic comparator obtained by the source user. 

Note that in the previous discussion, we assume all the explicit semantics sent by the source user can be successfully received and recovered by the destination user as an input to the semantic interpreter, i.e., the channel fading and additive noise in equation (\ref{eq_ReceivedEmbeddingGaussianNoise}) do not affect the relative distance between different entities or cause any misinterpretation of entities.  However, in many practical scenarios, it is possible that some visible entities and relations can be corrupted during the physical channel transmission. 
We thus propose two semantic interpretation/decoding schemes, soft and hard decoding, according to whether or not to apply the projection function-based solution learned from (\ref{eq_TransELoss}) before applying the path reasoning mechanism learned by the semantic interpreter. 
\begin{itemize}
    \item {\em Hard Interpretation:} The destination user will first recover every corrupted explicit semantics including both visible entity and relation from the received signals by using the projection function learned from (\ref{eq_TransELoss}) to match the corrupted entities and relations to the closest entities (or relations). 
    It will then apply the reasoning mechanism learned by the semantic interpreter to generate the possible reasoning paths. 
    
    \item {\em Soft Interpretation:} The destination user will not recover any corrupted entities and relations, but will directly apply (\ref{eq_interpreter}) to search for the optimal reasoning paths based on the corrupted explicit semantics obtained from its received signal. 
\end{itemize}

It can be observed that the above two semantic interpretation solutions have various advantages and disadvantages when being applied to different scenarios. In particular, hard interpretation performs better when most of the corrupted entities and relations can be successfully recovered, e.g., in high SNR scenarios. Meanwhile, the soft interpretation works better when the corrupted visible entities and relations are unrecoverable, e.g., in low SNR scenarios. We will give a more detailed discussion based on the simulation results in Section \ref{Section_NumericalResult}. 

\subsection{G-RML Algorithm and Theoretical Analysis}



Our proposed iSAC architecture consists of two phrases of implementation. In the communication phrase, the source user will first detect the explicit semantic terms from the source signal. It will then apply semantic encoder to convert the  explicit semantics into a low-dimensional signal to be sent to the physical channel. The destination user will first recover the explicit semantics from its received signal using a pre-trained projection function. It will then generate the implicit semantics based on the explicit semantics using a pre-trained semantic interpreter.  

The training phrase of iSAC architecture consists of a joint training algorithm, called G-RML algorithm, that jointly trains the semantic comparator and semantic interpreter. In particular,  
the semantic interpreter will try to generate a reasoning path extended from the recovered explicit semantics via a series of reasoning episodes based on the trained semantic reasoning network. Particularly, in each reasoning episode $t$, the semantic interpreter chooses an action, a set of relations to extend the semantic paths, based on the current state $\bs_t$. When the next relations are decided, the  semantic interpreter will update the path $\eta_t$ by concatenating the selected relation $r_{t+1}$ as well as the linked hidden entity $e_{t+1}$. By repeating the above process, the semantic reasoning paths will be sequentially extended until the maximum length of the reasoning path $L$ arrives. Note that the value of $L$ can be empirically selected based on the depth of the meaning in the previous communication between the source and destination users.

The detailed procedures of semantic interpreter during communication phrase as well as the training procedures of G-RML algorithm are 
described in Algorithms 1 and 2, respectively. 

\begin{algorithm}
\scriptsize
\caption{Semantic Interpreter (Communication Phrase)}
\label{Algorithm_Interpreter}
{\bf Input}: policy $\pi_\theta$, received explicit symbols (e.g., entities and/or relations) ${\bv}$ and maximum length $L$ of reasoning paths  \\
{\bf Output}: Set of possible implicit semantic reasoning paths $\hat{\mathcal{K}}$ \\
{\bf For} reasoning episode $j=0,1,2\ldots$ {\bf do}
\begin{itemize}
    \item Initialize $\eta_0 = \bv$, $\bs _0 = ( \bv, 0)$
    \item {\bf For } each step $t$ {\bf do}
    \begin{itemize}
        \item [-] Get action $a_t\leftarrow r^{t}\sim \pi_\theta (a_t|\bs_t)$
        \item [-]{\bf If} $r^t$ is valid {\bf Then}\\
        Choose its connected entity $e^t$
        \item [-] Update path $\eta_t\leftarrow {\rm concatenate} (\eta_t, r^t,e^t)$
        \item [-] Update state $\bs_{t+1}=(\eta^t, t)$
        \item [-] {\bf If} max number of hops is reached {\bf Then} \\Store path $\eta_t$ in $\hat{\mathcal{K}}$
    \end{itemize}
    \item {\bf End For}
\end{itemize}
\normalsize
\end{algorithm}

%

\begin{algorithm}
\scriptsize
\caption{G-RML Algorithm (Training Phrase)}
\label{Algorithm_JointTraining}
{\bf Input}: Observed visible entities $e^0$, expert semantic path set $\mathcal{K}$, initial policy network $\pi_{\theta_0}(a_0|s_0)$ and comparator network $\varpi_{\phi_0}(\bp^0)$ with parameters $\theta_0, \phi_0$, and max length of hops $L$\\
{\bf Output} Learned reasoning policy $\pi^*_{\theta}$ at the destination user\\
{\textbf{For}} each training iteration $t$ {\textbf{do}}
\begin{itemize}
    \item  Destination user generates a semantic path set $\hat{\mathcal{K}}_t$ from policy $\pi_{\theta_t}$ based on Algorithm \ref{Algorithm_Interpreter} to be sent to the source user;
    \item  Source user updates semantic comparator parameters $\phi_t$ to $\phi_{t+1}$ with the following gradient:
    \begin{equation}
    \centering
    \begin{aligned}
     &{\Psi_{{\hat \Pi}_{\pi_\theta}}}[\nabla _{\phi}\log \varpi_{\phi}(\eta)] +
     {\Psi_{\Pi}}[\nabla _{\phi}(1-\log \varpi_{\phi}(\eta))]
    \end{aligned}
    \end{equation}
    \item Destination user updates reasoning policy parameters $\theta_t$ to $\theta_{t+1}$ by minimizing the cost function with Monte Carlo Policy Gradient:
    \begin{equation}
        \begin{aligned}
         \mathbb{E}_{\bp_{t}^D\in  \hat{\mathcal{K}}}[\nabla_{\theta}\log(\pi_{\theta}(\ba|\bs)Q(\bs, \ba))], 
\mbox{where}\,\, Q(\bs, \ba) = \log (\varpi_{\phi_{t+1}}(\eta))
        \end{aligned}
    \end{equation}
{\textbf{End For}}
\end{itemize}
\normalsize
\end{algorithm}

We can prove that, despite the destination user cannot observe any expert semantic paths, by applying the proposed G-RML algorithm, the semantic interpreter can always generate the semantic paths that follow the same probability distribution as the paths generated by the true reasoning mechanism of the source user. 

\begin{theorem}
\label{MainConvergTheorem}
Suppose ${\Delta}|\Pi \neq \emptyset$ and ${\hat \Delta}|\hat \Pi_{\pi_\theta} \neq \emptyset$. 
Then, there always exists an optimal solution $\varpi^*_\phi$ of (\ref{eq_comparator}) under a given $\pi_\theta$. 
When $\pi_\theta$ converges to the optimal solution $\pi^*_\theta$ of (\ref{eq_interpreter}), the expected semantic distance between the expert paths and the path generated by the semantic interpreter at the destination user approaches zero, i.e.,  distribution $\hat \Pi_{\left(\pi_{\theta},L\right)}$ of generated semantic paths at the destination user approaches the distribution $\Psi_{{\Pi}}$ of expert paths.
\end{theorem}
\begin{IEEEproof}
See Appendix \ref{Appendix_TheoremProof}.
\end{IEEEproof}

The above theorem suggests that by learning by reasoning policy, the semantic interpreter at the destination user can always generate reasoning paths that follow the same distribution as the expert paths. 
We can also observe that our proposed solution does not require the destination user to estimate any reward function. Also, since the destination user always tries to learn a reasoning policy to map any given set of explicit semantics to the possible relations, our proposed solution scales well to more complex scenarios involving 
a large number of entities and relations. 

\section{Simulation Results and Analysis}
\label{Section_NumericalResult}

\subsection{Dataset and Simulation Setup}
We simulate the semantic-aware communication process between a source and destination user pair based on the entities and relations sampled from three commonly used real-world knowledge base datasets, NELL-995\cite{carlson2010toward} FB15k-237\cite{toutanova2015representing}, and WN18RR\cite{dettmers2018convolutional} where NELL-995 consists of $754,920$ unique entities and $200$ types of relations, FB15k-237 consists of $14,541$ unique entities and $237$ types of relations, and WN18RR consists of $40,943$ unique entities and $11$ types of relations. 
We first sample a set of sub-knowledge graphs (SKGs) from these real-world datasets to simulate the knowledge base of each user (source or destination user) and then generate expert paths from each SKGs using a two-sided breadth first search algorithm to simulate the source user who tends to express his/her meaning based on some reasoning paths. Motivated by the fact that most existing deep learning-based solution can only perform object identification and recognition and have limited performance when being applied to identify relations, in this section, we assume that the source user can only observe entities at the beginning of generated paths and all relations are implicit semantics that cannot be directly observed by source user.

We set the semantic interpreter as a fully-connected network consisting of two hidden layers, each followed by a Rectified Linear Unit (ReLU) and one output layer. The output of the interpreter is normalized using a softmax function. For the semantic comparator, we adopt a two-layer fully-connected network with one hidden layer and one output layer. The output layer is normalized by a sigmoid function while others are activated by ReLU. Our simulations are performed based on Tensorflow open source platform on a workstation with an Intel(R) Core(TM) i9-9900K CPU@3.60GHz, 128.0 GB RAM@2133 MHz, 2 TB HD, and two NVIDIA Corporation GP102 [TITAN X] GPUs.

\vspace{-0.2in}

\subsection{Numerical Results}

\begin{figure}
  \begin{minipage}[t]{0.24\textwidth}
   \centering
   \includegraphics[width=\textwidth]{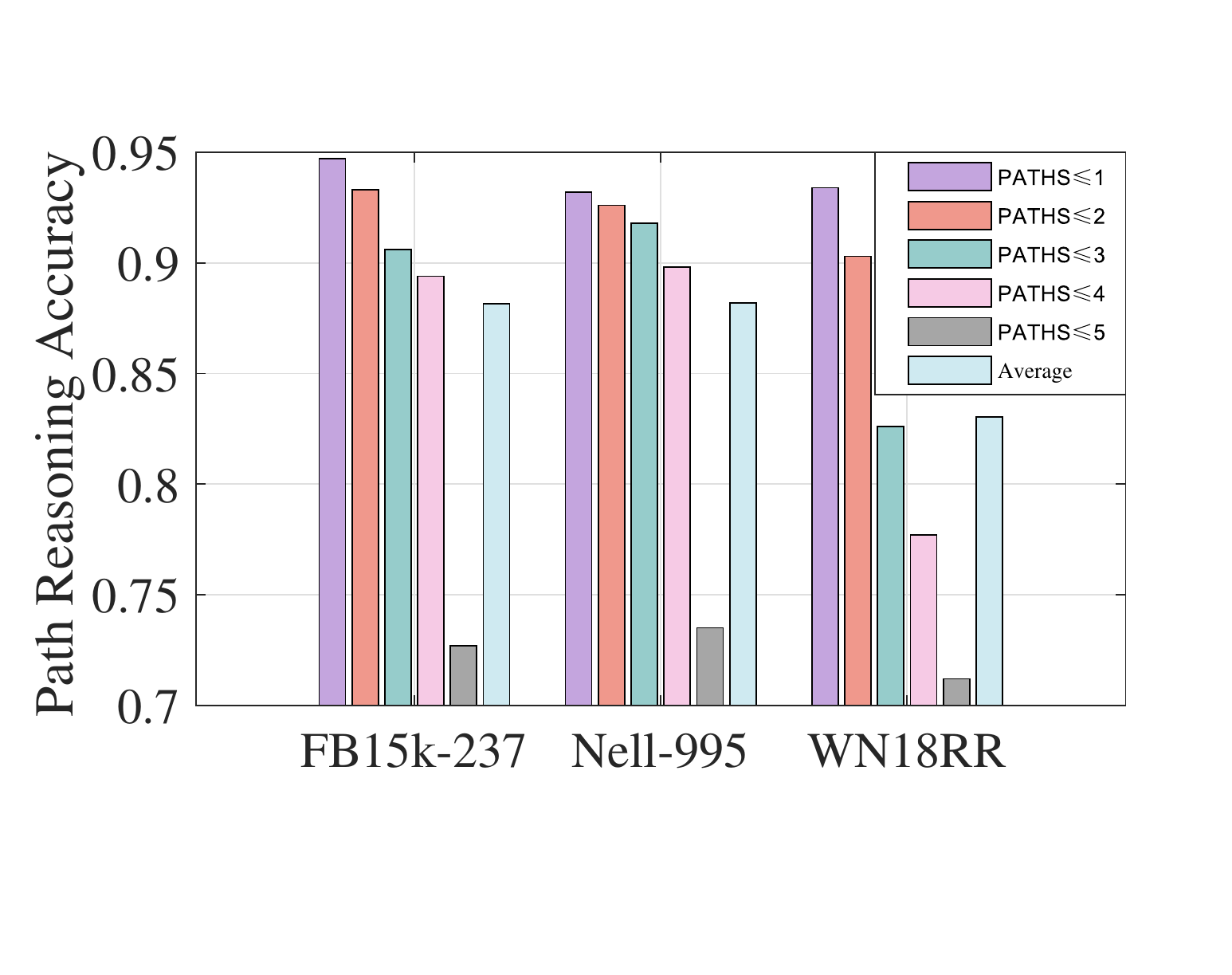}
   \vspace{-0.45in}
	 \caption{\footnotesize{Accuracy of implicit  semantic paths inferred by  destination user using  G-RML  under different path lengths.} }
   \label{Fig_AccuracyMaxLength}
  \end{minipage}
    \begin{minipage}[t]{0.24\textwidth}
	\centering
	\includegraphics[width=\textwidth]{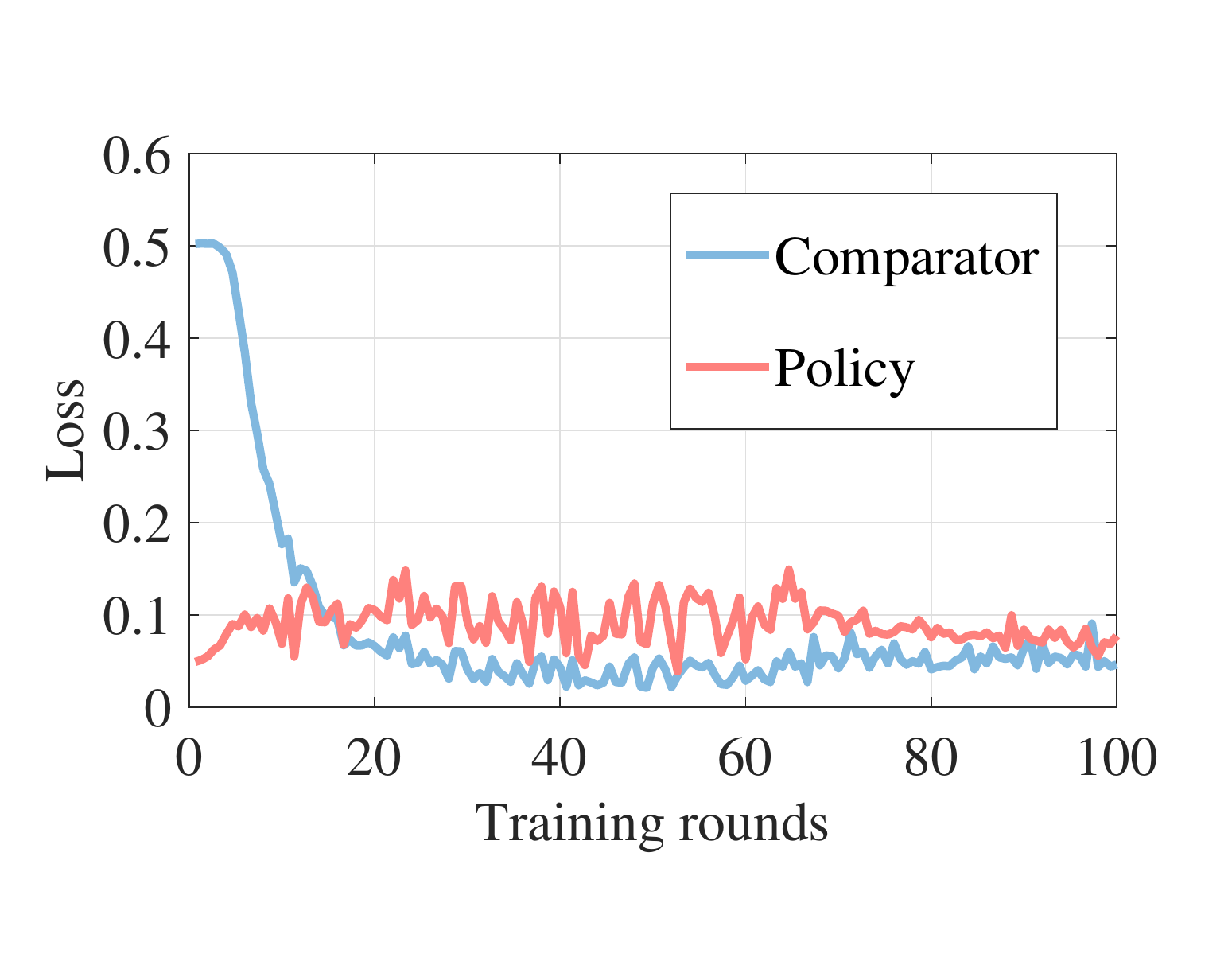}
	\vspace{-0.45in}
	\caption{\footnotesize{Loss of semantic comparator  and interpreter under different training  rounds.}}
	\label{Fig_convergence}
  \end{minipage}%
  
  \vspace{-0.15in}
  \begin{minipage}[t]{0.24\textwidth}
	\centering
	\includegraphics[width=\textwidth]{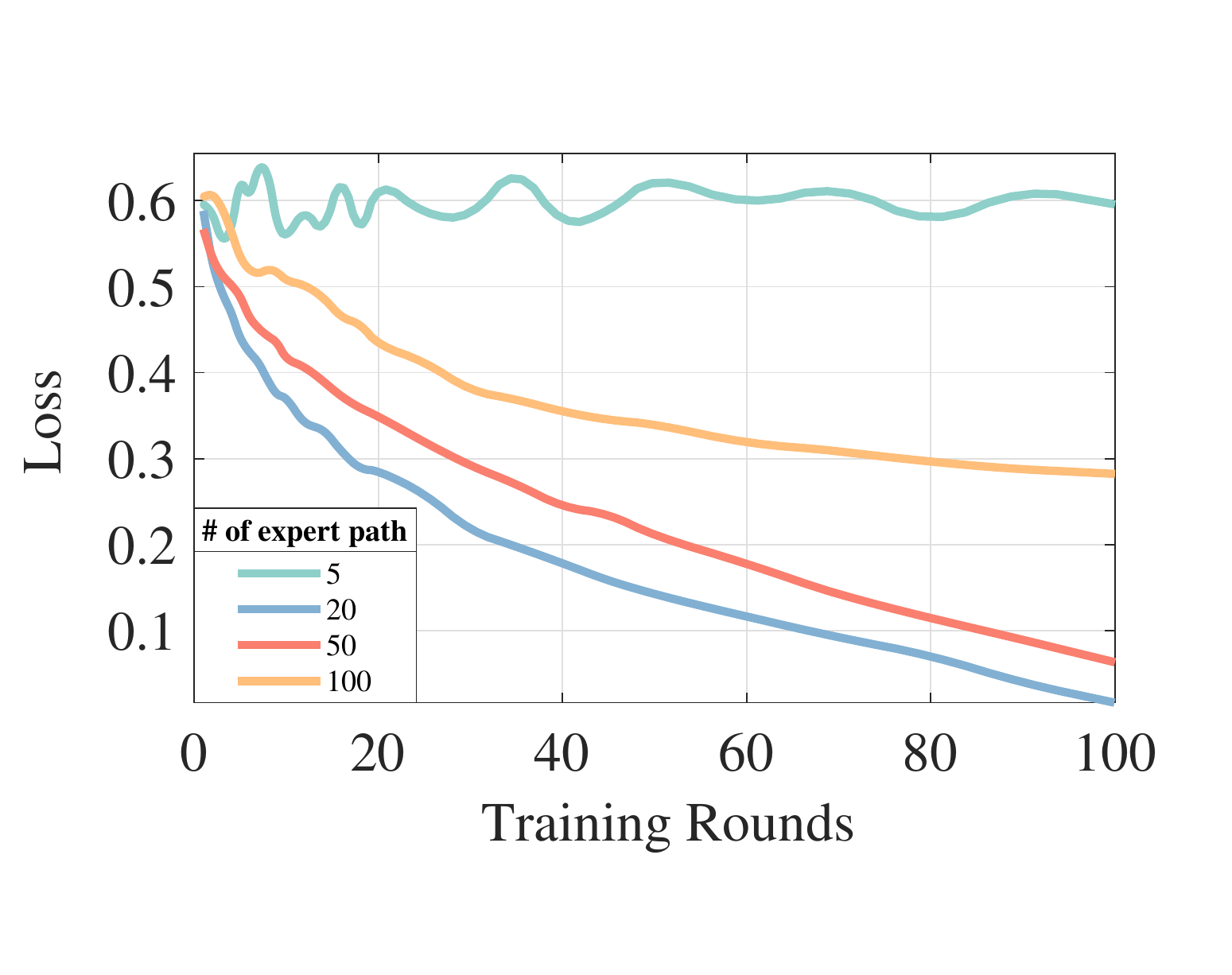}
	\vspace{-0.35in}
	\caption{\footnotesize{Loss of G-RML  when \\ being trained with different  numbers \\ of expert paths.}}
	\label{Fig_NumExpertPath}
  \end{minipage}%
    \begin{minipage}[t]{0.24\textwidth}
	\centering
	\includegraphics[width=\textwidth]{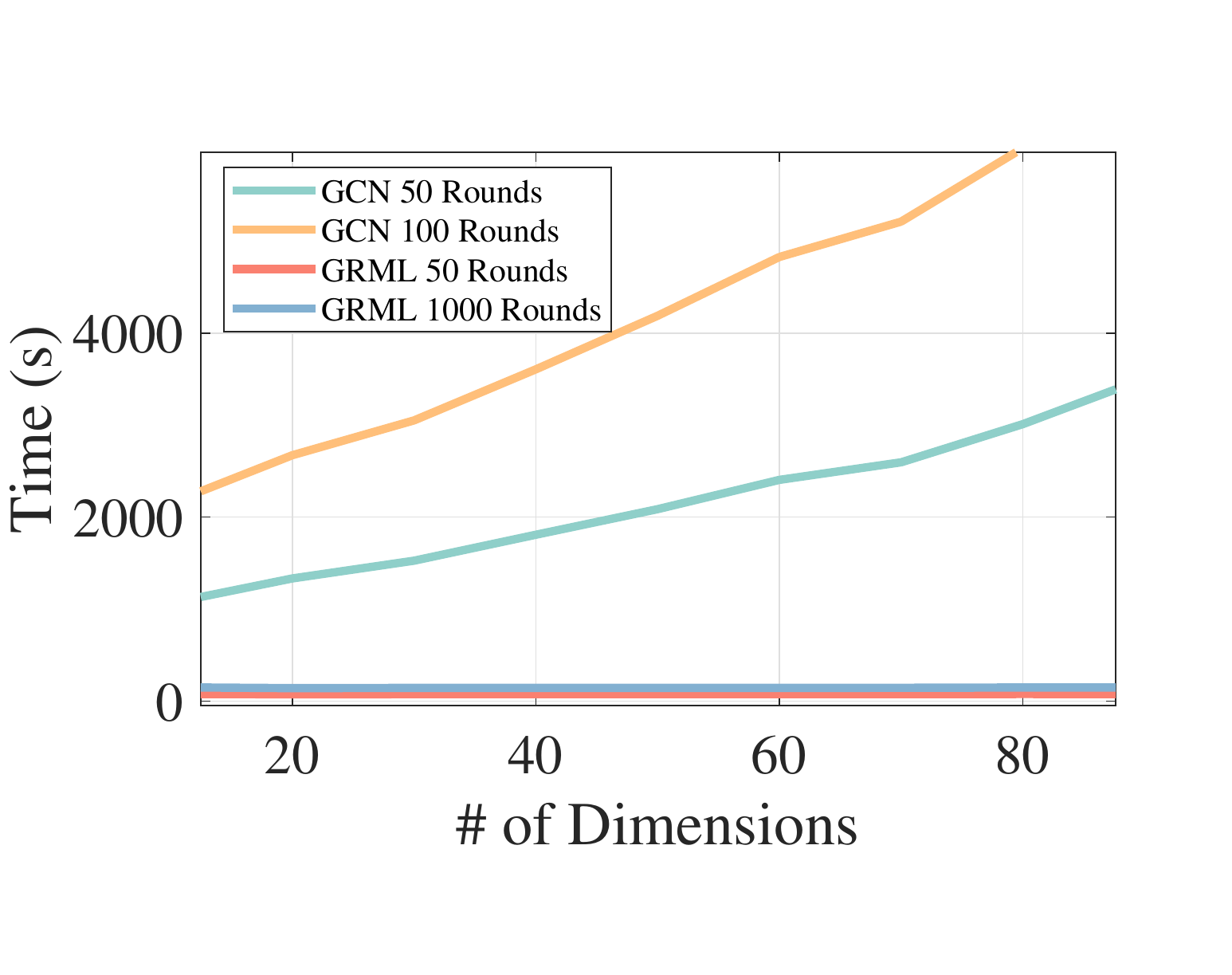}
	\vspace{-0.35in}
	\caption{\footnotesize{Comparison of the run time (sec) required for training semantic encoder based on our proposed solution and a GCN-based solution. }}
	\label{Fig_complexity}
  \end{minipage}%
  \vspace{-0.25in}
\end{figure}

As mentioned earlier, the main idea of G-RML is to train a reasoning policy to sequentially decide the next possible relations to extend the reasoning paths. In other words, the maximum length limit of the reasoning paths will directly affect the complexity and accuracy of our proposed G-RML algorithm. We therefore first evaluate the impact of the maximum length limits of the semantic reasoning paths on path reasoning accuracy of the semantic interpreter of the destination user, measured by the similarity between the expert paths and the synthetic paths generated by the semantic interpreter. In particular, in Fig. \ref{Fig_AccuracyMaxLength}, we consider the expert reasoning paths randomly generated from three different knowledge base datasets with different path length limits (from 1 to 5) and compare the resulting path reasoning accuracy achieved by G-RML. We can observe that the accuracy of the semantic interpreter decreases with the maximum lengths of expert paths for all three datasets. This means that the deeper the semantic reasoning paths the more difficult for the semantic interpreter to generate accurate paths that mimic the distribution of the expert paths. We however observe that, when the maximum length limit increases, the accuracy of decreasing rates of G-RML varies between different datasets. In particular, the accuracy decreasing rate for FB15k-237 and Nell-995 drops slowly when the maximum limits of the path length are between 1 and 4 and decrease sharply when the length limit increases from 4 to 5. In WN18RR, however, the accuracy decreasing rate changes almost linearly when the reasoning path length limit increases from 1 to 5. This is because the entities in FB15k-237 and Nell-995 are more closely related to each other, e.g., the average number of relations connected to each entity is higher, compared to WN18RR, and therefore when the length of the reasoning paths increases above 4, it is very difficult to find many expert paths with all the entities being different from each other.

In Fig. \ref{Fig_convergence}, we verify the convergence of G-RML by presenting the loss values of the semantic comparator and interpreter under different training rounds defined in (\ref{eq_comparator}) and (\ref{eq_interpreter}), respectively. We can observe that the loss functions of both comparator and interpreter approach relatively stationary values with only a limited number of training rounds (e.g., less than 50 rounds of training). This means that the communication overhead for training a relatively satisfactory model at the semantic interpreter can be limited to a moderate value.

In Fig. \ref{Fig_NumExpertPath}, we evaluate  the convergence of G-RML when the model has been trained with different numbers of expert paths. We can observe that when the number of expert paths is limited (e.g., 5), the proposed algorithm may not even converge when the number of training rounds becomes large. This is because, in this case, the limited number of sampled expert paths may not be able to capture the real distributions of the source user's semantic meanings. However, when the number of observable expert paths grows, the convergence performance will first increase and then slowly decrease. This is because the convergence of the reasoning mechanism learning solution is closely related to the diversity of the probability distribution of the observed expert paths. When the number of expert paths is limited (e.g., 5), the overall semantic diversity of different expert paths will also be limited which results in a faster convergence rate. However, when the total number of expert paths continues to grow to a large number (e.g., from 20 to 50), the convergence speed of G-RML becomes slower due to the increase in the diversity of the observed expert paths.

\blu{In Fig. \ref{Fig_complexity}, we evaluate the overall time required for training the semantic encoder using our proposed solution with 200 expert semantic paths and a 2-layer GCN-based solution based on  FB15k-237 dataset under different dimensional sizes of explicit semantics. We can observe that the overall training time of our proposed semantic encoding function is much shorter than that required to train a GCN-based solution. This is because the number of semantic entities involved in most knowledge base is very large (e.g., 14,541 entities in our considered dataset) which results in long computational delay during each iteration of GCN model training process. } 

 \begin{figure}
  \begin{minipage}[t]{0.24\textwidth}
   \centering
   \includegraphics[width=\textwidth]{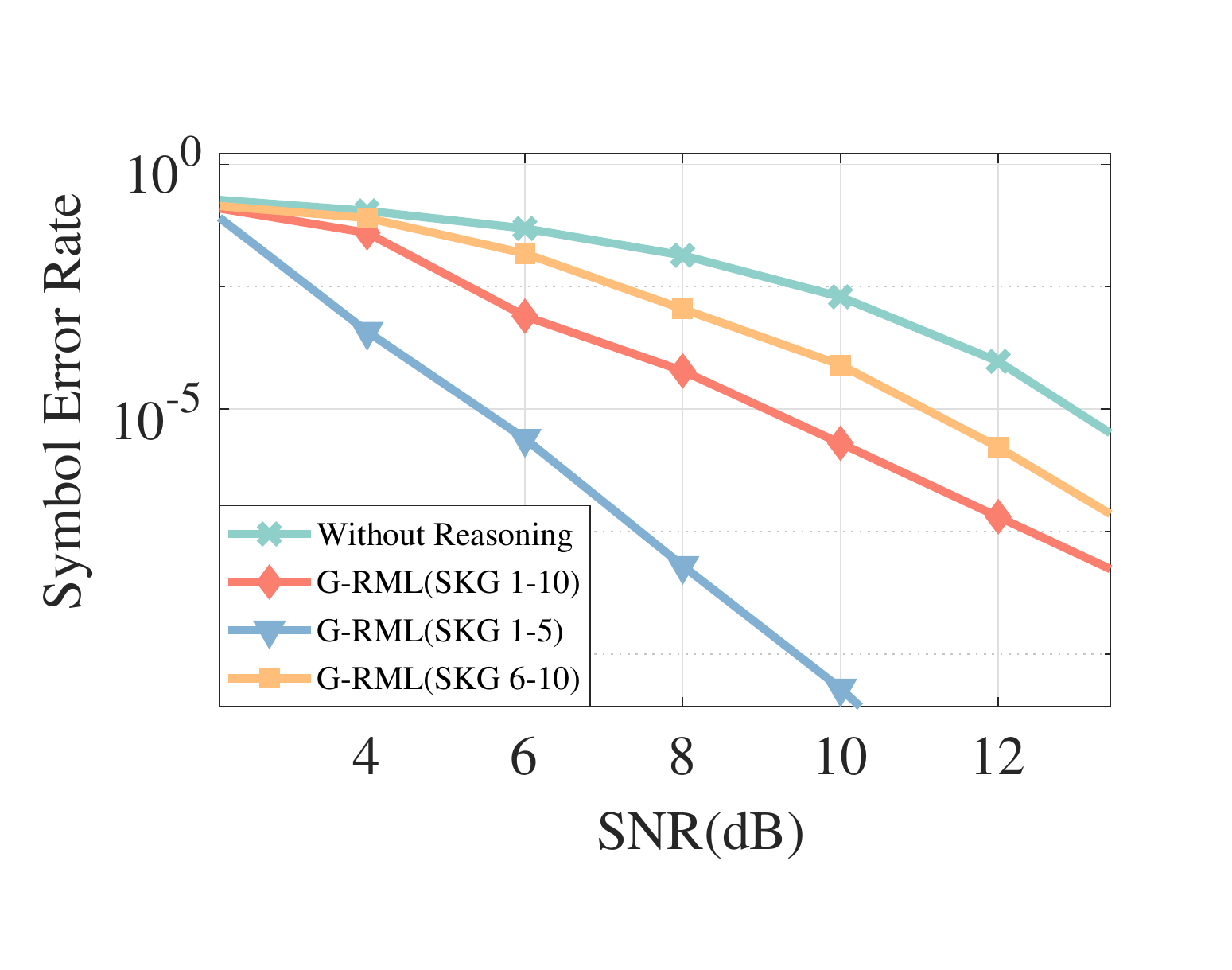}
   \vspace{-0.35in}
	 \caption*{\footnotesize{(a)}}
   \label{Fig_SNR noise embedding Nell-995}
  \end{minipage}
  \begin{minipage}[t]{0.24\textwidth}
	\centering
	\includegraphics[width=\textwidth]{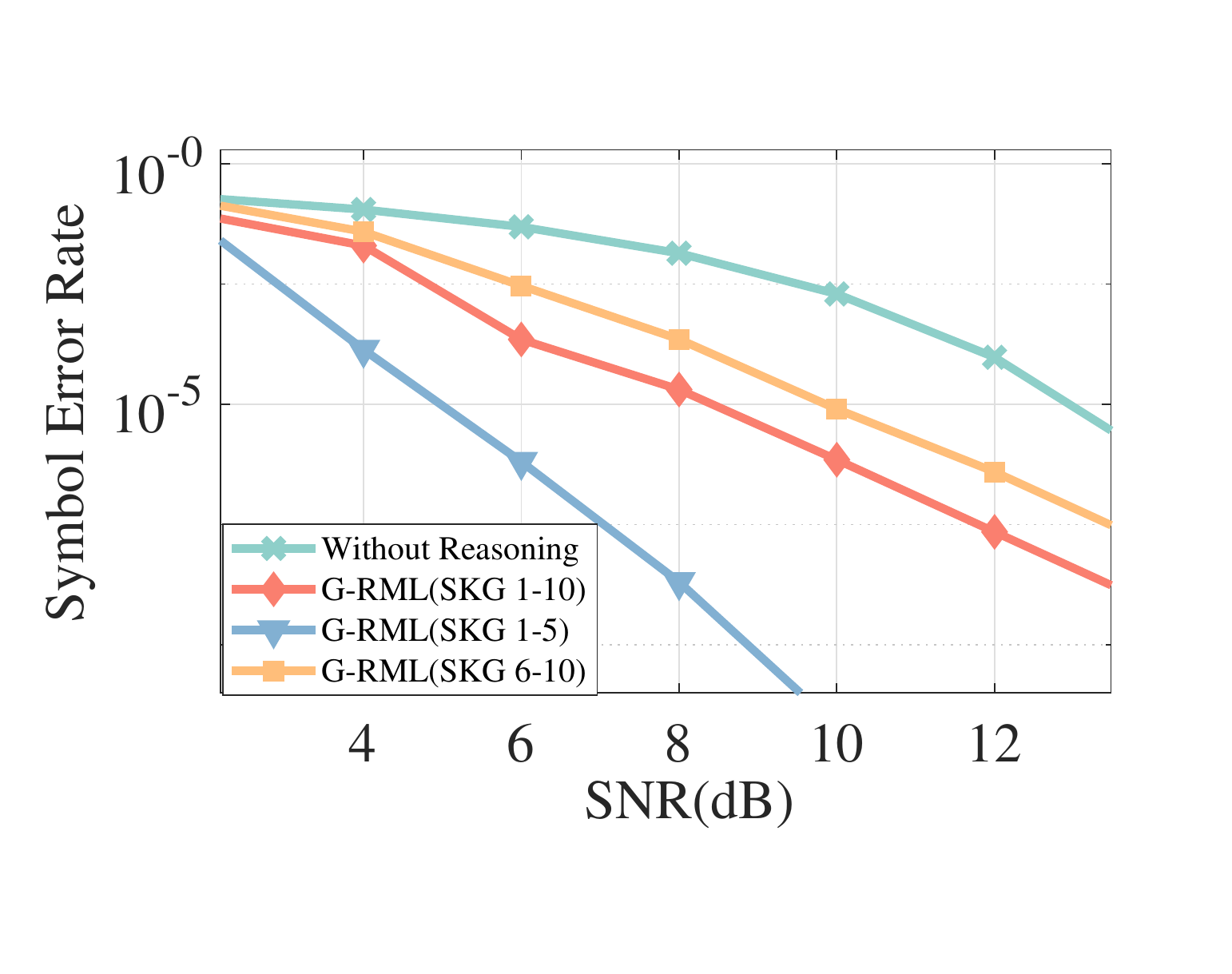}
	\vspace{-0.35in}
	\caption*{\footnotesize{(b)}}
	\label{Fig_SNR noise on TransE Nell-995}
  \end{minipage}
  \begin{minipage}[t]{0.24\textwidth}
    \centering
    \includegraphics[width=\textwidth]{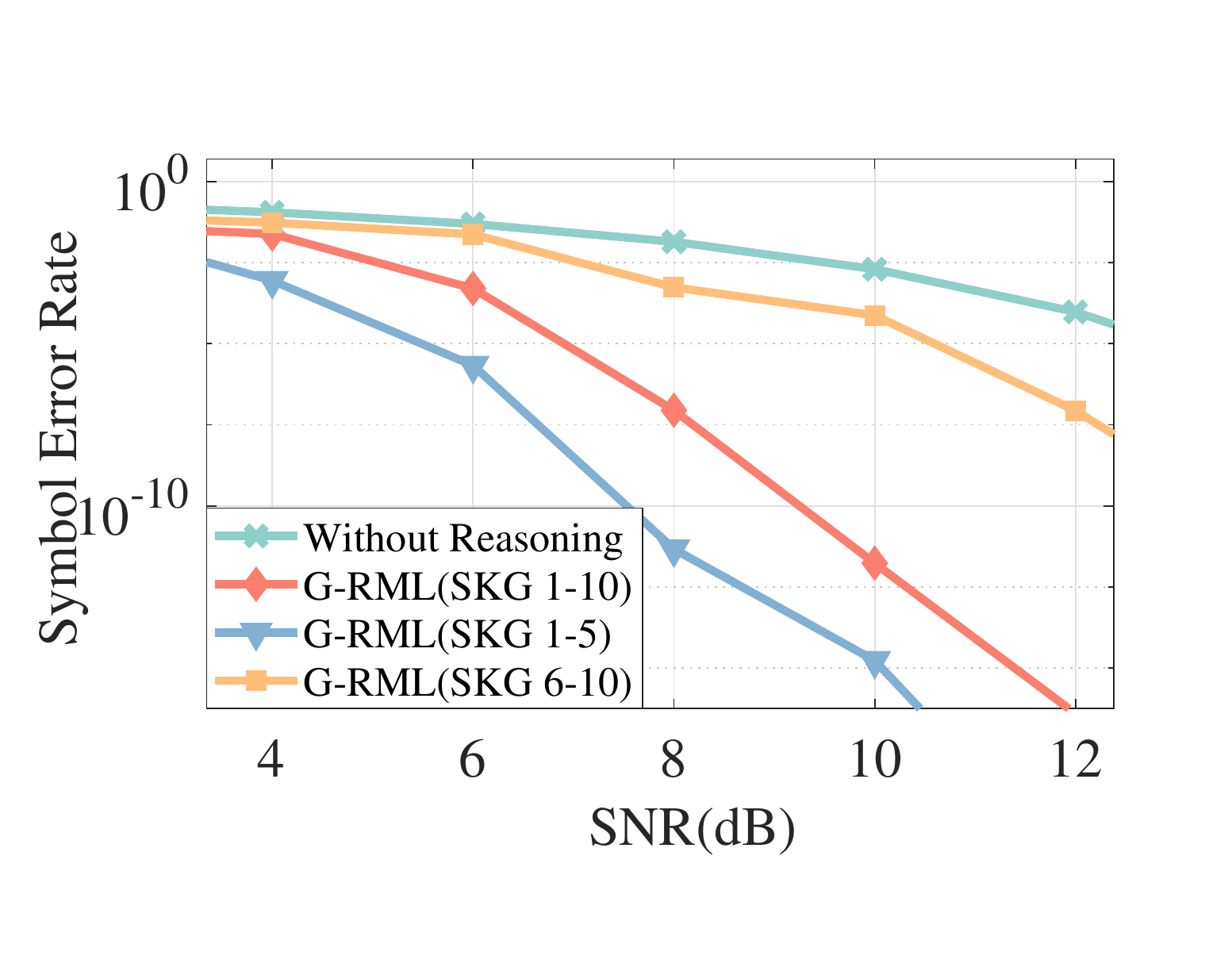}
    \vspace{-0.35in}
    \caption*{\footnotesize{(c)}}
    \label{Fig_SNR noise embedding WN18RR}
   \end{minipage}
   \begin{minipage}[t]{0.24\textwidth}
    \centering
    \includegraphics[width=\textwidth]{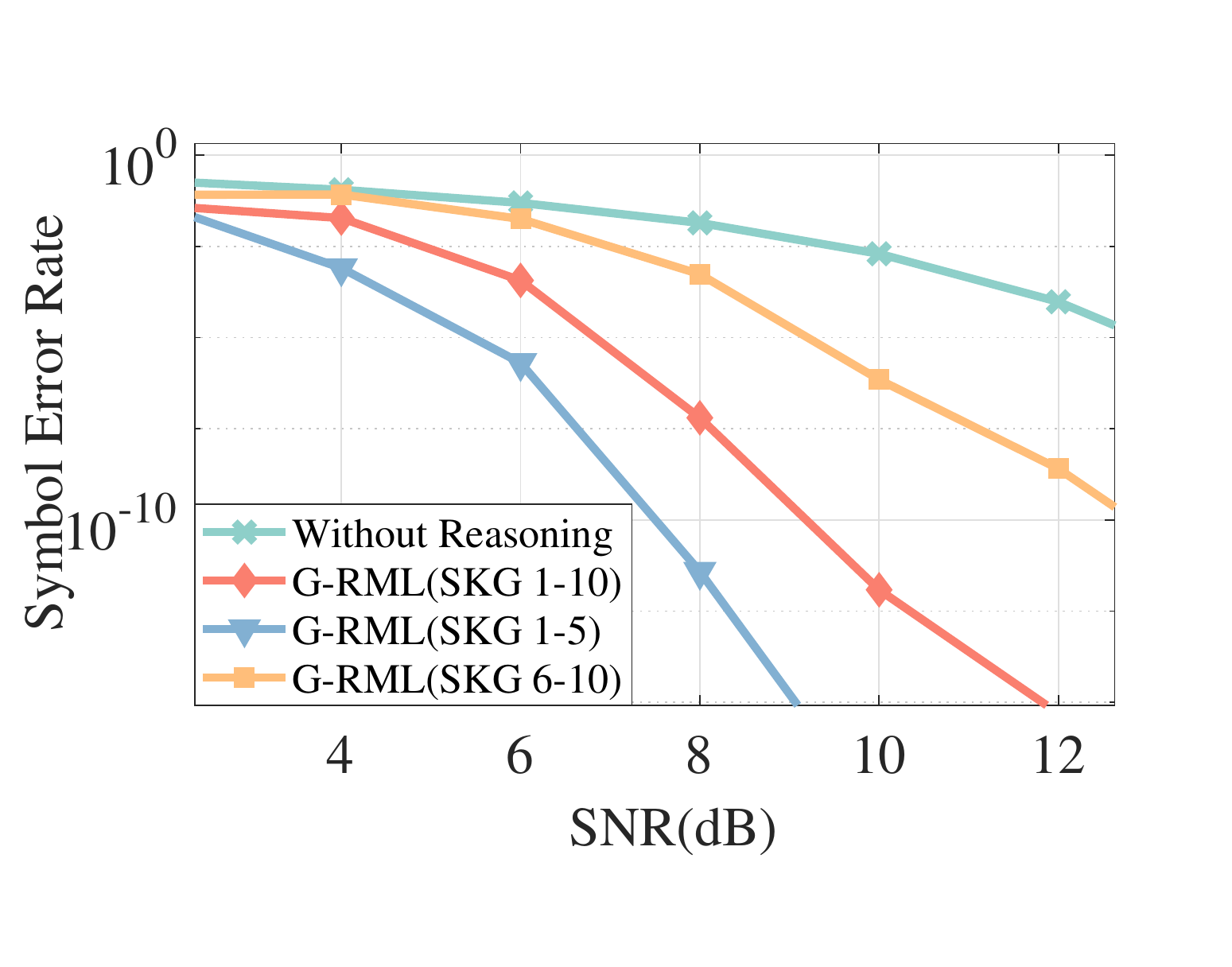}
    \vspace{-0.35in}
    \caption*{\footnotesize{(d)}}
    \label{Fig_SNR noise on TransE WN18RR}
    \end{minipage}
    \vspace{-0.15in}
      \caption{\footnotesize{Symbol (entity) error rate with different lengthed reasoning paths under different  SNRs based on SKGs sampled from Nell-995 when using (a) soft interpretation and (b) hard interpretation, WN18RR when using (c) soft interpretation and (d) hard interpretation.}}
      \label{Fig_SNRHardandSoft}
    \vspace{-0.3in}
   \end{figure}

In Fig. \ref{Fig_SNRHardandSoft}, we consider the lossy channel transmission of the semantic message and evaluate the semantic-aware communication performance of our proposed G-RML algorithm by investigating the symbol error rate of semantic-aware communication in an additive Gaussian noise channel when the semantic interpreter adopts both soft and hard interpretation approaches to recover the semantic meanings the symbols (entities) corrupted by the physical channel. 
%
%
%
As mentioned earlier, the accuracy of semantic reasoning is closely related to the structural feature of the considered SKGs. For example, the density of connections among entities may reflect the similarity of meaning of these entities, i.e., the higher the density of connections, the closer the meaning among these entities. To investigate the impact of meaning similarity of entities on the error correction performance, we consider 10 SKGs, each consists of an exclusive set of $75,492$ entities linked with different numbers of relations and reasoning paths. We then rank these 10 SKGs according to their connection densities from the highest to the lowest to simulate the communication involving entities with different levels of semantic meaning diversity, i.e., semantic diversities of SKGs 1 and 10 are the lowest and the highest, respectively. 

We also present the symbol error rate achieved by our proposed reasoning-based recovery solution with different combinations of SKGs, compared to the scenarios without semantic reasoning, in Fig. \ref{Fig_SNRHardandSoft} (a) and (b). We can observe that, compared to the traditional communication solution without any semantic reasoning, the G-RML can provide 18 dB and 26 dB improvement (when SNR=4 dB) in terms of semantic symbol error rate when adopting soft and hard interpretation schemes, respectively.  We can also observe that our proposed solution can significantly improve the robustness of communication, especially when the  semantic diversity of messages is limited. When the semantic diversity increases, the recovery performance of the semantic interpreter decreases. We can also observe that soft interpretation performs slightly better than the hard interpretation approach in low SNR scenario. However, when the SNR increases, the hard interpretation can achieve a higher recovery rate compared to the soft interpretation. To compare the robustness of our proposed reasoning-based recovery approach under different datasets, we present the symbol error rate when the SKGs  are sampled from WN18RR in Fig. \ref{Fig_SNRHardandSoft} (c) and (d). We can observe similar results as that in Nell-995. More specifically, compared to the communication solution without semantic reasoning, the G-RML can provide 23 dB and 31 dB improvement (when SNR=4 dB) in terms of semantic symbol error rate when adopting soft and hard interpretation schemes, respectively. An interesting observation is that compared to Nell-995, the hard interpretation achieves more performance improvement over soft interpretation, especially in high SNR scenarios for SKGs sampled from WN18RR. 

\begin{figure*}[htbp]
 \centering
 \vspace{-0.25in}
 \begin{minipage}{0.50\linewidth}
  \centering
\subfigcapskip=-5pt
  \subfigure[d=10]{
   \centering
   \includegraphics[width=1.4in]{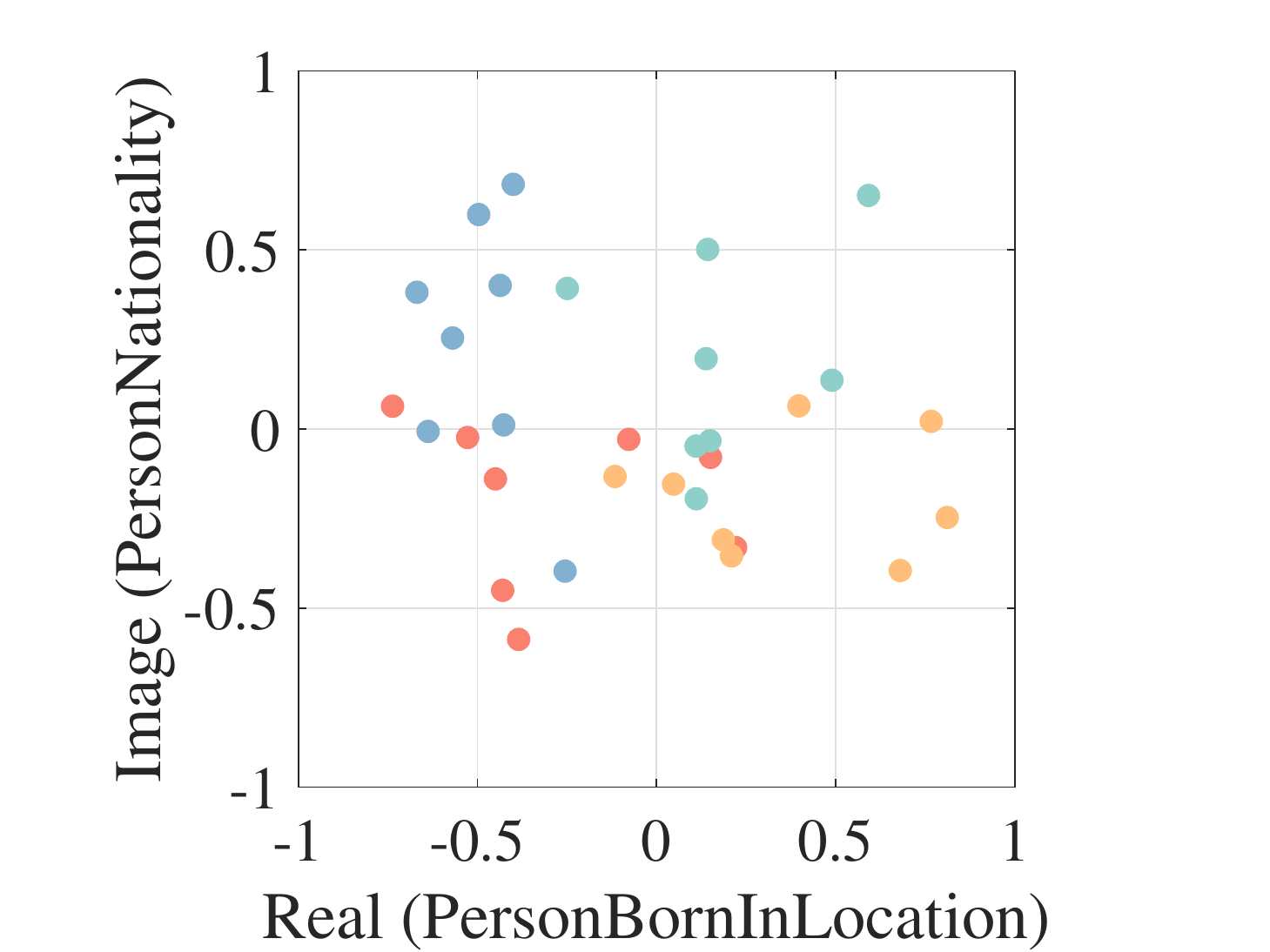}
}
  \subfigure[d=50]{
   \centering
   \includegraphics[width=1.4in]{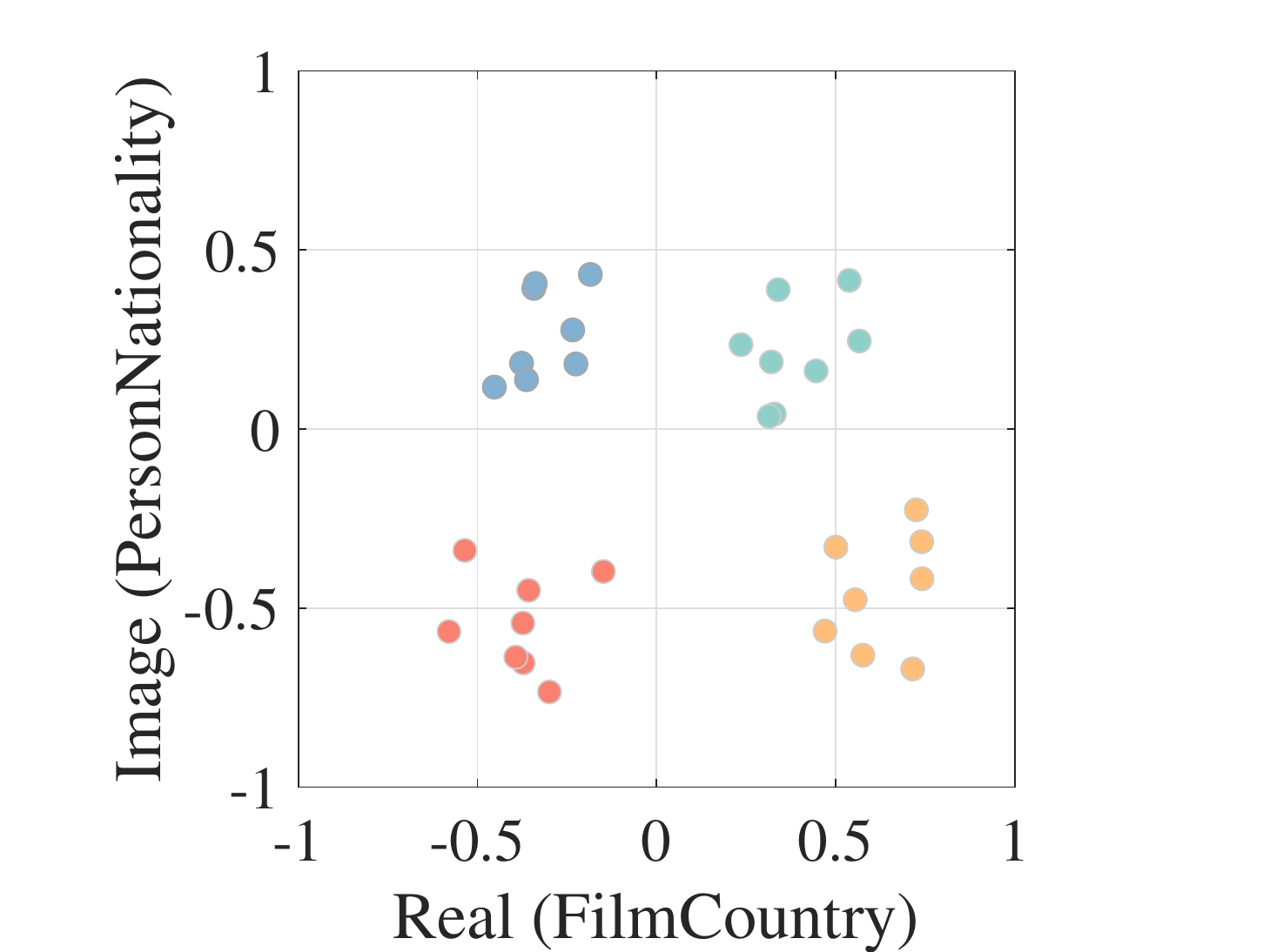}
}
  \vspace{-0.1in}
  \footnotesize
  \caption{\footnotesize{The same two-dimensional representations selected from (a) 10 and (b) 50-dimensional semantic constellation spaces converted from the SKG given in Fig.}}
  \label{Fig_SimSemanticConstellation}
 \end{minipage}
 \centering
 \hfill
\begin{minipage}{0.46\linewidth}
  \centering
\subfigcapskip=-7pt
  \subfigure[AWGN]{
   \centering
   \includegraphics[width=1.1in]{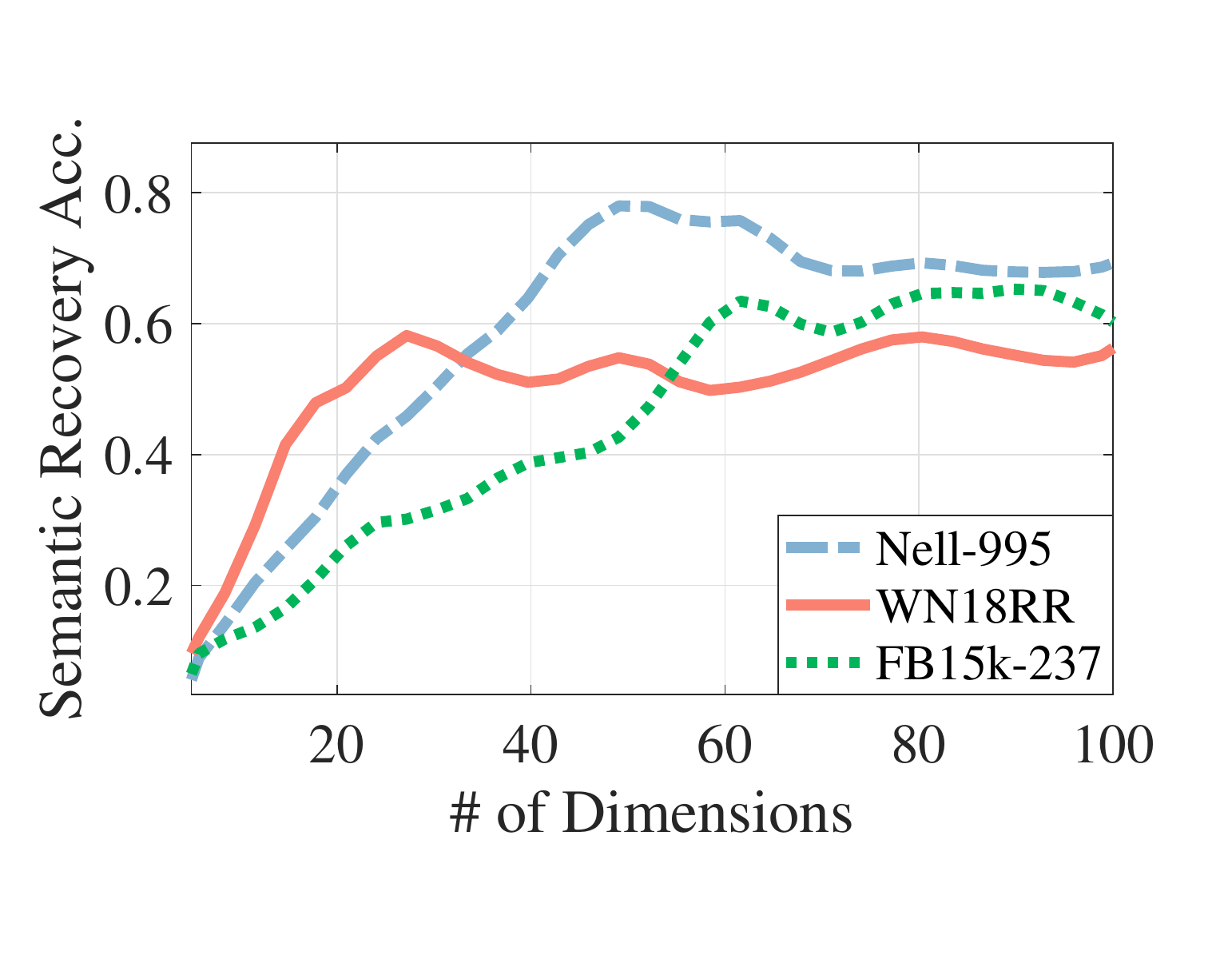}
  }
  \subfigure[Rayleigh]{
   \centering
   \includegraphics[width=1.1in]{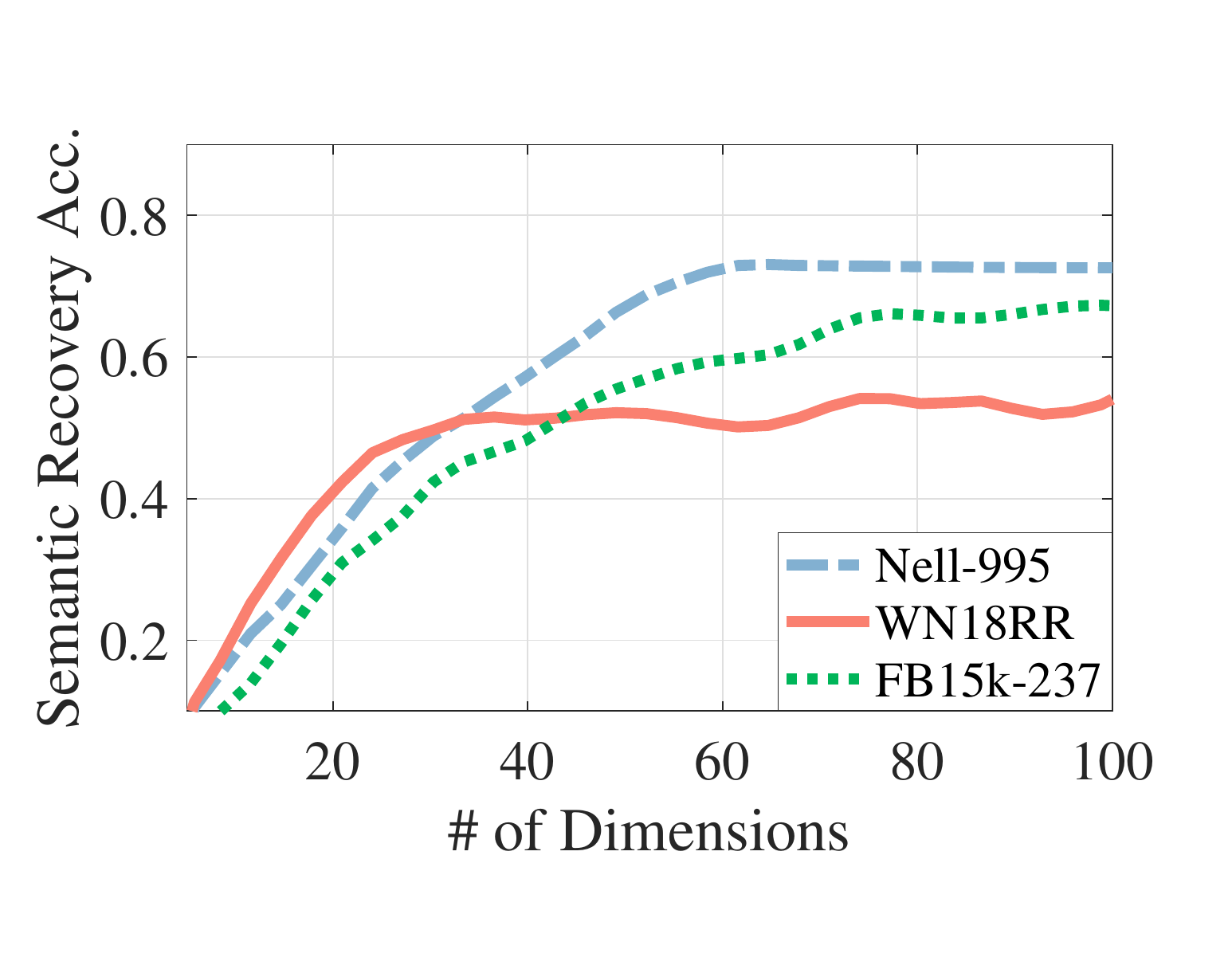}
  }
  \vspace{-0.1in}
  \footnotesize
  \caption{\footnotesize{Semantic recovery accuracy under different number of dimensions of semantic constellation transmitted in (a) AWGN channel and (b) Rayleigh fading channel (SNR=7dB).}}
\label{Fig_AccVSDimension}
 \end{minipage}
 \centering
 \vspace{-0.25in}
\end{figure*}

The key idea of semantic encoder is to convert the high-dimensional graphical presentation of explicit semantics into a low-dimensional semantic constellation space for efficient channel transmission. Generally speaking, reducing the number of dimensional size of the converted semantic constellation spaces will improve the  efficiency of semantic communication. It  will also result in low performance in semantic recovery and differentiation of semantic terms with different meaning.  
In Fig. \ref{Fig_SimSemanticConstellation}, we present the semantic constellation space representation of the SKG given in Fig. \ref{Fig_SemanticEncoding}(a) 
the high-dimensional SKG has been converted into low-dimensional semantic constellation spaces with different dimensional sizes. We can observe that entities associated with different classes cannot be separated well when the dimensional size of the converted semantic constellation space is small, e.g., when the dimensional size is 10 as shown in Fig. \ref{Fig_SimSemanticConstellation}(a). When the number of dimension size increases, the entities associated with different classes are separated further apart, e.g., when the dimensional size is 50 as shown in Fig. \ref{Fig_SimSemanticConstellation}(b). This is because more semantic information especially those capture the difference between different entities can be kept when the dimensional sizes of the semantic constellation space increases. In other words, the dimensional sizes of semantic representation need to  be carefully chosen to maintain the optimal balance between the transmission efficiency and the robustness of semantic recovery.  
To verify the impact of difference dimensional sizes of the semantic constellation space on the path reasoning accuracy, in Fig. \ref{Fig_AccVSDimension}, we compare the performance of G-RML under different sizes of dimensions of semantic constellation spaces when different SKGs have been considered for generating semantic messages. More specifically, we consider semantic constellations  communicated in two specific types of channel implementations including AWGN channel shown in Fig. \ref{Fig_AccVSDimension} (a) and Rayleigh fading channel shown in Fig. \ref{Fig_AccVSDimension} (b). In the latter scenario, we assume the channel fading coefficient follows a Rayleigh distribution. We can observe that, in the AWGN case, for SKGs sampled from all three datasets, there exist optimal values of semantic dimensional sizes that achieve the highest performance in terms of accuracy of semantic reasoning, i.e., for SKGs samples from Nell-995, WN18RR and FB15k-237, the optimal dimensional sizes for semantic constellation space are given by 46, 24, and 62, respectively. This observation once again verifies the communication efficiency of our proposed solution, i.e., compared to sending all the expert semantic paths with all the entities and relations from the source to destination users, our proposed iSAC framework only requires 12, 23 and 31 transmissions of semantic constellations, each represented by a 2-dimensional complex semantic constellation representation. For the Rayleigh fading case, we 
can observe that, the optimal values of semantic dimensional sizes that achieves the highest semantic recovery accuracy for our considered knowledge datasets, Nell-995, WN18RR and FB15k-237, are 66, 34, 74, respectively, resulting in around 16-41\% of increases in the number of semantic constellation transmissions. In other words, under the same SNR, the semantic-aware communication in the Rayleigh fading channel requires more transmission overhead for the optimal semantic information recovery. However, even in this case, the communication efficiency of our proposed iSAC is still much higher than sending the complete expert path information through the channel.  

\begin{figure}[htbp]
 \vspace{-0.2in}
  \begin{minipage}[t]{0.24\textwidth}
   \centering
   \includegraphics[width=\textwidth]{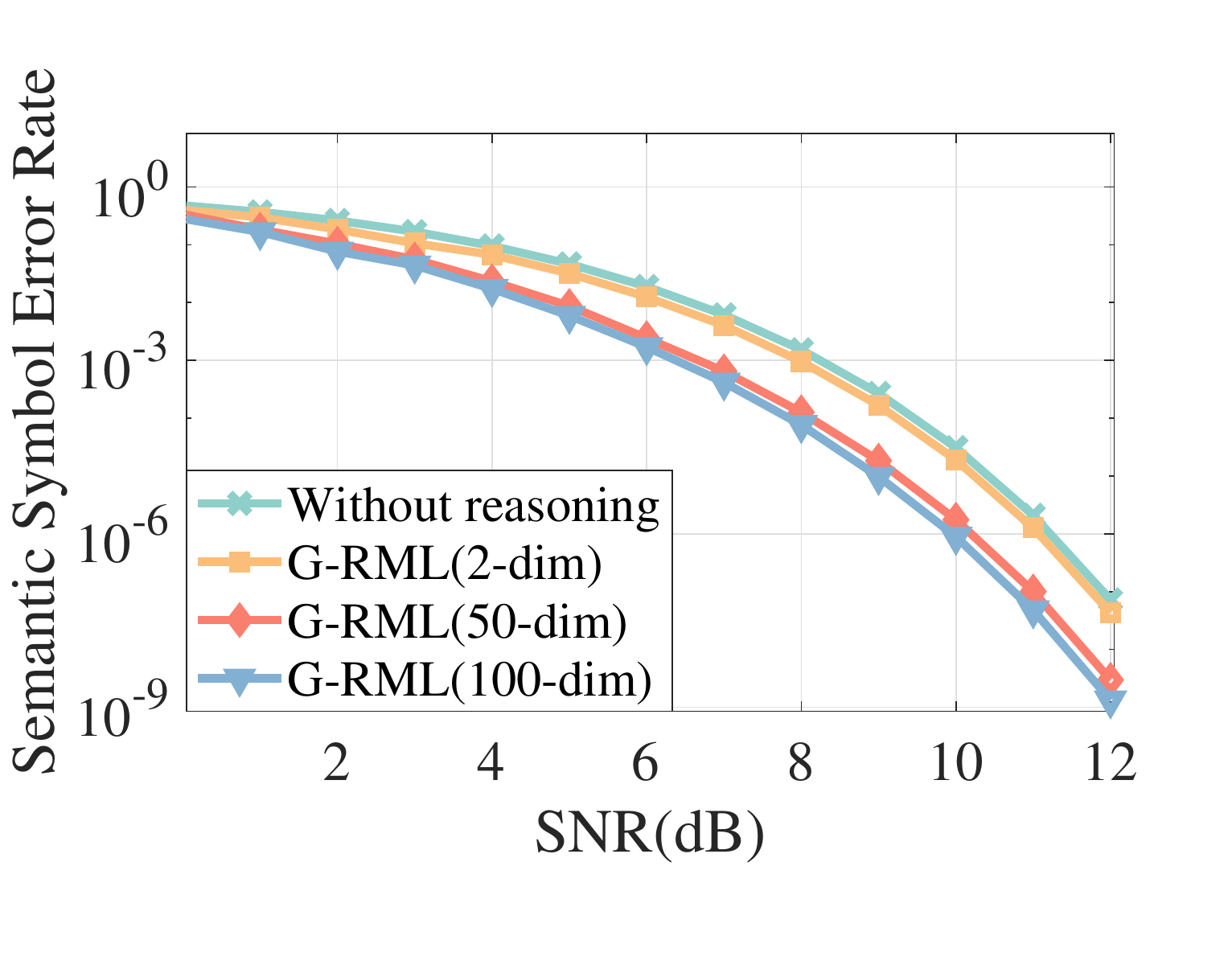}
   \vspace{-0.35in}
	 \caption*{\footnotesize{(a)}}
  \end{minipage}
  \begin{minipage}[t]{0.24\textwidth}
	\centering
	\includegraphics[width=\textwidth]{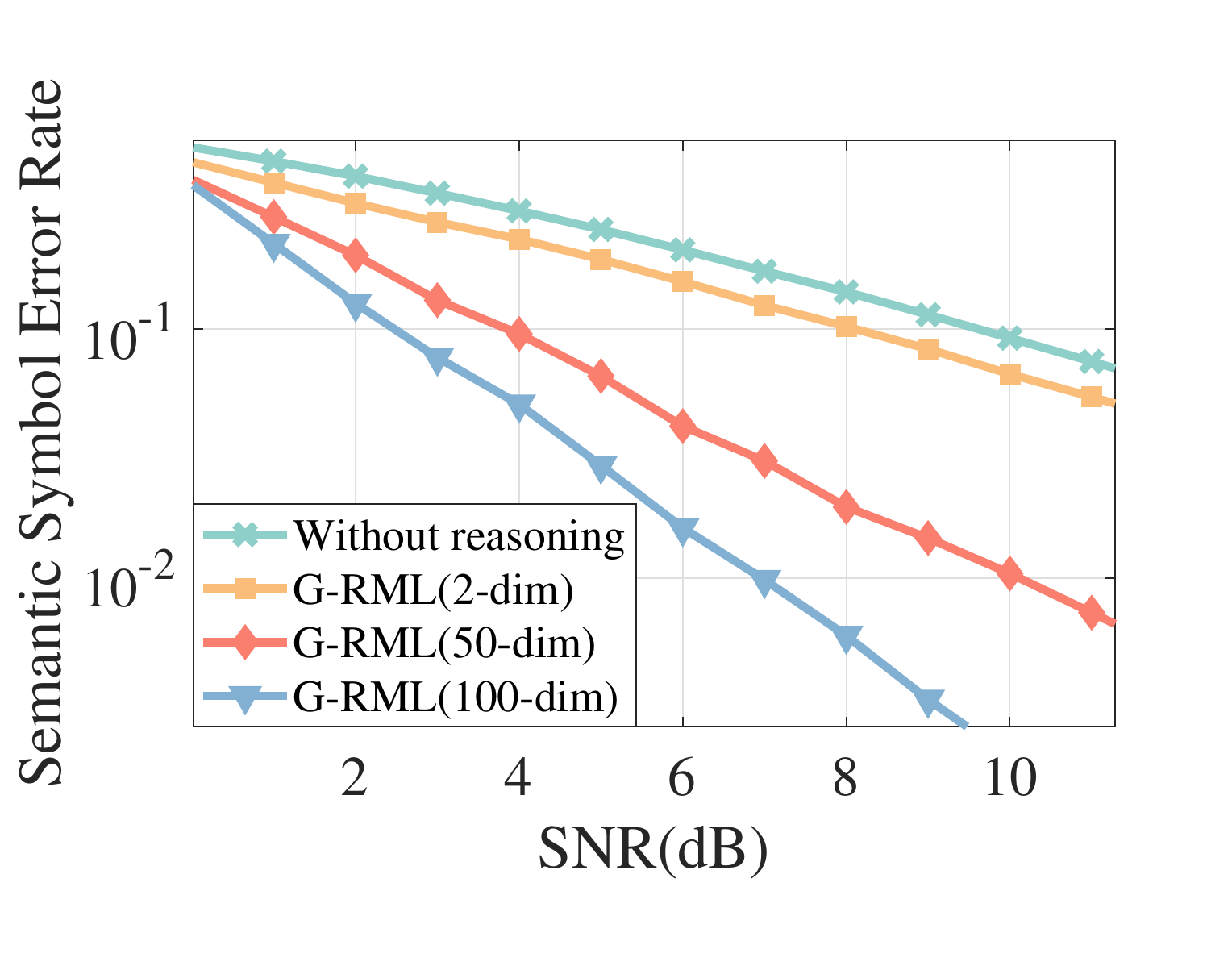}
	\vspace{-0.35in}
	\caption*{\footnotesize{(b)}}
  \end{minipage}
  \begin{minipage}[t]{0.24\textwidth}
	\centering
	\includegraphics[width=\textwidth]{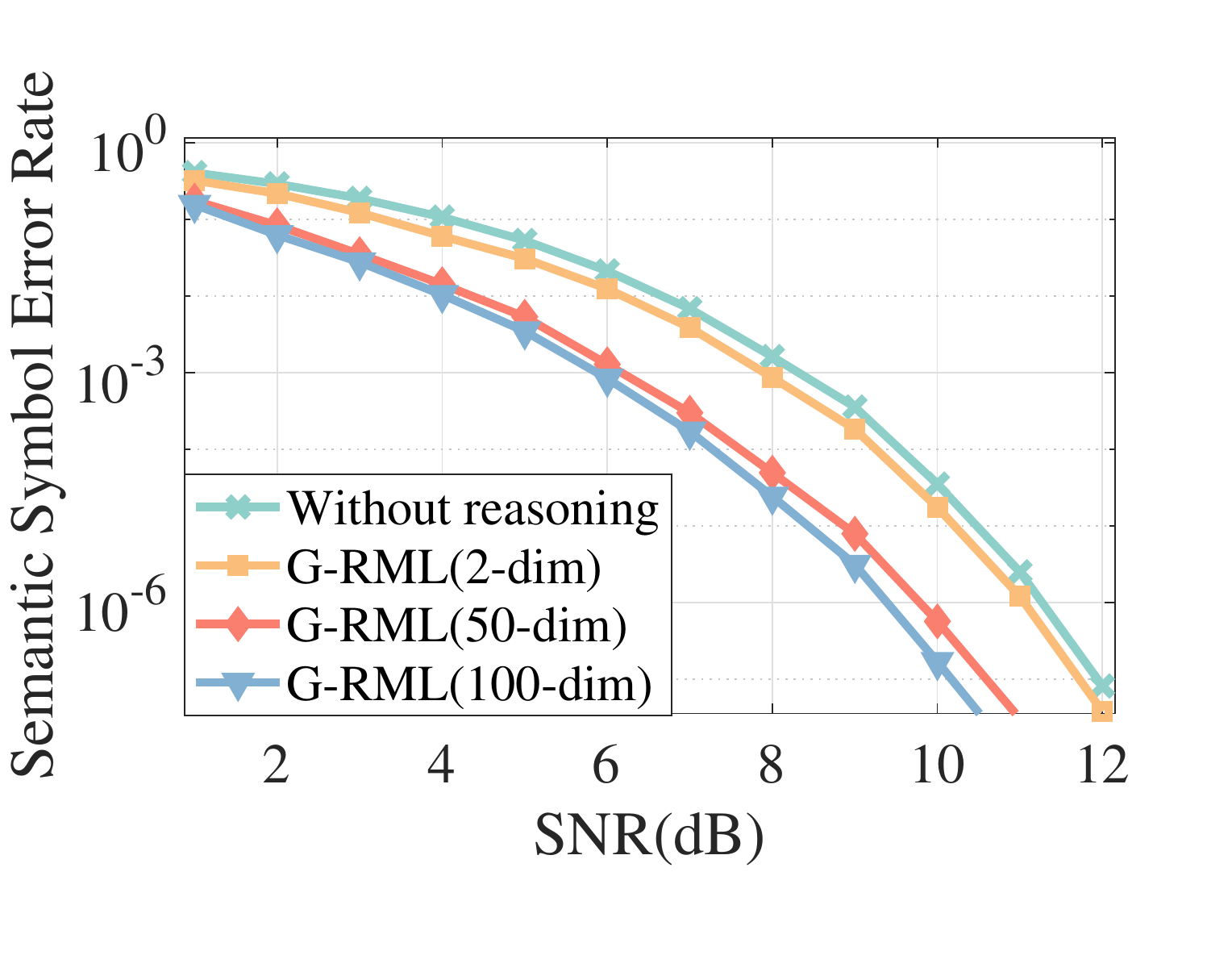}
	\vspace{-0.35in}
	\caption*{\footnotesize{(c)}}
  \end{minipage}
  \begin{minipage}[t]{0.24\textwidth}
	\centering
	\includegraphics[width=\textwidth]{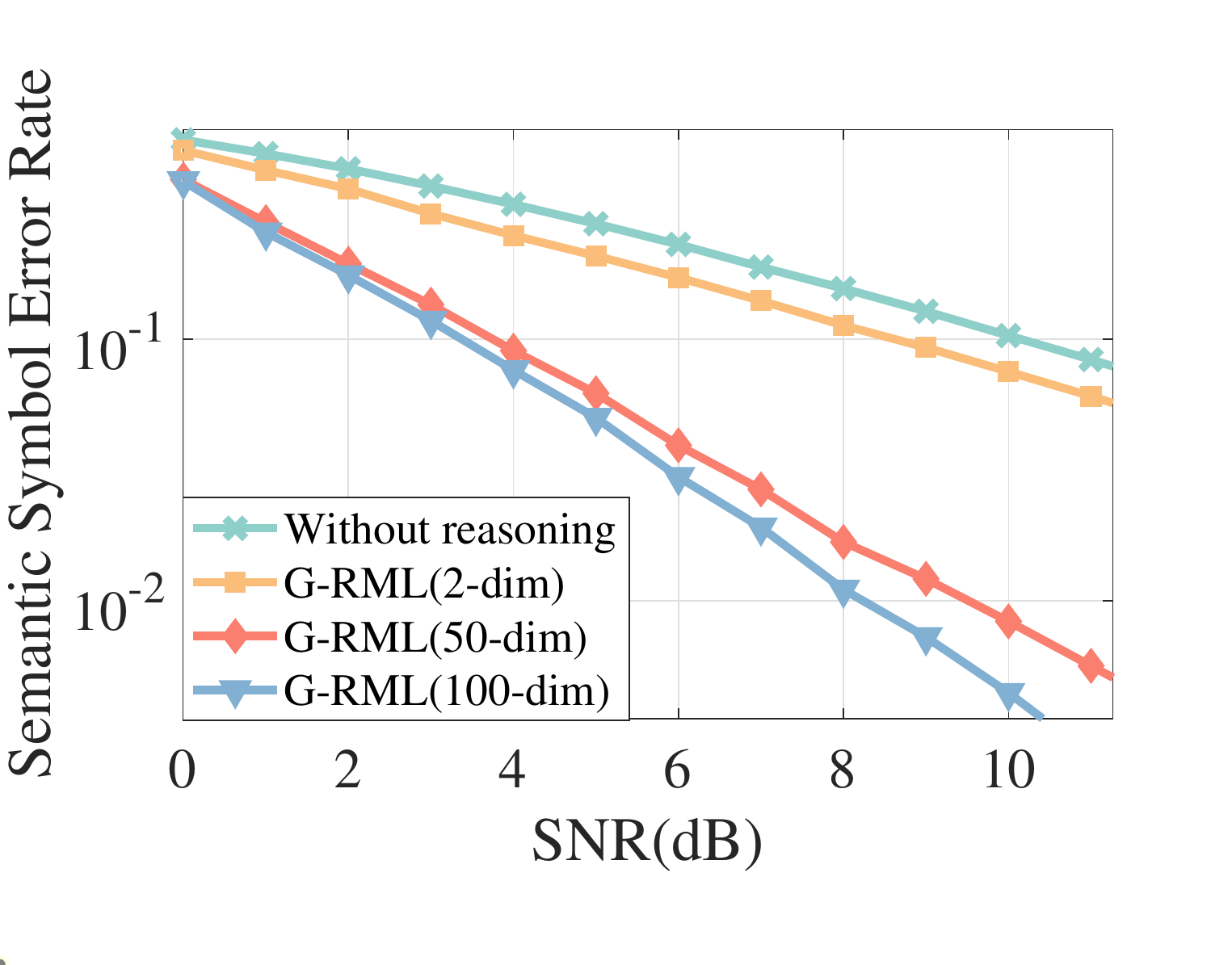}
	\vspace{-0.35in}
	\caption*{\footnotesize{(d)}}
  \end{minipage}

    \vspace{-0.1in}
      \caption{\footnotesize{Semantic symbol error rate after edge recovery in different dimensions based on FB15k-237 under different dimensions with (a) AWGN channel, (b) Rayleigh fading channel; based on Nell-995 under different dimensions with (c) AWGN channel, (d) Rayleigh fading channel. }}
      \label{Fig_dimensionrecovery}
    \vspace{-0in}
   \end{figure}

{In Fig. \ref{Fig_dimensionrecovery}, we investigate the impact of the dimensional size of the transmit semantic constellation space on the accuracy of semantic symbol recovery under 
different SNRs. 
We consider two knowledge datasets, FB15k-237 and Nell-995, respectively. For each dataset, we consider semantic encoding solutions when the explicit semantics have been converted into semantic constellations with different dimensional sizes, including 2-, 50-, and 100-dimensions labeled as 2-dim, 50-dim, and 100-dim, respectively.   
}
%
%
In Fig. \ref{Fig_dimensionrecovery} (a), we can observe that our proposed G-RML solution provides significant reduction in the semantic symbol error rate, compared to the existing communication solution without semantic reasoning. For example, when the SNR is 4dB, our proposed semantic encoding solutions with 2-, 50- and 100- semantic dimensional sizes provide 1dB, 3dB, 12dB, and 15dB improvement, respectively, in the AWGN channel case.  
In the Rayleigh fading case shown in Fig. \ref{Fig_dimensionrecovery} (b), our proposed semantic encoding solution once again achieves promising results in terms of semantic error rate. For example, in the 4dB of SNR, transmitting semantic constellations with 2-, 50- and 100- dimensional sizes achieve 1dB, 2dB, 8dB, and 14dB reduction of the semantic symbol error rate. 
Similarly, for dataset Nell-995 as shown in Fig. \ref{Fig_dimensionrecovery} (c) and (d), we can observe that the semantic error rate is closely related to the selected semantic dimensional sizes for semantic constellation transmission, e.g., in 4dB of SNR, 2-, 50- and 100- semantic dimensional sizes can provide 1dB, 4dB, 17dB, and 18dB reduction in the semantic symbol error rates in the AWGN channel, and  1dB, 2dB, 10dB, and 17dB reduction in the Rayleigh fading channel.  

 \begin{figure}[htbp]
    \vspace{-0.2in}
  \begin{minipage}[t]{0.24\textwidth}
   \centering
   \includegraphics[width=\textwidth]{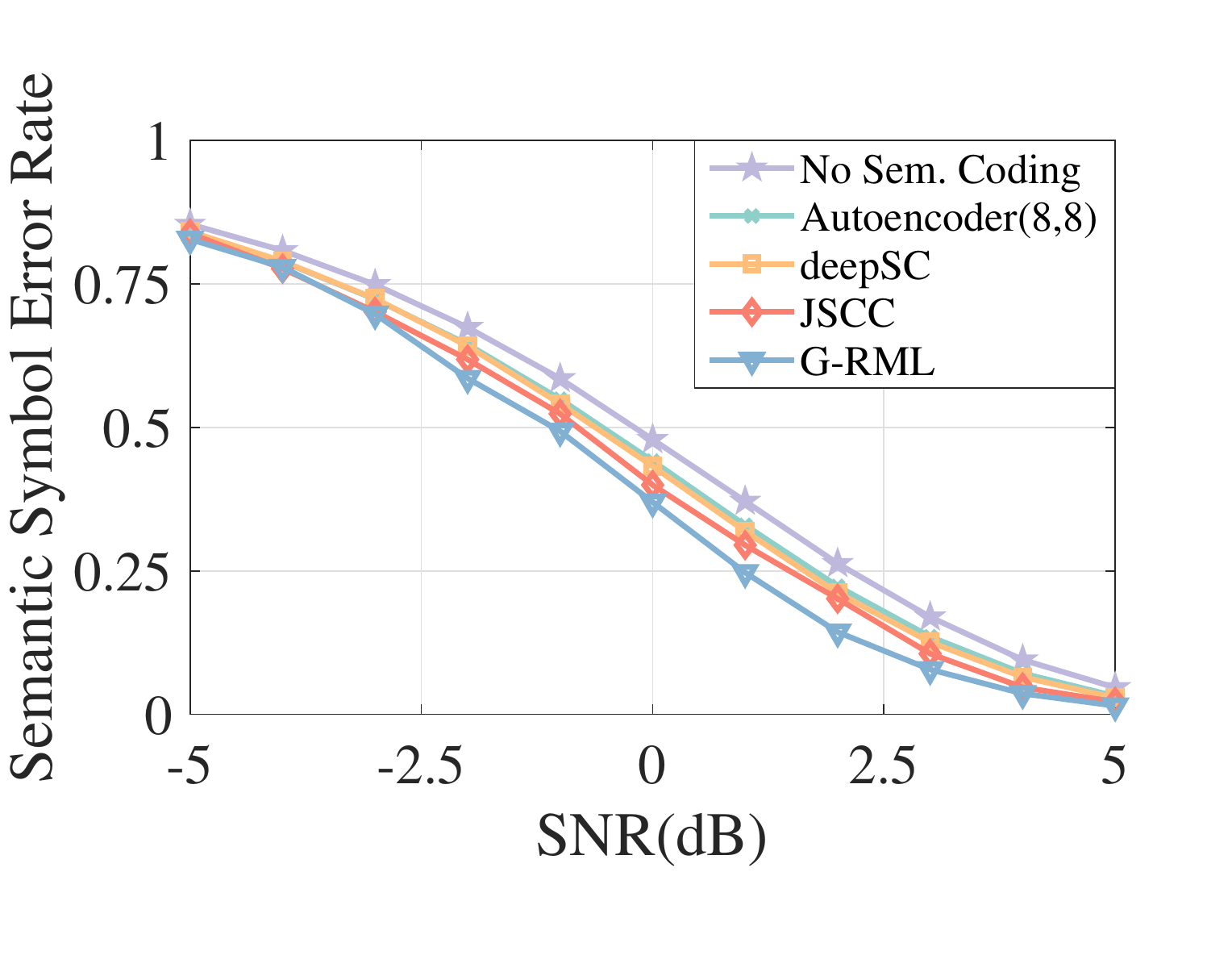}
   \vspace{-0.35in}
	 \caption*{\footnotesize{(a)}}
  \end{minipage}
  \begin{minipage}[t]{0.24\textwidth}
	\centering
	\includegraphics[width=\textwidth]{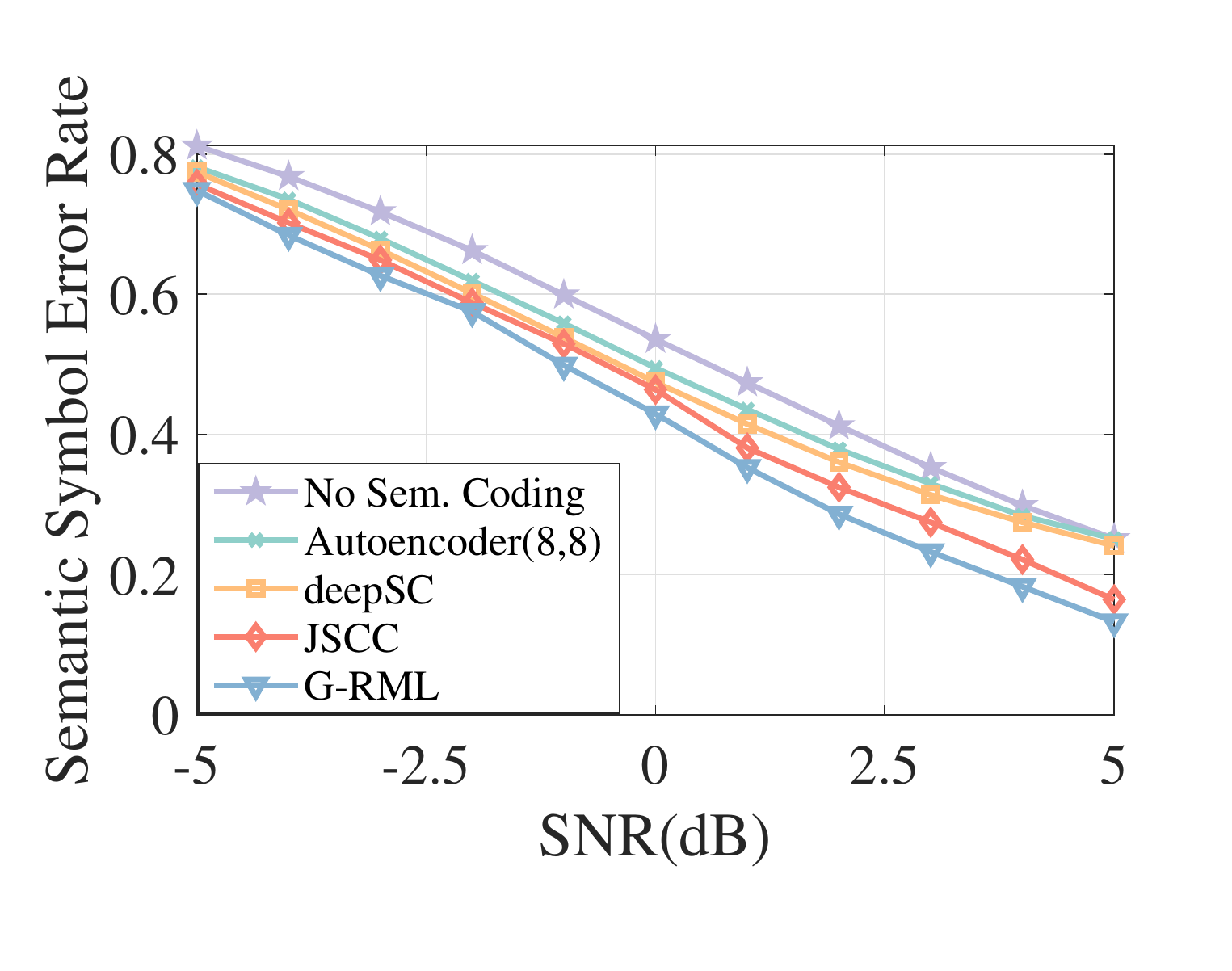}
	\vspace{-0.35in}
	\caption*{\footnotesize{(b)}}
  \end{minipage}
  \begin{minipage}[t]{0.24\textwidth}
	\centering
	\includegraphics[width=\textwidth]{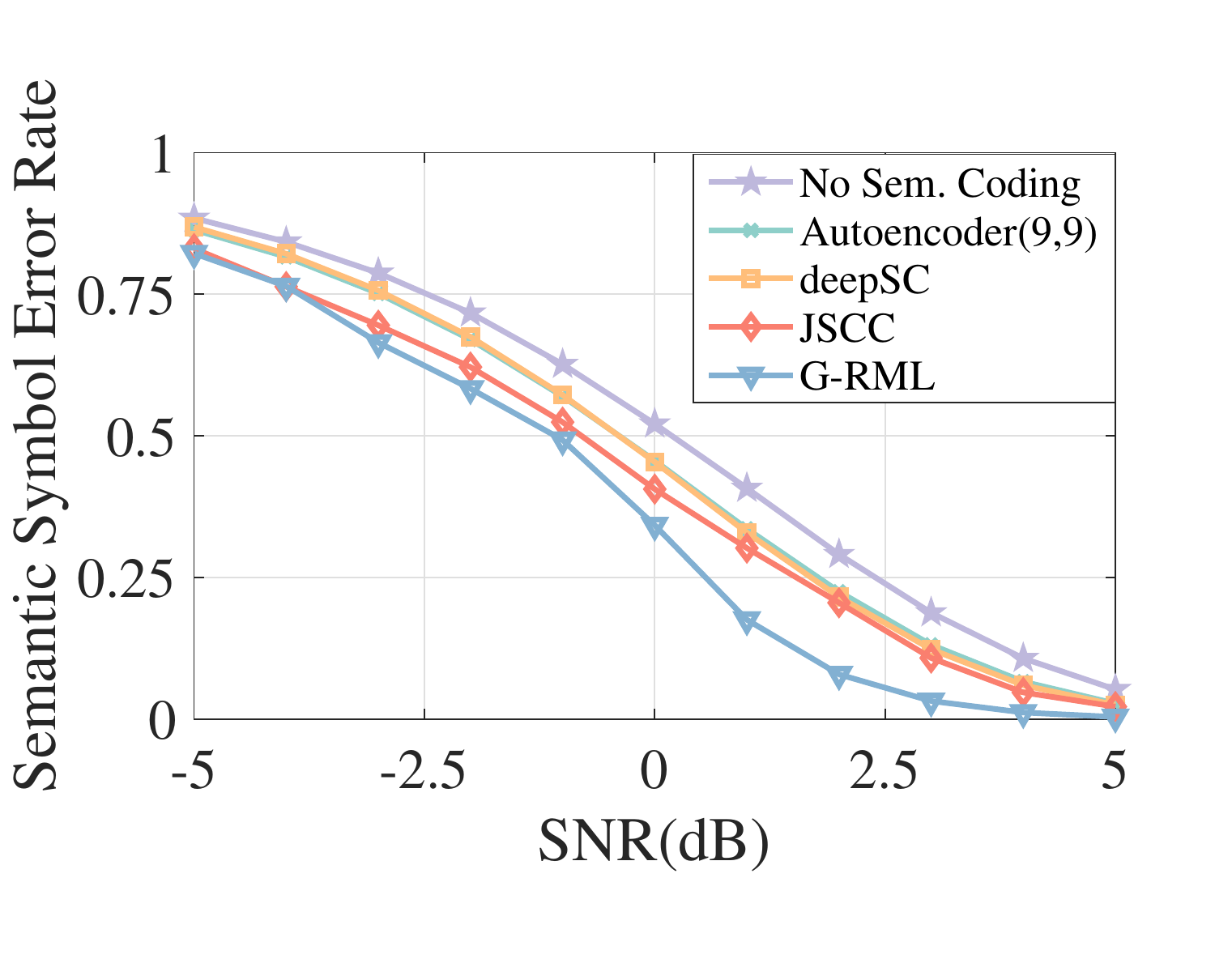}
	\vspace{-0.35in}
	\caption*{\footnotesize{(c)}}
  \end{minipage}
  \begin{minipage}[t]{0.24\textwidth}
	\centering
	\includegraphics[width=\textwidth]{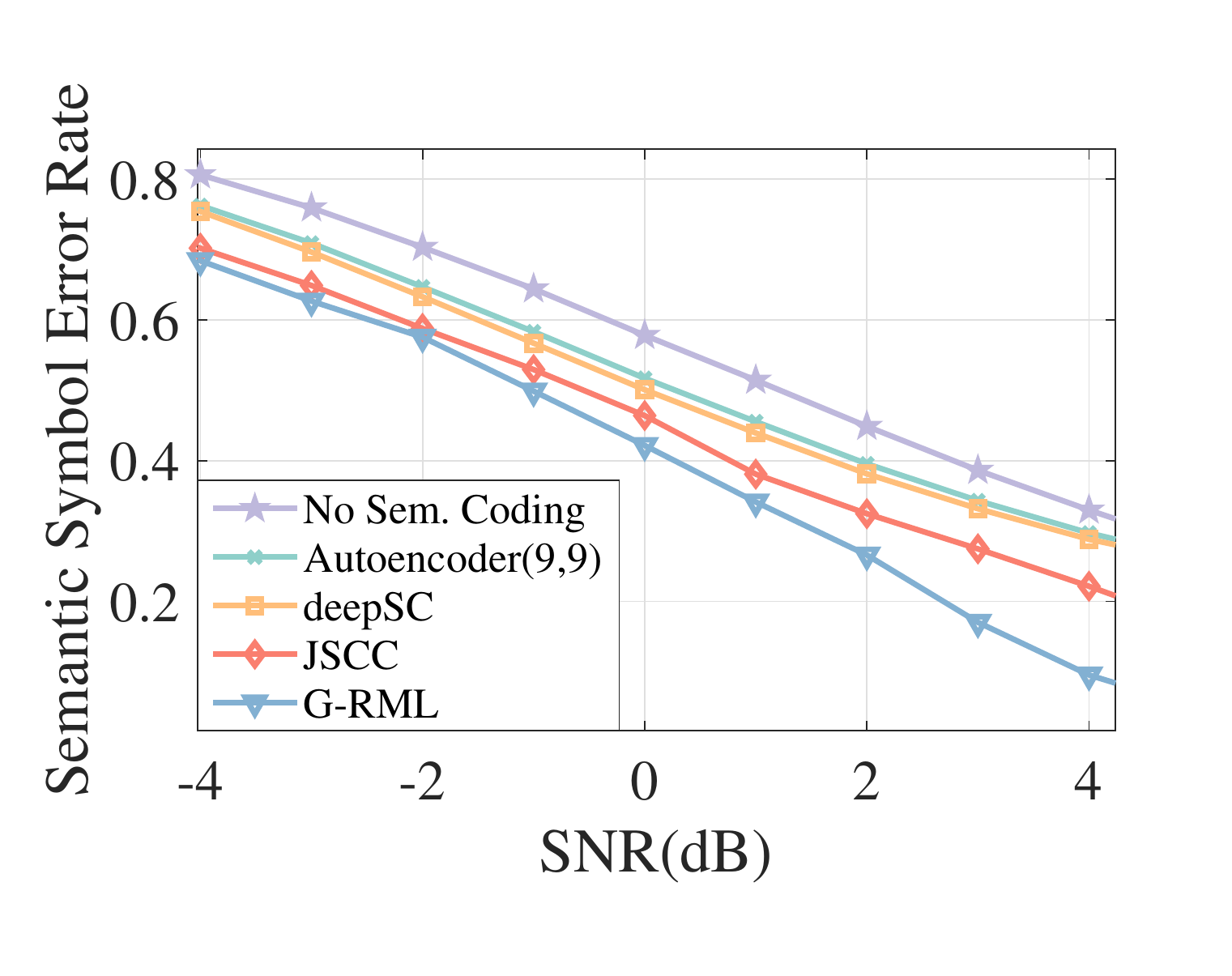}
	\vspace{-0.35in}
	\caption*{\footnotesize{(d)}}
  \end{minipage}

    \vspace{-0.1in}
      \caption{\footnotesize{The semantic symbol error rate of different schemes versus the SNR of semantic noise based on FB15k-237 through (a) the AWGN channel and (b) Rayleigh fading channel; based on Nell-995 through (c) the AWGN channel and (d) Rayleigh fading channel}}
      \label{Fig_schemescompare}
    \vspace{-0.2in}
   \end{figure}

In Fig. \ref{Fig_schemescompare}, we compare our proposed G-RML with previously proposed semantic coding algorithms including Autoencoder \cite{o2017introduction}, deepSC \cite{xie2021deep}, and JSCC \cite{bourtsoulatze2019deep} under different knowledge datasets. 
We can observe that our proposed G-RML achieves the lowest semantic error rate than all these existing solutions.  More specifically, in Fig. \ref{Fig_schemescompare} (a), we can observe that when the source signal is sampled from dataset FB15k-237, semantic coding algorithms Autoencoder, deepSC, JSCC, and G-RML provides 0.71dB, 0.88dB, 1.57dB and 2.25dB improvements, respectively, over traditional non-semantic coding solutions when communicating over AWGN channel with 0dB of SNR (the signal power and noise power are equal). For the Rayleigh fading case, as shown in Fig. \ref{Fig_schemescompare} (b), semantic coding algorithms deepSC, JSCC and G-RML provide 0.69dB, 1.06dB, 1.25dB and 1.93dB improvement, respectively, in the semantic symbol error rate when the SNR is equal to 0dB. 
Similarly, for dataset NELL-995 shown in Fig. \ref{Fig_schemescompare} (c) and (d), we can observe that semantic coding algorithms Autoencoder, deepSC, JSCC, and G-RML provide 1.16dB, 1.20dB, 2.17dB, and 3.65dB reduction in the semantic symbol error rate when transmitting in the AWGN channel (SNR=0 dB) and achieve 0.96dB, 1.24dB, 1.91dB and 2.74dB reduction in the semantic symbol error rate when sending in the Rayleigh fading channel (SNR=0 dB). 

\section{Conclusion}
\label{Section_Conclusion}
In this paper, we have proposed a comprehensive framework for representing, modeling, and interpreting implicit semantic meaning among users. We have first introduced  a novel graph-inspired structure to represent both explicit and implicit meaning of messages. 
We have then developed a novel implicit semantic-aware communication architecture, iSAC, in which a reasoning mechanism can be trained at the destination user with the help of the source user to automatically generate the possible reasoning paths that best represent the implicit meaning of the message. We have also developed a generative imitation-based reasoning mechanism learning (G-RML) framework that allows the destination user to imitate the reasoning process observed by the source users. We have proved that, by applying G-RML, the destination user will learn a reasoning mechanism to generate reasoning paths that follow the same probability distribution as the expert paths. Numerical results show that the proposed solution achieves up to  92\% accuracy of implicit meaning interpretation at a destination user. 

\appendices

\section{Proof of Proposition \ref{Proposition_discriminator}}
\label{Proof_Proposition_discriminator}
Let us now prove that the optimal semantic comparator that solves (\ref{eq_comparator}) is the semantic distance between $\rho_\Pi$ and $\rho_{\hat \Pi_{\pi_\theta}}$ given in (\ref{eq_JSD}). In particular, we need to prove that
\begin{eqnarray}
\varpi _{\phi}^{*}=\frac{\rho _{\Pi}}{\rho _{\Pi}+\rho _{\hat \Pi_{\pi _{\theta}}}}.
\end{eqnarray}


To prove the above results, we can show that, for any given ${\pi}_{\theta}$, the semantic comparator is trained to maximize the value function:
\begin{equation}
\begin{aligned}
\begin{aligned}
V\left( {\pi _{\theta}},\varpi _{\phi} \right) &=\int_{{\eta \sim {\Psi_{{\Pi}}}}}p\left( \eta \right) \log \left( \varpi _{\boldsymbol{\phi }}\left( \eta \right) \right) \,d\eta \\ &+\int_{{\eta \sim {\Psi_{{\hat \Pi}_{\pi_\theta}}}}}p\left( \eta \right) \log \left( 1-\varpi _{\phi}\left( \eta \right) \right) \,d\eta 
\end{aligned}
\end{aligned}
\label{value_function}
\end{equation}

For any $\eta$  under the constraint of $\Delta \left( \eta \right) >0$ and $\hat \Pi_{(\pi_\theta, L)}(\eta)>0$, we have $f\left( \varpi _{\phi}\left( \eta \right) \right) =\Delta \left( \eta \right) \log \left( \varpi _{\phi}\left( \eta \right) \right) +\hat \Pi_{\left(\pi_{\theta},L\right)}\left( \eta \right) \log \left( 1-\varpi _{\phi}\left( \eta \right) \right) $. 

Since $f\left( \varpi _{\phi}\left( \eta \right) \right)$ is convex, it is maximized  when the first derivative $f'\left( \varpi _{\phi}\left( \eta \right) \right)$ $=\frac{\rho_\Pi}{\varpi _{\phi}\left( \eta \right)}+\frac{\rho _{\hat \Pi_{\pi _{\theta}}}}{\varpi _{\phi}\left( \eta \right) -1}$ is 0. We can then show that $V\left( \pi _{\theta},\varpi _{\phi} \right) $ is maximized when $\varpi_{\phi} $ is given by
\begin{equation}\label{optimal comparator} 
\varpi _{\phi}^{*}=\frac{\rho _{\Pi}}{\rho _{\Pi}+\rho _{\hat \Pi_{\pi _{\theta}}}}
\end{equation}
This concludes the proof.

\section{Proof of Theorem \ref{MainConvergTheorem}}
\label{Appendix_TheoremProof}

To prove Theorem \ref{MainConvergTheorem}, we first derive the optimal solution  of the semantic comparator $\varpi_{\phi}$ and then prove that the semantic interpreter $\pi_{\theta}$ can always converge under the given optimal solution of $\varpi^*_{\phi}$. Since we have already derived the optimal semantic comparator in Proposition \ref{Proposition_discriminator}, we now only need to optimal  semantic interpreter  as follows. 


\begin{proposition}\label{global optimal}
Given the optimal $\varpi^* _{\phi}$, the global optimal  solution of  $V\left( \pi _{\theta},\varpi^* _{\phi} \right) $ is achieved when $\hat \Pi_{\left(\pi_{\theta},L\right)}= \pi$.
\end{proposition}

\begin{IEEEproof}
Suppose $\varpi_{\phi}$ is optimal. According to Proposition 1, (\ref{value_function}) can be rewritten as follows:  
\begin{eqnarray}
\begin{aligned}
	&V'\left( \pi _{\theta},\varpi _{\phi} \right) =V\left( \pi _{\theta},\varpi _{\phi} \right) \\&=\mathbb{E}_{\eta \sim {\Psi_{{\Pi}}}}\left[ \log \varpi _{\phi}\left( \eta \right) \right] +\mathbb{E}_{\eta \sim {\Psi_{{\hat \Pi}_{\pi_\theta}}}}\left[ 1-\log \varpi _{\phi}\left( \eta \right) \right]\\
	&=\mathbb{E}_{\eta \sim {\Psi_{{\Pi}}}}\left[ \frac{\rho _{\Pi}}{\rho _{\Pi}+\rho _{\hat \Pi_{\pi _{\theta}}}} \right] +\mathbb{E}_{\eta \sim {\Psi_{{\hat \Pi}_{\pi_\theta}}}} \left[ 1 - \frac{\rho _{\Pi}}{\rho _{\Pi}+\rho _{\hat \Pi_{\pi _{\theta}}}} \right]
\end{aligned}
\label{V'}
\end{eqnarray}

Suppose $\pi=\hat \Pi_{\left(\pi_{\theta},L\right)}$ and $V'(\hat{\pi}_{\theta})=\mathbb{E}_{\eta \sim {\Psi_{{\Pi}}}}[\log{\frac{1}{2}}]+\mathbb{E}_{\eta \sim {\Psi_{{\hat \Pi}_{\pi_\theta}}}}[\log{1-\frac{1}{2}}]=-\log2-\log2=-\log 4$, which is the minimum of (\ref{V'}). And we found (\ref{V'}) can be written as follows:
\begin{eqnarray}
\begin{aligned}
    V'(\pi)&= \mathbb{E}_{\eta \sim {\Psi_{{\Pi}}}}\left[ \log \frac{\rho _{\Pi}}{\rho _{\Pi}+\rho _{\hat \Pi_{\pi _{\theta}}}} \right]-\log2  \\&+\mathbb{E}_{\eta \sim {\Psi_{{\hat \Pi}_{\pi_\theta}}}} \left[ 1 - \frac{\rho _{\Pi}}{\rho _{\Pi}+\rho _{\hat \Pi_{\pi _{\theta}}}} \right]-\log2 \\ 
   &= -\log 4 + D_{\rm KL}\left(\rho _{\Pi}\|\frac{ \rho _{\Pi}+\rho _{\hat \Pi_{\pi _{\theta}}}}{2}\right)  \\&+D_{\rm KL}\left(\rho _{\hat \Pi_{\pi _{\theta}}}\|\frac{ \rho _{\Pi}+\rho _{\hat \Pi_{\pi _{\theta}}}}{2}\right)
\end{aligned}
\end{eqnarray}
where $D_{\rm KL}$  is the Kullback-Leibler divergence. 

The Jenson-Shannon divergence between distribution $\pi $ and $\hat \Pi_{\left(\pi_{\theta},L\right)}$ is given by,
\begin{equation}
\begin{aligned}
\begin{aligned}
 D_{\rm JS}(\pi ||\hat \Pi_{\left(\pi_{\theta},L\right)}) &= [D_{\rm KL}(\pi||\hat \pi +(\hat \Pi_{\left(\pi_{\theta},L\right)}/2)\\&+D_{\rm KL}(\Pi_{(\pi_{\theta},L)}||\pi +\hat \Pi_{\left(\pi_{\theta},L\right)}/2)]/2
\end{aligned}
\end{aligned}
\end{equation}

We can then rewrite (\ref{V'}) as:
\begin{equation}
\begin{aligned}
	V'\left( \pi _{\theta},\varpi _{\phi} \right) &=-\log 4+ D_{\rm KL}\left(\rho _{\Pi}\|\frac{ \rho _{\Pi}+\rho _{\hat \Pi_{\pi _{\theta}}}}{2}\right) \\ &+D_{\rm KL}\left(\rho _{\hat \Pi_{\pi _{\theta}}}\|\frac{ \rho _{\Pi}+\rho _{\hat \Pi_{\pi _{\theta}}}}{2}\right) \\
	&=-\log 4+2\cdot D_{\text{JS}}\left( \rho _{\Pi} || \rho _{\hat \Pi_{\pi _{\theta}}} \right)\\
\end{aligned}
\end{equation}

So we can rewrite (\ref{V'}) as $V(\pi_{\theta})'=-\log 4 + 2\cdot D_{\rm JS}(\rho _{\Pi} || \rho _{\hat \Pi_{\pi _{\theta}}})$.  Obviously, $V'(\pi)$ gets its optimal solution with the value of $-\log 4$ when $\rho _{\Pi} =\rho _{\hat \Pi_{\pi _{\theta}}}$, which means the {semantic interpreter} can generate  paths that follows the same distribution as the expert path distribution $\pi$. 
This concludes the proof.
\end{IEEEproof}



We can now prove Theorem \ref{MainConvergTheorem} based on Proposition 1. In particular, suppose $V(\pi_{\theta}, \varpi_{\phi})$  $=C(\hat \Pi_{\left(\pi_{\theta},L\right)}, \varpi_{\phi})$ is a convex function of $\hat \Pi_{\left(\pi_{\theta},L\right)}$. Then, given the optimal comparator $\varpi^*_{\phi}$, we can use gradient descent to update $\hat \Pi_{\left(\pi_{\theta},L\right)}$. Since $C(\hat \Pi_{\left(\pi_{\theta},L\right)}, \varpi^*_{\phi})$ is convex in $\hat \Pi_{\left(\pi_{\theta},L\right)}$ with a global optima as proven in proposition \ref{global optimal}, with sufficiently updates of $\Pi_{(\pi_{\theta}, L)}$, $\rho _{\hat \Pi_{\pi _{\theta}}}$ 
converges to $\rho _{\Pi}$. 
This concludes the proof.

\bibliographystyle{IEEEtran}
\bibliography{mybib}

\begin{IEEEbiography}[{\includegraphics[width=1.1in,height=1.3in,clip,keepaspectratio]{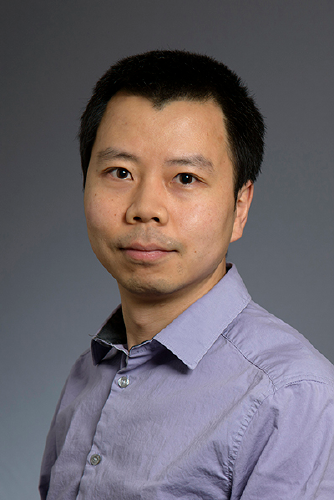}}]{Yong Xiao} (Senior Member, IEEE) received his B.S. degree in electrical engineering from China University of Geosciences, Wuhan, China in 2002, M.Sc. degree in telecommunication from Hong Kong University of Science and Technology in 2006, and his Ph. D degree in electrical and electronic engineering from Nanyang Technological University, Singapore in 2012. He is now a professor in the School of Electronic Information and Communications at the Huazhong University of Science and Technology (HUST), Wuhan, China. He is also with Peng Cheng Laboratory, Shenzhen, China and Pazhou Laboratory (Huangpu), Guangzhou, China. He is the associate group leader of the network intelligence group of IMT-2030 (6G promoting group) and the vice director of 5G Verticals Innovation Laboratory at HUST. Before he joins HUST, he was a  research assistant professor in the Department of Electrical and Computer Engineering at the University of Arizona where he was also the center manager of the Broadband Wireless Access and Applications Center (BWAC), an NSF Industry/University Cooperative Research Center (I/UCRC) led by the University of Arizona. His research interests include machine learning, game theory, distributed optimization, and their applications in cloud/fog/mobile edge computing, green communication systems, wireless communication networks, and Internet-of-Things (IoT).
\end{IEEEbiography}

\begin{IEEEbiography}[{\includegraphics[width=1.1in,height=1.3in,clip,keepaspectratio]{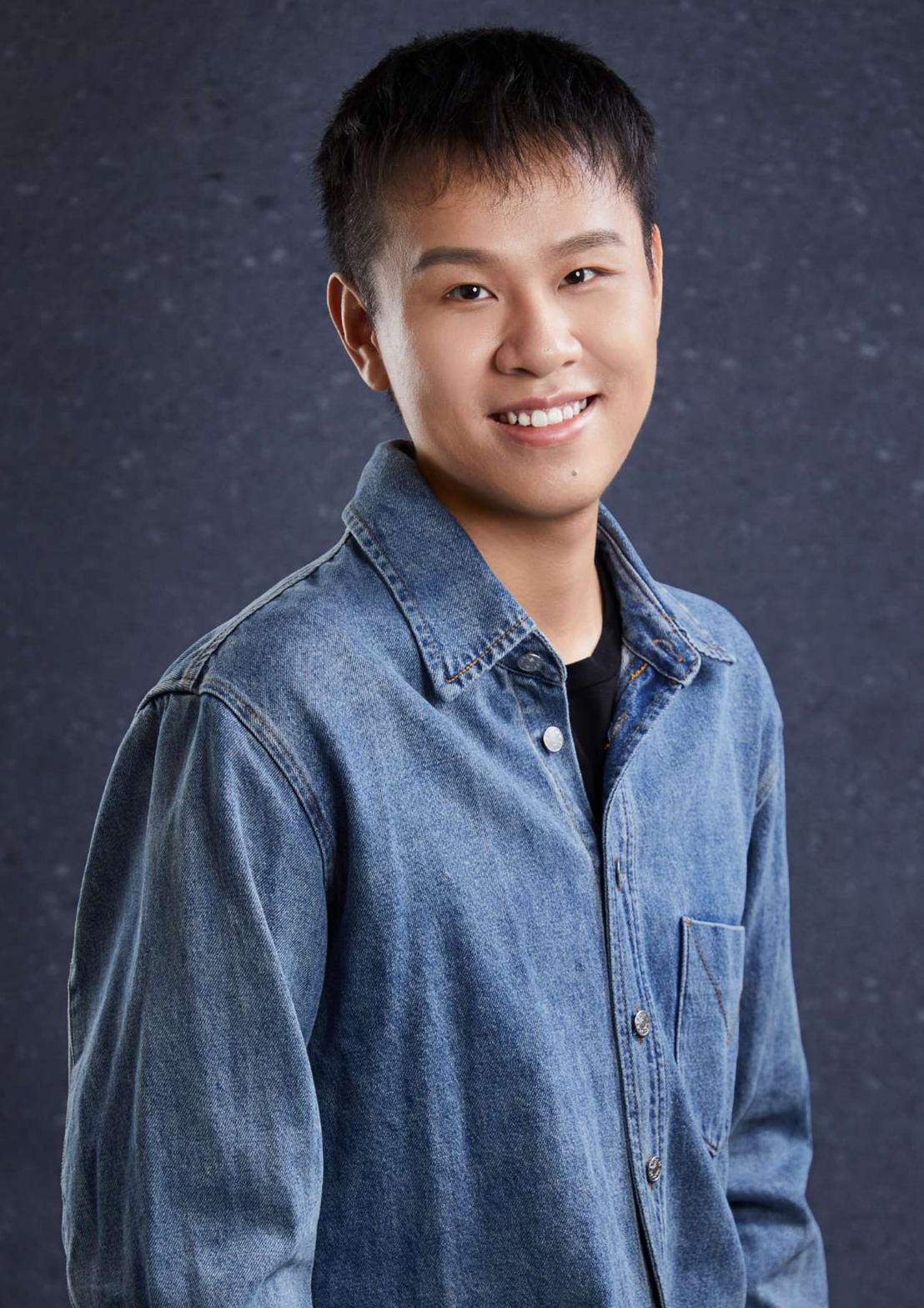}}]{Yiwei Liao} (Student Member, IEEE) received his B.S. degree in information engineering from Huazhong University of Science and Technology, Wuhan, China in 2017, and M.S. degree in Northeastern University, MA, US in 2020. He is currently pursuing his PhD in the school of electronic information and communications at the Huazhong University of Science and Technology, Wuhan, China. His research interest includes network AI and next generation communication technology, bridging the gap between traditional engineering practices and cutting-edge computational methodologies. 
\end{IEEEbiography}

\begin{IEEEbiography}[{\includegraphics[width=1.1in,height=1.3in,clip,keepaspectratio]{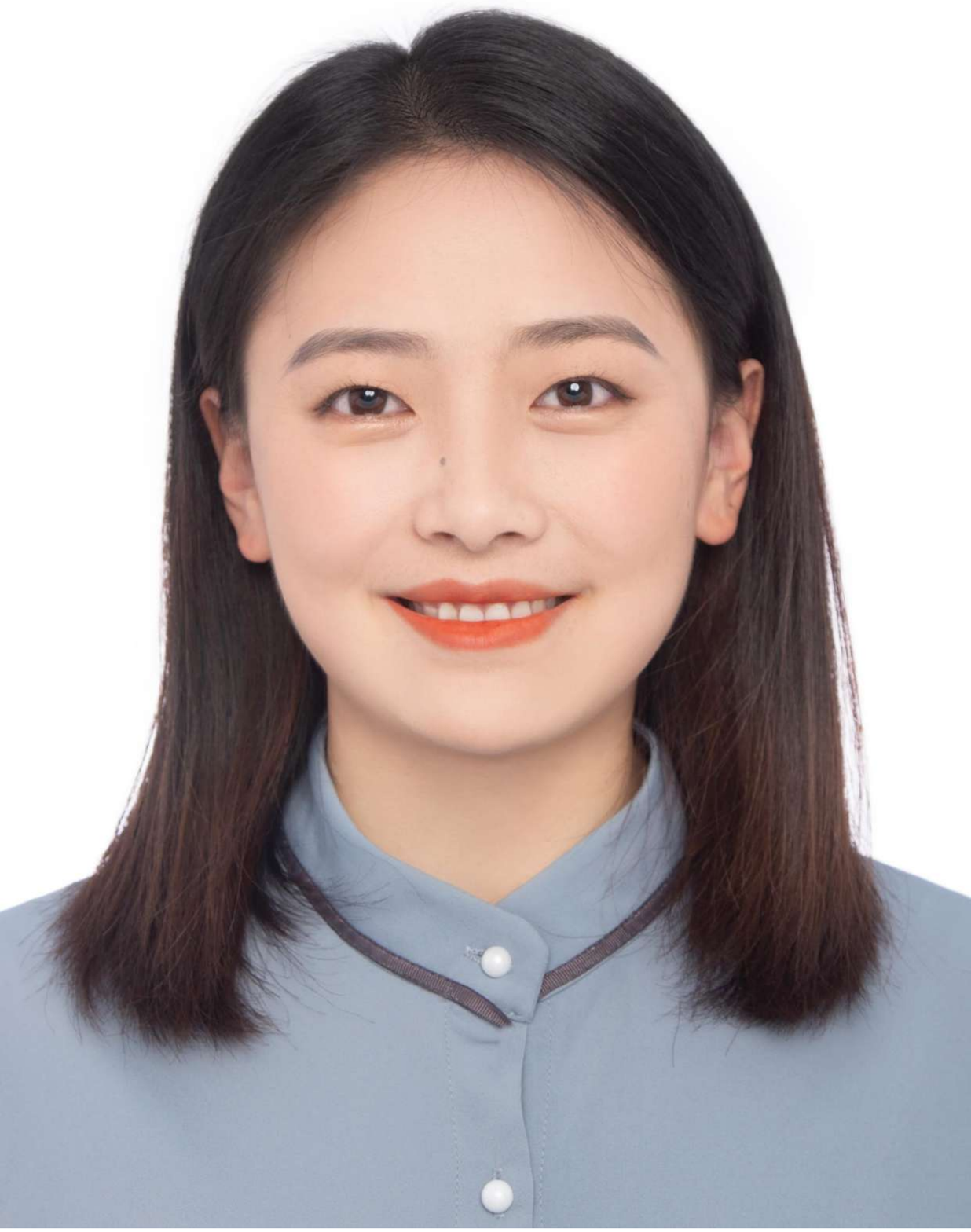}}]{Yingyu Li} (Member, IEEE) received the B.Eng. degree in electronic information engineering and the Ph.D. degree in circuits and systems from the Xidian University, Xi'an, China, in June 2012 and September 2018, respectively. From September 2014 to September 2016, she was a Research Scholar with the Wireless Networking, Signal Processing and Security Lab, Department of Electronic Computer Engineering, University of Houston, USA. She was a postdoctoral researcher in the School of Electronic Information and Communications at Huazhong University of Science and Technology from October 2018 to November 2021. Since December 2021, she has been an Associate Professor at the School of Mechanical Engineering and Electronic Information, China University of Geosciences (Wuhan). Her research interests include machine learning and artificial intelligence for next-generation wireless networks, federated edge intelligence, green/low-carbon communication networks, distributed optimization, semantic communications, and intelligent Internet of Things. 

\end{IEEEbiography}

\begin{IEEEbiography}[{\includegraphics[width=1.1in,height=1.3in,clip,keepaspectratio]{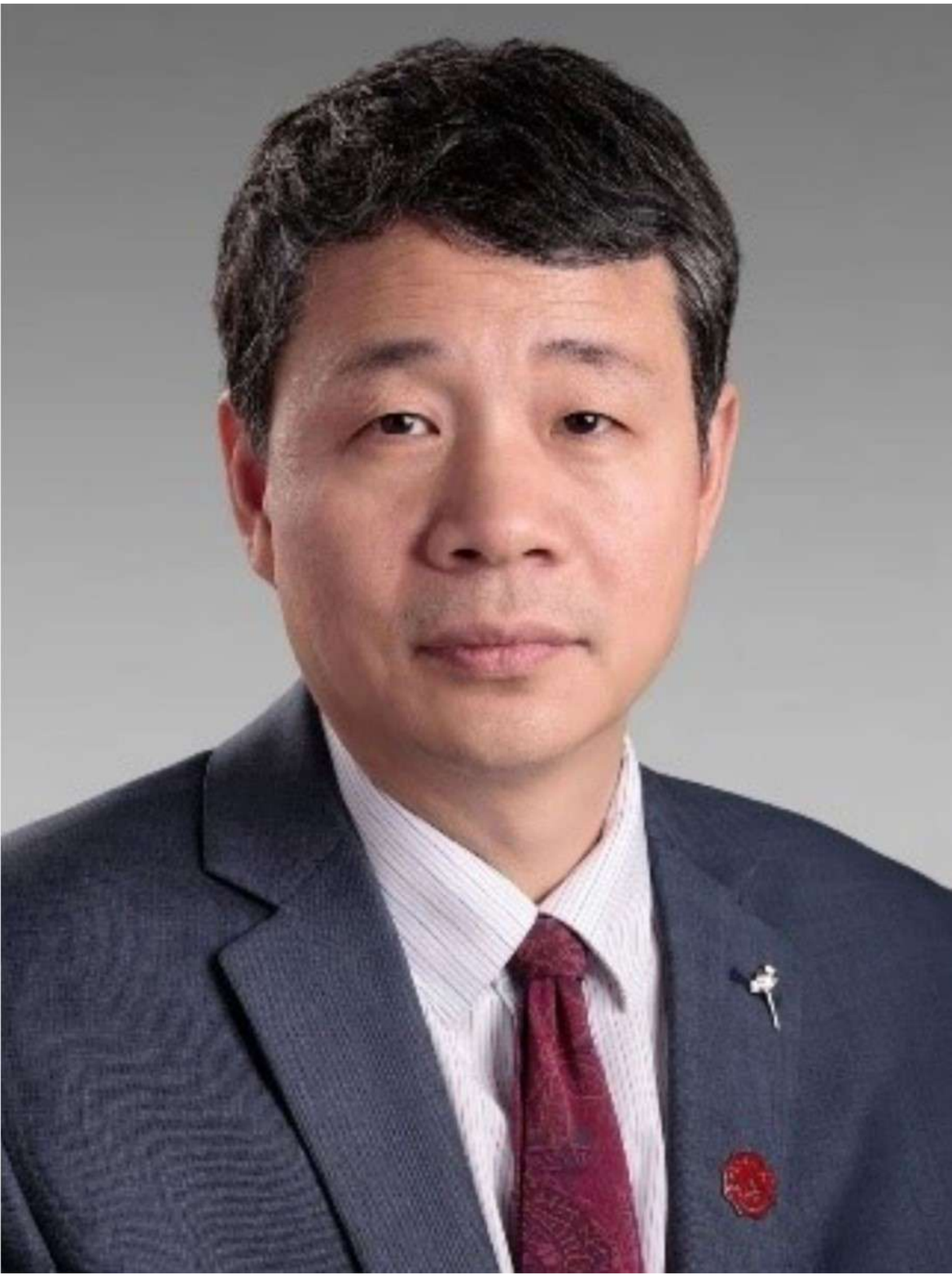}}]{Guangming Shi} (Fellow, IEEE) received the M.S. degree in computer control, and the Ph.D. degree in electronic information technology from Xidian University, Xi’an, China, in 1988, and 2002, respectively. He was the vice president of Xidian University from 2018 to 2022. Currently, he is  the Vice Dean of Peng Cheng Laboratory and a Professor with the School of Artificial Intelligence, Xidian University. He is an IEEE Fellow, the chair of IEEE CASS Xi’an Chapter, senior member of ACM and CCF, Fellow of Chinese Institute of Electronics, and Fellow of IET. He was awarded Cheung Kong scholar Chair Professor by the ministry of education in 2012. He won the second prize of the National Natural Science Award in 2017. His research interests include Artificial Intelligence, Semantic Communications, and Human-Computer Interaction. 
\end{IEEEbiography}

\begin{IEEEbiography}[{\includegraphics[width=1.1in,height=1.3in,clip,keepaspectratio]{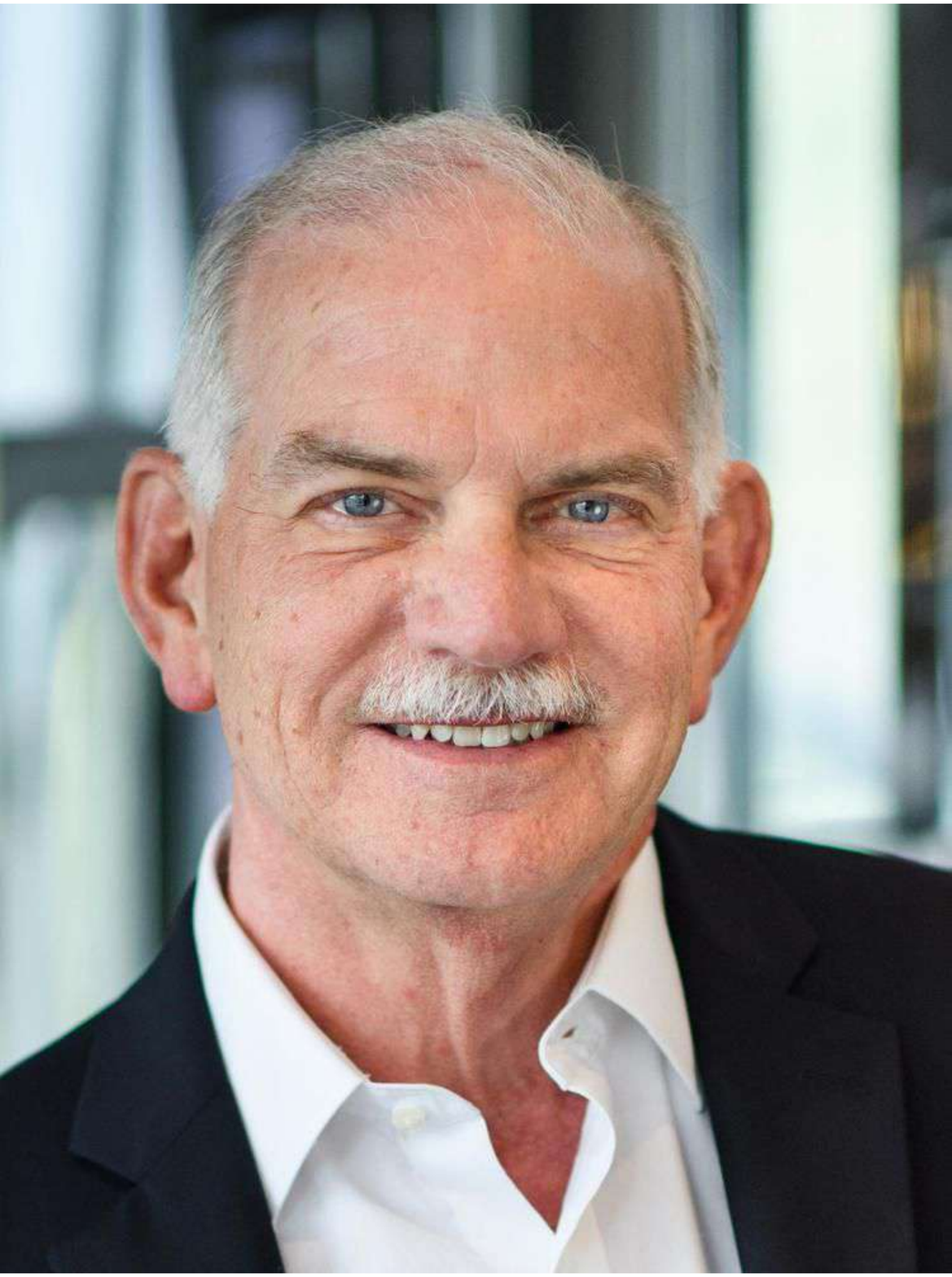}}]{H. Vincent Poor} (Fellow, IEEE) received the Ph.D. degree in EECS from Princeton University in 1977.  From 1977 until 1990, he was on the faculty of the University of Illinois at Urbana-Champaign. Since 1990 he has been on the faculty at Princeton, where he is currently the Michael Henry Strater University Professor. During 2006 to 2016, he served as the dean of Princeton’s School of Engineering and Applied Science. He has also held visiting appointments at several other universities, including most recently at Berkeley and Cambridge. His research interests are in the areas of information theory, machine learning and network science, and their applications in wireless networks, energy systems and related fields. Among his publications in these areas is the recent book Machine Learning and Wireless Communications.  (Cambridge University Press, 2022). Dr. Poor is a member of the National Academy of Engineering and the National Academy of Sciences and is a foreign member of the Chinese Academy of Sciences, the Royal Society, and other national and international academies. He received the IEEE Alexander Graham Bell Medal in 2017.
\end{IEEEbiography}

\begin{IEEEbiography}[{\includegraphics[width=1.1in,height=1.3in,clip,keepaspectratio]{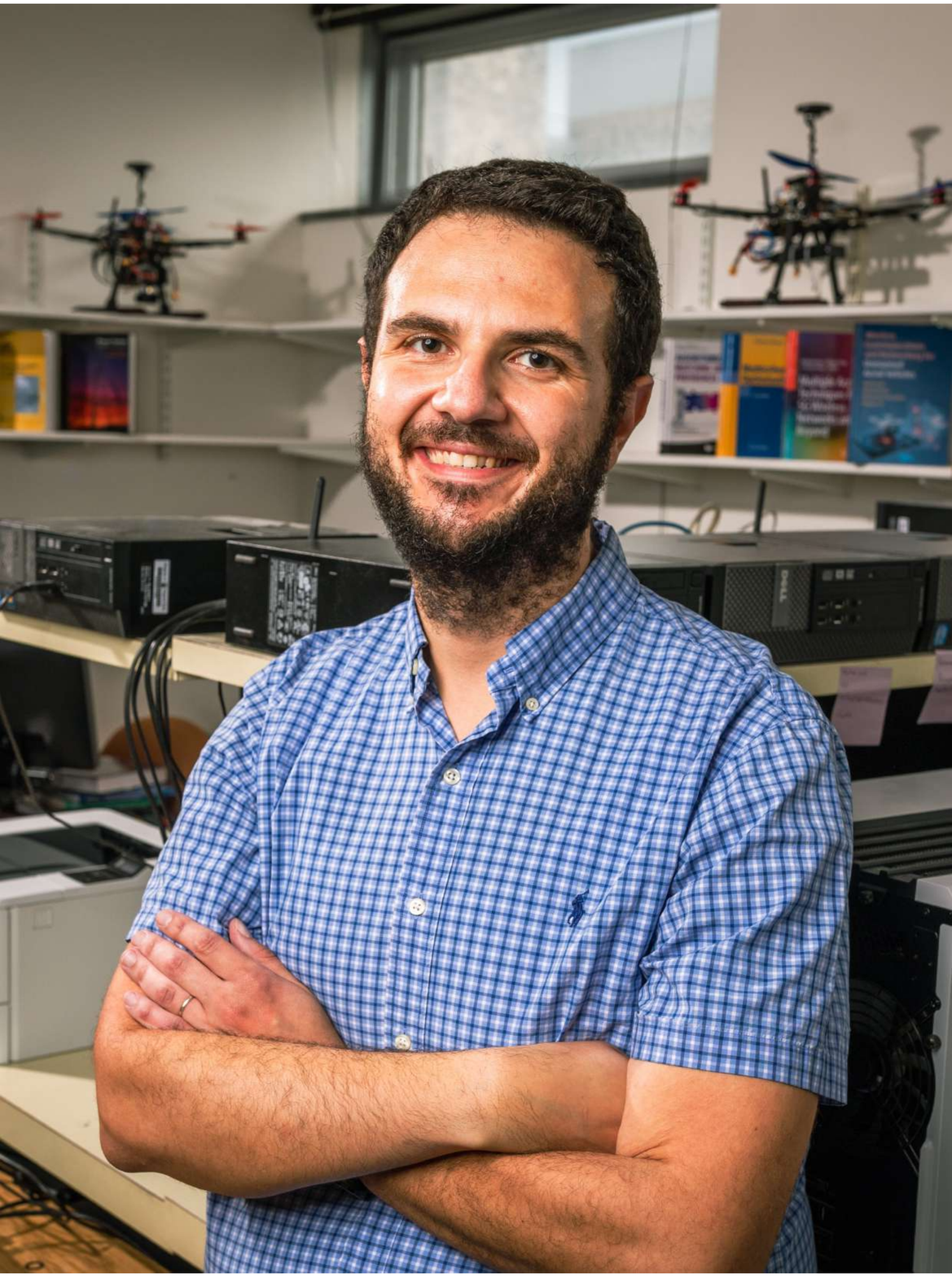}}]{Walid Saad} (Fellow, IEEE) received his Ph.D degree from the University of Oslo, Norway in 2010. He is currently a Professor at the Department of Electrical and Computer Engineering at Virginia Tech, where he leads the Network sciEnce, Wireless, and Security (NEWS) laboratory. He is also with Cyber Security Systems and Applied AI Research Center, Lebanese American University, Beirut, LebanonHe is also the Next-G Wireless Faculty Lead at Virginia Tech's Innovation Campus. His research interests include wireless networks (5G/6G/beyond), machine learning, game theory, security, UAVs, semantic communications, cyber-physical systems, and network science. Dr. Saad is a Fellow of the IEEE. He is also the recipient of the NSF CAREER award in 2013, the AFOSR summer faculty fellowship in 2014, and the Young Investigator Award from the Office of Naval Research (ONR) in 2015. He was the (co-)author of eleven conference best paper awards at IEEE WiOpt in 2009, ICIMP in 2010, IEEE WCNC in 2012, IEEE PIMRC in 2015, IEEE SmartGridComm in 2015, EuCNC in 2017, IEEE GLOBECOM (2018 and 2020), IFIP NTMS in 2019, IEEE ICC (2020 and 2022). He is the recipient of the 2015 and 2022 Fred W. Ellersick Prize from the IEEE Communications Society, of the IEEE Communications Society Marconi Prize Award in 2023, and of the IEEE Communications Society Award for Advances in Communication in 2023. He was also a co-author of the papers that received the IEEE Communications Society Young Author Best Paper award in 2019, 2021, and 2023. Other recognitions include the 2017 IEEE ComSoc Best Young Professional in Academia award, the 2018 IEEE ComSoc Radio Communications Committee Early Achievement Award, and the 2019 IEEE ComSoc Communication Theory Technical Committee Early Achievement Award. From 2015-2017, Dr. Saad was named the Stephen O. Lane Junior Faculty Fellow at Virginia Tech and, in 2017, he was named College of Engineering Faculty Fellow. He received the Dean's award for Research Excellence from Virginia Tech in 2019. He was also an IEEE Distinguished Lecturer in 2019-2020. He has been annually listed in the Clarivate Web of Science Highly Cited Researcher List since 2019. He currently serves as an Area Editor for the IEEE Transactions on Network Science and Engineering and the IEEE Transactions on Communications. He is the Editor-in-Chief for the IEEE Transactions on Machine Learning in  Communications and Networking.

\end{IEEEbiography}

\begin{IEEEbiography}[{\includegraphics[width=1.1in,height=1.3in,clip,keepaspectratio]{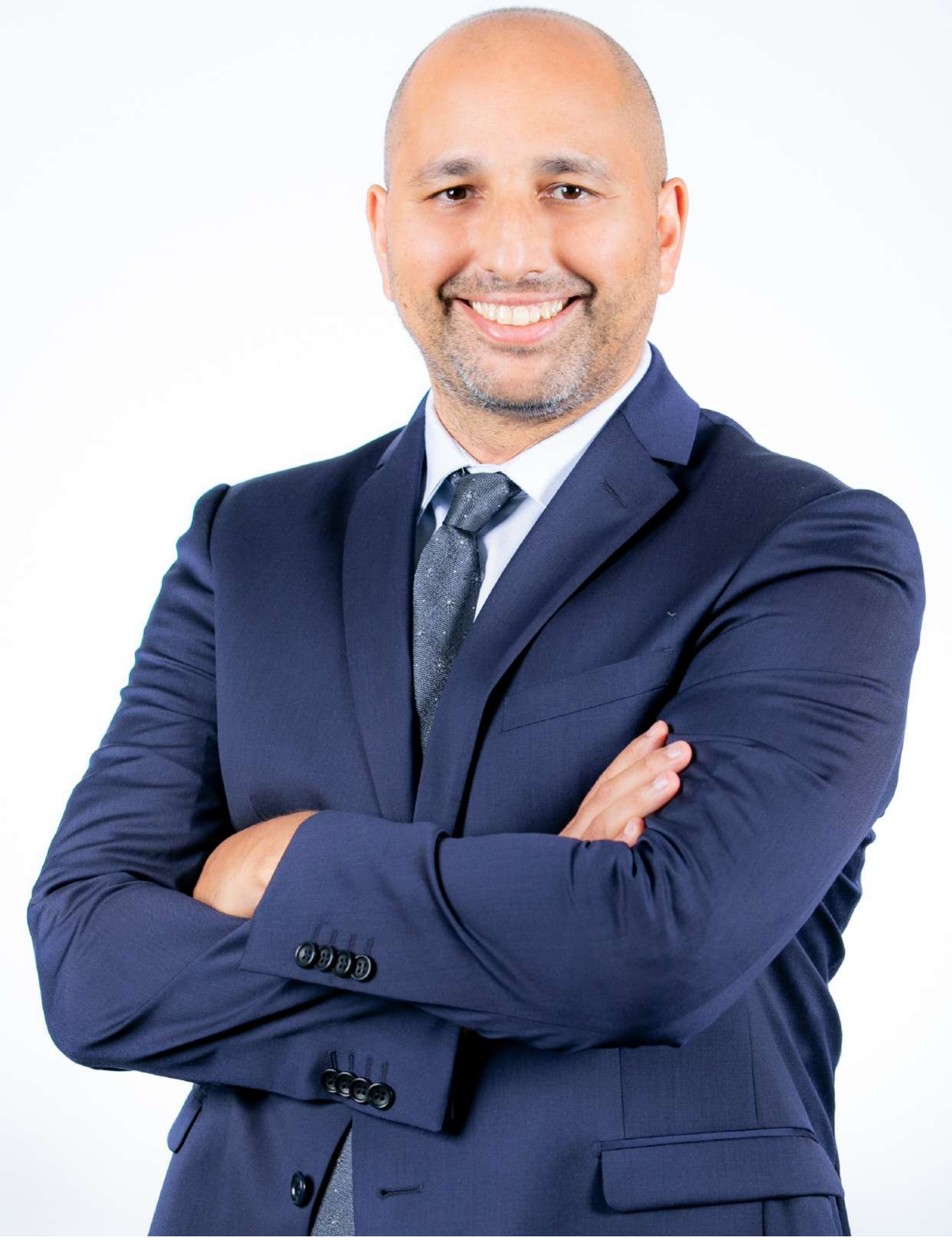}}]{M\'erouane Debbah} (Fellow, IEEE) is Professor at Khalifa University of Science and Technology in Abu Dhabi. He received the M.Sc. and Ph.D. degrees from the Ecole Normale Supérieure Paris-Saclay, France. He was with Motorola Labs, Saclay, France, from 1999 to 2002, and then with the Vienna Research Center for Telecommunications, Vienna, Austria, until 2003. From 2003 to 2007, he was an Assistant Professor with the Mobile Communications Department, Institut Eurecom, Sophia Antipolis, France. Since 2007, he is Full Professor at CentraleSupelec, Gif-sur-Yvette, France. From 2007 to 2014, he was the Director of the Alcatel-Lucent Chair on Flexible Radio. From 2014 to 2021, he was Vice-President of the Huawei France Research Center. He was jointly the director of the Mathematical and Algorithmic Sciences Lab as well as the director of the Lagrange Mathematical and Computing Research Center. From 2021 to 2023,  he was Chief Researcher at the Technology Innovation Institute  and  leading the AI \& Digital Science Research centers at the Technology Innovation Institute.  He was also Adjunct Professor with the Department of Machine Learning at the Mohamed Bin Zayed University of Artificial Intelligence in Abu Dhabi.  Since 2023, he is a  Professor at  Khalifa University of Science and Technology in Abu Dhabi and founding director of the 6G center. He has managed 8 EU projects and more than 24 national and international projects. His research interests lie in fundamental mathematics, algorithms, statistics, information, and communication sciences research. He holds more than 50 patents. He is an IEEE Fellow, a WWRF Fellow, a Eurasip Fellow, an AAIA Fellow, an Institut Louis Bachelier Fellow and a Membre émérite SEE. He was a recipient of the ERC Grant MORE (Advanced Mathematical Tools for Complex Network Engineering) from 2012 to 2017. He was a recipient of the Mario Boella Award in 2005, the IEEE Glavieux Prize Award in 2011, the Qualcomm Innovation Prize Award in 2012, the 2019 IEEE Radio Communications Committee Technical Recognition Award and the 2020 SEE Blondel Medal. He received more than 30 best paper awards, among which the 2007 IEEE GLOBECOM Best Paper Award, the Wi-Opt 2009 Best Paper Award, the 2010 Newcom++ Best Paper Award, the WUN CogCom Best Paper 2012 and 2013 Award, the 2014 WCNC Best Paper Award, the 2015 ICC Best Paper Award, the 2015 IEEE Communications Society Leonard G. Abraham Prize, the 2015 IEEE Communications Society Fred W. Ellersick Prize, the 2016 IEEE Communications Society Best Tutorial Paper Award, the 2016 European Wireless Best Paper Award, the 2017 Eurasip Best Paper Award, the 2018 IEEE Marconi Prize Paper Award, the 2019 IEEE Communications Society Young Author Best Paper Award, the 2021 Eurasip Best Paper Award, the 2021 IEEE Marconi Prize Paper Award, the 2022 IEEE Communications Society Outstanding Paper Award, the 2022  ICC Best paper Award, the 2022 IEEE GLOBECOM Best Paper Award, 2022 IEEE TAOS TC Best GCSN Paper Award, the  2022 IEEE International Conference on Metaverse Best Paper Award, the 2023 IEEE Communications Society Fred W. Ellersick Prize, the 2023 ICC best paper award  as well as the Valuetools 2007, Valuetools 2008, CrownCom 2009, Valuetools 2012, SAM 2014, and 2017 IEEE Sweden VT-COM-IT Joint Chapter best student paper awards. He is an Associate Editor-in-Chief of the journal Random Matrix: Theory and Applications. He was an Associate Area Editor and Senior Area Editor of the IEEE TRANSACTIONS ON SIGNAL PROCESSING from 2011 to 2013 and from 2013 to 2014, respectively. From 2021 to 2022, he served as an IEEE Signal Processing Society Distinguished Industry Speaker. 
\end{IEEEbiography}

\begin{IEEEbiography}[{\includegraphics[width=1.1in,height=1.3in,clip,keepaspectratio]{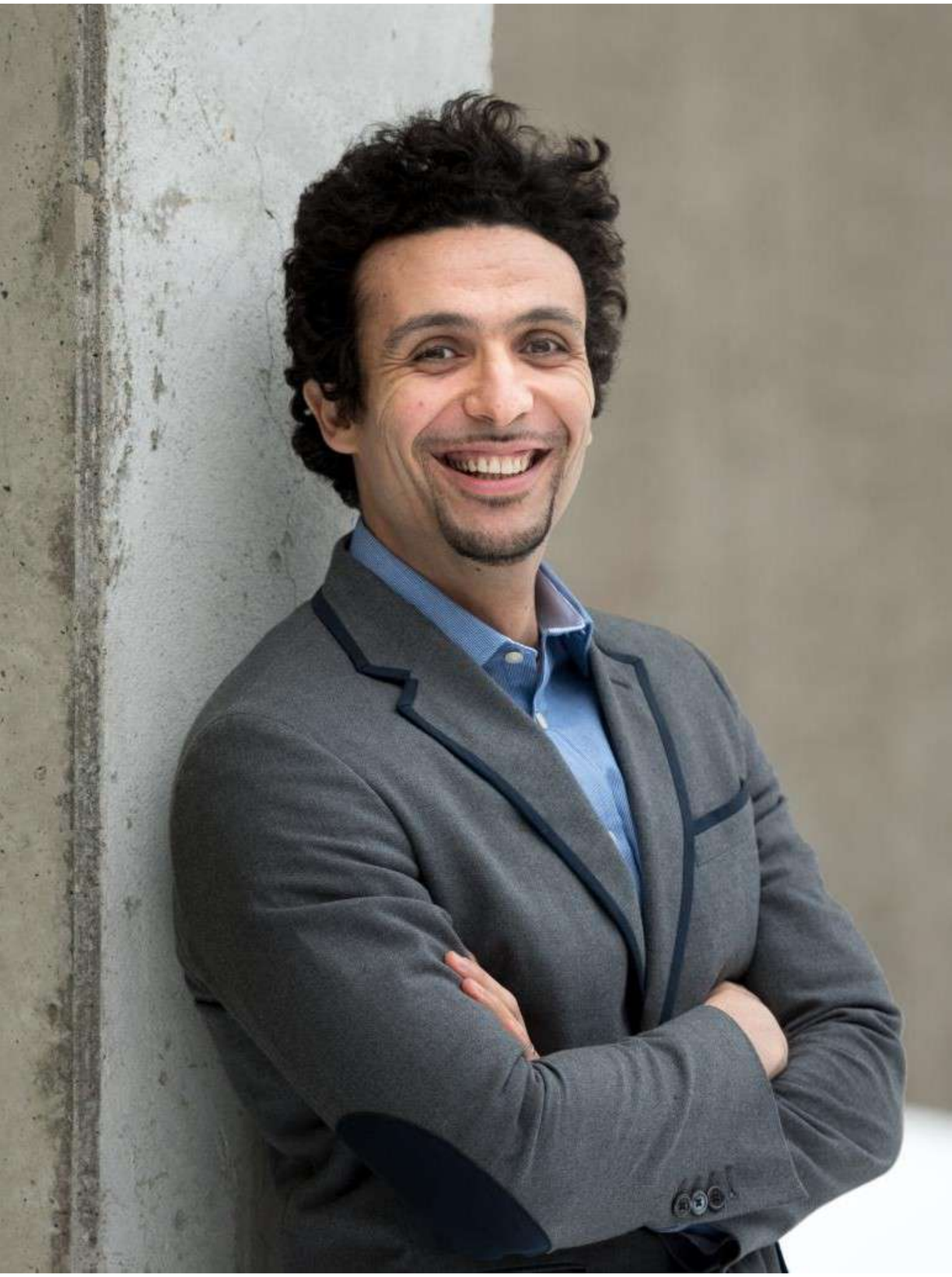}}]{Dr Mehdi Bennis} (Fellow, IEEE) is a full (tenured) Professor at the Centre for Wireless Communications, University of Oulu, Finland and head of the intelligent connectivity and networks/systems group (ICON). His main research interests are in radio resource management, game theory and distributed AI in 5G/6G networks. He has published more than 200 research papers in international conferences, journals and book chapters. He has been the recipient of several prestigious awards including the 2015 Fred W. Ellersick Prize from the IEEE Communications Society, the 2016 Best Tutorial Prize from the IEEE Communications Society, the 2017 EURASIP Best paper Award for the Journal of Wireless Communications and Networks, the all-University of Oulu award for research, the 2019 IEEE ComSoc Radio Communications Committee Early Achievement Award and the 2020 Clarviate Highly Cited Researcher by the Web of Science. Dr Bennis is an editor of IEEE TCOM and Specialty Chief Editor for Data Science for Communications in the Frontiers in Communications and Networks journal.
\end{IEEEbiography}

\end{document}